\documentclass[%
reprint,
showpacs,
showkeys,
amsmath,amssymb,
aps,
prd,
]{revtex4-1}

\usepackage{graphicx}
\usepackage{pgffor}
\usepackage{capt-of}
\usepackage{soul}
\usepackage{lipsum}

\newcommand{\diff}{\mathrm{d}}
\def\beq{\begin{equation}}
\def\eeq{\end{equation}}
\def\bea{\begin{eqnarray}}
\def\eea{\end{eqnarray}}

\let\phi=\varphi

\let\phi=\varphi
\let\rho=\varrho

\newcommand{\uvozovky}[1]{``#1''}

\begin{document}

\frenchspacing

\author{Zden\v{e}k Stuchl\'{\i}k}
\email{zdenek.stuchlik@fpf.slu.cz}
\author{Martin Blaschke}
\email{martin.blaschke@fpf.slu.cz}
\author{Jan Schee}
\email{jan.schee@fpf.slu.cz}

\affiliation{%
Institute of Physics and Research Centre of Theoretical Physics and Astrophysics,
Silesian University in Opava,\\
Bezru\v{c}ovo n\'am.~13, CZ-746\,01 Opava, Czech Republic%
}

\title[Collisions in Kerr--Newman spacetimes] {Particle collisions and optical effects in the mining Kerr--Newman spacetimes}

\begin{abstract}
We study ultra-high-energy particle collisions and optical effects in the extraordinary class of mining braneworld Kerr-Newman (KN) naked singularity spacetimes, predicting extremely high efficiency of Keplerian accretion, and compare the results to those related to the other classes of the KN naked singularity and black hole spacetimes. We demonstrate that in the mining KN spacetimes the ultra-high centre-of-mass energy occurs for collisions of particles following the extremely-low-energy stable circular geodesics of the \uvozovky{mining regime}, colliding with large family of incoming particles, e.g., those infalling from the marginally stable counter-rotating circular geodesics. This is qualitatively different situation in comparison to the standard KN naked singularity or black hole spacetimes where the collisional ultra-high centre-of-mass energy can be obtained only in the near-extreme spacetimes. We also show that observers following the stable circular geodesics of the mining regime can register extremely blue-shifted radiation incoming from the Universe, and see strongly deformed sky due to highly relativistic motion along such stable orbits. The strongly blue-shifted radiation could be thus a significant source of energy for such orbiting observers.
\end{abstract}


\maketitle

\section{Introduction}
The~higher-dimensional String Theory and particularly M-theory \cite{Hor:Wit:1996b:,Hor:Wit:1996:}, becoming effectively 4D at low enough energies, inspired the~so called braneworld models with the~observable universe being a~3-brane on which the~standard particle-model fields are confined, while gravity enters the~extra spatial dimensions \cite{Ark:Dim:Dva:1998:}. In the~braneworld models, gravity can be localized near the~3D brane in the~bulk space with a~non-compact, infinite size extra dimension with the~warped spacetime satisfying the~5D Einstein equations \cite{Ran:Sun:1999b:} with the~non-compact dimension assumed to be related to the~M-theory. 

When the~5D Einstein equations at the~bulk space are constrained to the 3D brane, modified 4D Einstein equations arise \cite{Shi:Mae:Sas:2000:}, solution of which is quite complex in the~presence of the~matter stress energy tensor, e.g., in the~case of models of neutron stars \cite{Ger:Maa:2001:,Hla-Stu:2011:JCAP:,Stu-Hla-Urb:2012:GRG:}, but it is simple for the vacuum solutions related to the braneworld black holes and naked singularities. The~spherically symmetric and static braneworld black holes can be described by the Reissner--Nordstr{\" o}m geometry (RN) \cite{Dad:2000:}, the axially symmetric and stationary braneworld rotating black holes can be described by the Kerr--Newman (KN) geometry \cite{Ali-Gum:2005:CLAQG:}, where the influence due to the tidal effects from the bulk is simply represented by a single parameter called tidal charge due to the~similarity of the~effective stress-energy tensor of the~tidal effects of the~bulk space and the~stress-energy tensor of the~electromagnetic field \cite{Dad:2000:}. The~tidal charge parameter can be both positive and negative \cite{Dad:2000:,Ali-Gum:2005:CLAQG:}, while in the~standard general relativity the~square of the~electric charge occurs giving thus only positively valued parameter. 

There exist many studies of the RN or KN black hole and naked-singularity geodesic motion \cite{deF:1974:aap,Stu:1980:BAC:,Stu:1981:,Stu:Hle:Tru:2011:,Stu-Sche:2012:CLAQG:,Stu:1981:,Stu:Cal:1991:,Stu-etal:1998:PHYSR4:,Stu:Hle:2000:,Stu-Hle:2002:ActaPhysSlov:,Pug-Que-Ruf:2011:PHYSR4:,Pug:Que:Ruf:2013:} that can be straightforwardly applied for the~braneworld black holes and naked singularities with positive tidal charge. Moreover, the~astrophysically relevant studies of the braneworld black holes (with both positive and negative tidal charges) were presented in a~number of papers related to the~optical effects \cite{Sche:Stu:2009:,Sche:Stu:2009b:,Nun:2010:PHYSR4:,Eir-Ama:2012:PHYSR4:}, or the~accretion phenomena \cite{Boh-Har-Lob:2008:CLAQG:,Kot-Stu-Tor:2008:CLAQG:,Stu-Kot:2009:GRG:,Ali-Tal:2009:prd:}. The astrophysical relevance of the KN naked singularity spacetimes can be expected in connection to the so called superspinars -- the Kerr (and KN) superspinars were proposed as possible remnants of the early phases of the existence of the Universe governed by the String theory \cite{Gim-Hor:2009:PhysLetB:}. Exterior of the Kerr (and KN) superspinars is assumed to be described by the Kerr (and KN) naked singularity geometry, their interior is assumed to be described by a solution of the String theory, while the matching surface has to be located above the region of causality violation of the external spacetime; the Kerr superspinars can be stable against perturbations \cite{Gim-Hor:2009:PhysLetB:,Stu-Sche:2010:CLAQG:,Nak-etal:2017:}. 

Quite recently, we have studied the Keplerian accretion and its efficiency for whole variety of the braneworld KN black hole and naked singularity spacetimes \cite{Bla-Stu:2016:PHYSR4:}. Surprisingly, we have found an extraordinary class of the braneworld KN naked singularity spacetimes, having the~braneworld dimensionless tidal charge $b \in (1/4,1)$ and the~dimensionless spin $a \in \left(2\sqrt{b}-\sqrt{b(4b-1)},2\sqrt{b}+\sqrt{b(4b-1)}\right)\,$, that demonstrate existence of an infinitely deep gravitational well centered at the stable photon circular orbit. Such spacetimes enable mining of unlimited energy due to the~Keplerian accretion, therefore, we call them mining KN naked singularity spacetimes. Of course, the mining KN naked singularity spacetimes have to be unstable with respect to the \uvozovky{mining} Keplerian accretion; the \uvozovky{perpetuum-mobile} character of the mining process has to be stopped when the assumption of the test particle motion of the accreting matter is violated due to the increasing (binding) energy of the accreting matter \cite{Bla-Stu:2016:PHYSR4:}. Note that the same conclusions as for the braneworld mining spacetimes hold for the standard KN spacetimes containing an electric charge $Q$; all the results can be obtained with simple transition $b \to Q^2$. 

In the present paper, we study two astrophysically important phenomena for particles (observers) following in the braneworld mining KN spacetimes the circular geodesics of the mining regime (i.e., near the stable photon circular orbit). First, we study the so called Banados, Silk and West (BSW) collisional processes demonstrated by an extremely large Centre-of-Mass (CM) energy \cite{Ban-Sil-Wes:2009:}. We demonstrate that the BSW effect is allowed for all members of the family of mining KN naked singularities, if one of the colliding particles is orbiting very close to the stable photon circular geodesic, contrary to the case of the KN black holes or ordinary naked singularities where the BSW effect is efficient only for the extreme or near-extreme spacetimes. We compare the obtained results to those related to ordinary braneworld KN black hole and naked singularity spacetimes, with both the positive and negative tidal charge parameter $b$. We thus study the collisional processes in the case of the ordinary naked singularities for the known model of collisions at (and near) the special radius of $r=M$ when ultra-high CM energy can be obtained for all the colliding particles that could reach this radius from large distances \cite{Stu:Hle:Tru:2011:,Stu-Sche:2012:CLAQG:,Stu-Sche:2013:CLAQG:}; we study in detail the simplest situation of collisions of \uvozovky{radially} incoming and outgoing (after reaching a turning point of the radial motion) particles, discussing specially the case that was not treated in previous studies, namely, dependence of the CM energy on the distance from which the colliding particles start to fall. For these purposes, we present also a detailed study of the equatorial radial motion of particles with vanishing angular momentum in the braneworld KN spacetimes. For the braneworld KN black hole spacetimes the collisional processes were treated in \cite{Sha-etal:2013:}; we do not repeat this discussion, making only short comparison to the naked singularity cases. 

Second, we study simple optical effects related to frequency shift of radiation emitted (or received) by the observers orbiting the KN naked singularities, especially in the mining regime, and appearance of the sky for observers orbiting the braneworld KN naked singularities. We give dependence of the frequency shift of the principal null congruence (PNC) photons (for physical importance of this special photon family see, e.g., \cite{Bic-Stu:1976:BAC:}) radiated from the stable circular geodesics, as this simple case can give fundamental knowledge on the intensity of optical effects in the deep gravitational field of naked singularities (black holes) \cite{Stu:1980:BAC:}. For observers following the special family of the \uvozovky{mining} orbits we show that blueshift of the cosmic microwave background (CMB) radiation can be extremely large, enabling for such observers to obtain sufficient energy supply related to the blue-shifted CMB radiation, implying thus possibility of a new variant of intelligent life survival in the cold expanding Universe, which could represent an additional possibility to the known mechanism related to the observers orbiting Kerr black holes \cite{Tho-etal:2015:,Tho,O}. These simple illustrative calculations are complemented by detailed modelling of the sky (related to the CMB radiation) as seen by observers following the circular geodesics in the deepest parts of the gravitational well in vicinity of the innermost stable circular geodesics of KN spacetimes.

\section{Braneworld Kerr--Newman geometry}

In the Boyer-Lindquist coordinates $(t,r,\theta,\varphi)$ and the geometric units $(c=G=1)$, the~line element of a braneworld Kerr-Newman black hole or naked singularity, representing solution of the Einstein equations constrained to the 3D-brane, reads \cite{Ali-Gum:2005:CLAQG:,Dad:2000:} 

\begin{widetext}
\begin{eqnarray}
\mkern-106mu
\mathrm{d}s^2= - \left(1-\frac{2Mr - b}{\Sigma }\right)\mathrm{d}t^2 &-& \frac{2a(2Mr - b)}{\Sigma}\,\mathrm{sin}^2\theta\,\mathrm{d}t\,\mathrm{d}\phi + \frac{\Sigma}{\Delta}\,\mathrm{d}r^2
\ \nonumber \\
 &+& \Sigma\, \mathrm{d}\theta^2 + \left(r^2 + a^2 + \frac{2Mr - b}
{\Sigma}\,a^2\mathrm{sin}^2\theta\right)
\mathrm{sin}^2\theta\, \mathrm{d}\phi^2\, , \label{Metrika}
\end{eqnarray}
\end{widetext}
with 
\begin{eqnarray}
\Delta = r^2 -2Mr +a^2 + b\, ,\\ 
\Sigma = r^2 + a^2\mathrm{cos}^2\theta\, ,
\end{eqnarray}

where $M$ is the mass parameter of the spacetime, $a=J/M$ is the~specific angular momentum of the spacetime with internal angular momentum $J$, and the~braneworld tidal charge parameter $b$ represents imprint of the non-local, tidal, gravitational effects of the~bulk space \cite{Ali-Gum:2005:CLAQG:}. 

The~form of the~metric (\ref{Metrika}) is identical to that of the~standard Kerr--Newman solution of the~4D Einstein--Maxwell equations, with squared electric charge $Q^2$ being replaced by the~tidal charge $b$ \cite{Gravitation}. 
We can separate three cases: 
\begin{itemize}
\item[a) ] $b=0$ corresponding to the standard Kerr metric.  
\item[b) ] $b>0$ corresponding the standard KN metric. 
\item[c) ] $b<0$ corresponding to the non-standard KN metric with negative tidal effects. 
\end{itemize}
It should be stressed that in the~braneworld Kerr--Newman spacetimes the~geodesic structure is relevant also for the~motion of electrically charged particles, as there is no electromagnetic field related to the tidal charge of the spacetimes. Of course, the case (b) can be equally considered for the~analysis of the~uncharged particle motion in the~standard electrically charged Kerr--Newman spacetimes. 

We put in the following $M=1$ for simplicity. The~spacetime parameters $a$ and $b$, and the time $t$ and radial $r$ coordinates become then dimensionless. Equivalently, we express the relevant quantities in units of $M$ making redefinitions: $a/M \to a$, $b/M \to b$, $t/M \to t$ and $r/M \to r$. 

Separation between the~black hole and naked singularity spacetimes is given by the~relation 
\beq
     a^2 + b = 1 
\eeq
determining the extreme black holes with coinciding horizons. The~condition $0 < a^2+b < 1$ governs the black hole spacetimes with two distinct event horizons, the~condition $a^2+b < 0$ governs black hole spacetimes with only one distinct event horizon at $r>0$. For $a^2+b > 1$, the~KN spacetimes describe naked singularities. 

For positive tidal charges, the~black hole spin has to satisfy the relation $a^2<1$, as in the~standard Kerr--Newman spacetimes, but for negative tidal charges the black holes can violate the~well know Kerr limit, having $a^2>1$ \cite{Sche:Stu:2009:,Stu-Kot:2009:GRG:}. The~physical \uvozovky{ring} singularity of the~braneworld rotating black holes and naked singularities is located at $r=0$ and $\theta = \pi/2$, as in the~Kerr spacetimes. Behavior of the Kretschman scalar around the ring singularity, properties of the ergosphere (ergoregion in the naked singularity spacetimes) and the causality violation region of the braneworld KN spacetimes have been discussed in \cite{Bla-Stu:2016:PHYSR4:}. In the following, we shall consider the particle and photon motion above the event horizon in the black hole cases, and above the causality violation region in the naked singularity cases. 

It is convenient to express physical quantities in local reference frames. To describe the physical processes in the rotating Kerr--Newman spacetimes, the family of locally non-rotating frames (LNRF) corresponding to zero angular momentum observers (ZAMO) is most convenient \cite{Bar-Pre-Teu:1972:ApJ:}. The vectors of the LNRF tetrad are given by the relations 
\beq\label{LNRF}
\mathrm{\bf e}^{(t)} = \left ( \omega^2g_{\phi\phi} - g_{tt}\right ) ^{\frac{1}{2}}{\bf d}t\ ,
\end{equation}
\begin{equation}
\mathrm{\bf e}^{(\phi)} = (g_{\phi\phi})^{\frac{1}{2}}({\bf d}\phi - \omega{\bf d}t)\, ,
\end{equation}
\begin{equation}
\mathrm{\bf e}^{(r)} = \left(\frac{\Sigma }{\Delta }\right)^{\frac{1}{2}}{\bf d}r\ ,
\end{equation}
\beq
\mathrm{\bf e}^{(\theta)} = \Sigma^{\frac{1}{2}}{\bf d}\theta\ ,
\eeq
where the~angular velocity of the~LNRF relative to distant observers $\omega$ reads
\beq\label{omega}
\omega =-\frac{g_{t\phi}}{g_{\phi\phi}}= \frac{a(2r-b)}{\Sigma(r^2 + a^2) + (2r - b)\,a^2\,\mathrm{sin^2\theta}}\, .
\eeq
Convenience of the LNRF is clearly demonstrated for the particles freely falling from rest at infinity that is purely radial namely in the family of LNRFs \cite{Stu-Bic-Bal:1999:GRG:}. 

\subsection{Geodesics and Carter's equations}

In the braneworld Kerr spacetimes, the separated first order differential equations of the~geodesic motion take the form \cite{Car:1968:PRD:,Car:1973:BlaHol:} 

\begin{eqnarray}
 	\Sigma\frac{\diff r}{\diff w}&=&\pm\sqrt{R(r)},\label{ce7}\\
 	\Sigma\frac{\diff \theta}{\diff w}&=&\pm\sqrt{W(\theta)},\label{ce8}\\
 	\Sigma\frac{\diff \varphi}{\diff w}&=&-\frac{P_W}{\sin^2\theta}+\frac{a P_R}{\Delta},\label{ce9}\\
 	\Sigma\frac{\diff t}{\diff w}&=&-a P_W + \frac{(r^2+a^2)P_R}{\Delta},\label{ce10}
\end{eqnarray}
where 
\begin{eqnarray}
  R(r)&=&P^2_R-\Delta(m^2 r^2 + \tilde{K}),\label{ce11}\\
  W(\theta)&=&(\tilde{K}-a^2m^2\cos^2\theta)-\left(\frac{P_w}{\sin\theta}\right)^2,\label{ce12}\\
  P_R(r)&=&\tilde{E}(r^2+a^2)-a\tilde{\Phi},\label{ce13}\\
  P_W(\theta)&=&a\tilde{E}\sin^2\theta - \tilde{\Phi}.\label{ce14}
\end{eqnarray}
The so called Carter equations presented above contain four constants of motion: the rest energy $m$, the~energy (related to the time Killing vector field) $\tilde{E}$, the~axial angular momentum (related to the axial Killing vector field) $\tilde{\Phi}$, and the~constant of motion connected to the total angular momentum (related to the Killing tensor field) $\tilde{K}$ that is usually replaced by the~constant $\tilde{Q}=\tilde{K}-(a\tilde{E}-\tilde{\Phi)}^2$, since for the motion in the~equatorial plane ($\theta=\pi/2$) there is $\tilde{Q}=0$. These equations can be integrated and expressed in terms of the elliptic integrals \cite{Gravitation,Kra:2005:CLAQG:,Kra:2007:CLAQG:}. The Carter equations have been generalized to the motion in the Kerr--Newman-de Sitter spacetimes \cite{Car:1973:BlaHol:,Stu:1983:BAC:,Stu-Hle:1999:PHYSR4:,Stu:Hle:2000:,Kra:2005:CLAQG:,Kra:2014:GRG:}. 

For the geodesic motion of photons, $m=0$ in the Carter equations. In the standard KN spacetimes, analysis of the photon motion has been presented in \cite{Stu:1981b:BAC,Bal-Bic-Stu:1989:BAC:,Pug:Que:Ruf:2013:} and it can be directly applied to the case of the braneworld Kerr-Newman spacetimes with positive tidal charge. Extension to the KN spacetimes with negative tidal charges can be found in \cite{Sche:Stu:2009b:,Sche:Stu:2009:}. 

Analysis of the test particle geodesic motion in the KN spacetimes can be found in \cite{Pug-Que-Ruf:2011:PHYSR4:}. Detailed classification of the braneworld KN spacetimes according to the properties of the circular geodesics related to the Keplerian accretion has been presented quite recently in \cite{Bla-Stu:2016:PHYSR4:}. 

\section{Circular geodesics}

In the Kerr and KN spacetimes, the circular geodesic motion is possible in the equatorial plane only \cite{Bar-Pre-Teu:1972:ApJ:}. Except the~rest energy $m$, two integrals of the motion are relevant as $\tilde{Q}=0$: 
\begin{equation}
U_t = -\,E\, , \; U_\phi =\,L\, ,
\end{equation}
where $U_\alpha = g_{\alpha\nu} dx^\nu/d\tau\,$ denotes the particle covariant 4-velocity, with $\tau $ being the~affine parameter. The motion constant $E=\tilde{E}/m$ is identified as the specific energy at infinity, i.e., energy related to the~rest energy, and the motion constant $L=\tilde{\Phi}/m$ as the specific angular momentum at infinity. 

For circular geodesics, the conditions  
\begin{equation}\label{podminky}
R(r)=0\,\quad \mathrm{and}\,\quad \partial_r R(r)=0\, 
\end{equation}
have to be satisfied simultaneously. For the~spacetime line element of the~braneworld KN spacetimes given by (\ref{Metrika}) \cite{Ali-Gum:2005:CLAQG:,Dad-Kal:1977:,Stu-Kot:2009:GRG:}, with the assumption of $M=1$ and  $a>0$, we obtain the~radial profiles of the specific energy $E$, specific axial angular momentum $L$, and the angular velocity related to infinity $\Omega$ in the~form: 
\begin{eqnarray}
E &=& \frac{r^2-2r +b \pm a\sqrt{r-b}}{r\sqrt{r^2-3r +2b\pm 2a\sqrt{r-b}}}\, ,\label{E} \\
L &=& \pm\frac{\sqrt{r-b}\left(r^2+a^2 \mp 2a\sqrt{r-b}\right)\mp ab}{r\sqrt{r^2-3r +2b\pm 2a\sqrt{r-b}}}\, ,\label{L} \\
\Omega  &=& \pm \frac{1}{\frac{r^2}{\sqrt{r-b}}\pm a}\, ,\label{Om}
\end{eqnarray}
where the upper and lower signs refer to two families of solutions. We refer to these two families as the upper sign family, and the lower sign family \cite{Bla-Stu:2016:PHYSR4:}. At large distances, the upper family orbits are corotating, while the lower family orbits are counterrotating with respect to rotation of the spacetime. This separation is valid in the~whole region above the event horizon of the KN black hole spacetimes, but in some of the KN naked singularity spacetimes the upper family orbits become counterrotating close to the naked singularity as demonstrated in \cite{Stu:1980:BAC:,Bla-Stu:2016:PHYSR4:}. 

Equations (\ref{E})--(\ref{Om}) imply two restrictions on the existence of circular geodesics: 
\begin{eqnarray}
r^2-3 r +2b\pm 2a\sqrt{r-b} \geq 0\, ,\\
r\geq b\, .
\end{eqnarray}
The~first condition determines in the equality limit the~photon circular geodesics -- positions of circular orbits of test particles are limited by the circular geodesics of massless particles. The~second condition is relevant in the~KN spacetimes with positive tidal charge $b$ only, if we restrict attention to the~region of positive radii. In some of the KN spacetimes its equality limit determines the innermost circular orbits. 

The specific energy of particles following the circular geodesics related to the LNRF ($E_{\mathrm{LNRF}}$) is given by the projection of the 4-velocity on the timelike vector of the~frame: 
\begin{eqnarray}\label{e25}
& &E_{\mathrm{LNRF}}=U^{(t)} = U^\mu\mathbf{e}^{(t)}_{\mu} = \left(\frac{dt}{d\tau}\right) \mathbf{e}^{(t)}_{t}  \\ \nonumber
&=& \frac{r^2\pm a\sqrt{r-b}}{\sqrt{r^4+a^2\left(r^2+2r-b\right)}}\frac{\sqrt{\Delta}}{\sqrt{r^2-3r+2b\pm 2a\sqrt{r-b}}}\, .
\end{eqnarray}
The~locally measured particle energy must be always positive for the~particles in the positive-root states assumed here. The locally measured energy is negative for the particles in the negative-root states that are physically irrelevant in the context of our study \cite{Bic-Stu-Bal:1989:BAC:}. For particles in the~positive-root states the~time evolution vector is oriented to future, i.e., $\mathrm{d}t/\mathrm{d}\tau > 0$, while particles in the negative-root states have past oriented time vectors, $\mathrm{d}t/\mathrm{d}\tau < 0$.

In Eq. (\ref{e25}), the +(-) sign corresponds to the upper (lower) family orbits. It is immediately clear from Eq. (\ref{e25}) that for the upper sign family orbits, demonstrating the mining effect, there is always $E_{\mathrm{LNRF}} > 0$ as $r^2+a \sqrt{r-b}>0\, .$ For the lower family orbits, we can also show that in the range of validity, above the photon circular orbit $(r>r_{\mathrm{ph-}})$, there is also $r^2-a\sqrt{r-b}>0$ and $L_{\mathrm{LNRF}}>0\, .$

In fact, the LNRF energy of particles following the circular geodesics diverges on the~photon circular orbits. For more details see \cite{Bla-Stu:2016:PHYSR4:}. 

\subsection{Photon circular geodesics and marginally stable circular geodesics}

Loci of the photon circular geodesics in the Kerr and KN spacetimes are determined by the relation \cite{Bar:1973:BlaHol:,Bal-Bic-Stu:1989:BAC:,Sche:Stu:2009b:}  
\begin{equation}\label{jedno}
r^2-3r+2b\pm 2a\sqrt{r-b}=0\, 
\end{equation}
that implies the same reality condition on the~radius of the photon circular orbit $r_{\mathrm{ph}}$ as the~one that follows from equations (\ref{E})-(\ref{Om}): 
\begin{equation}
r_{\mathrm{ph}}\geq b\, .
\end{equation} 
The solution of the~equation (\ref{jedno}) can be expressed in the~form 
\begin{equation}
a = a_{\mathrm{ph}}(r,b) \equiv \pm \frac{\left(3r-r^2-2b\right)}{2\sqrt{r-b}}\, .
\end{equation}

For given $a>0$ and given $b$, the~function $a_{\mathrm{ph}}(r,b)$ determines radius of both the corotating and counterrotating photon circular orbits. The zeros of the function $a_{\mathrm{ph}}(r,b)$ are located at radii giving photon circular orbits in the Reissner-Nordstron spacetimes \cite{Stu-Hle:2002:ActaPhysSlov:,Pug-Que-Ruf:2011:PHYSR4:} 
\begin{equation}
r_{\mathrm{ph}\pm}=\frac{1}{2}\left(3\pm\sqrt{9-8b}\right)\, .
\end{equation}
The local extrema of the function $a_{\mathrm{ph}}(r,b)$ are located at $r=1$ and at $r=4b/3 \,$. They correspond to the extreme KN black holes when 
\begin{equation}\label{aphbb}
a_{\mathrm{ph-e}}(r=1,b)=\sqrt{1-b}\, ,
\end{equation} 
and to the KN spacetimes satisfying the condition 
\begin{equation}\label{aphbb2}
a_{\mathrm{ph-ex}}(r=4b/3,b) = \pm\frac{\sqrt{b}}{3\sqrt{3}}\left(8b-9\right)\, .
\end{equation} 
Detailed discussion of the photon circular geodesics in the braneworld KN spacetimes can be found in \cite{Bla-Stu:2016:PHYSR4:}. 

Loci of the~stable circular geodesics are governed by the~condition 
\begin{equation}\label{R2}
\frac{\partial^2 R(r,a,b,E,L)}{\partial r ^2}\leq 0\, . 
\end{equation} 
For the marginally stable circular orbits the equality must hold along with the relations for the energy and angular momentum of the circular geodesics -- such orbits correspond to an inflexion point of the effective potential of the test particle motion \cite{Bla-Stu:2016:PHYSR4:}. 

\begin{figure*}[t]
\begin{center}
\begin{minipage}{.5\linewidth}
\centering
\includegraphics[width=\linewidth]{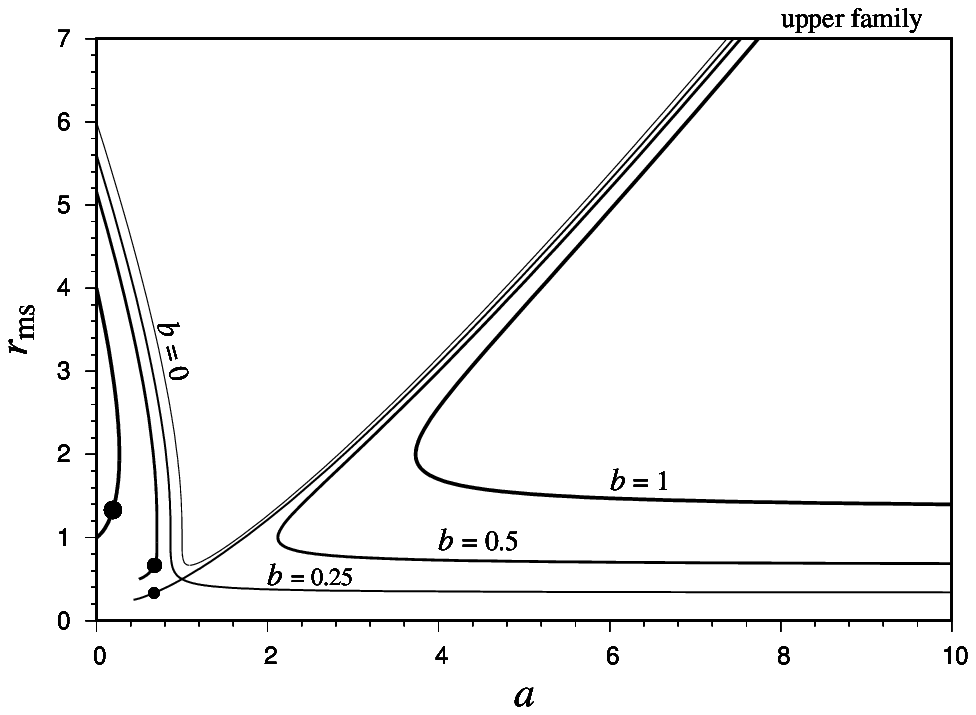}
\end{minipage}\hfill
\begin{minipage}{.5\linewidth}
\centering
\includegraphics[width=\linewidth]{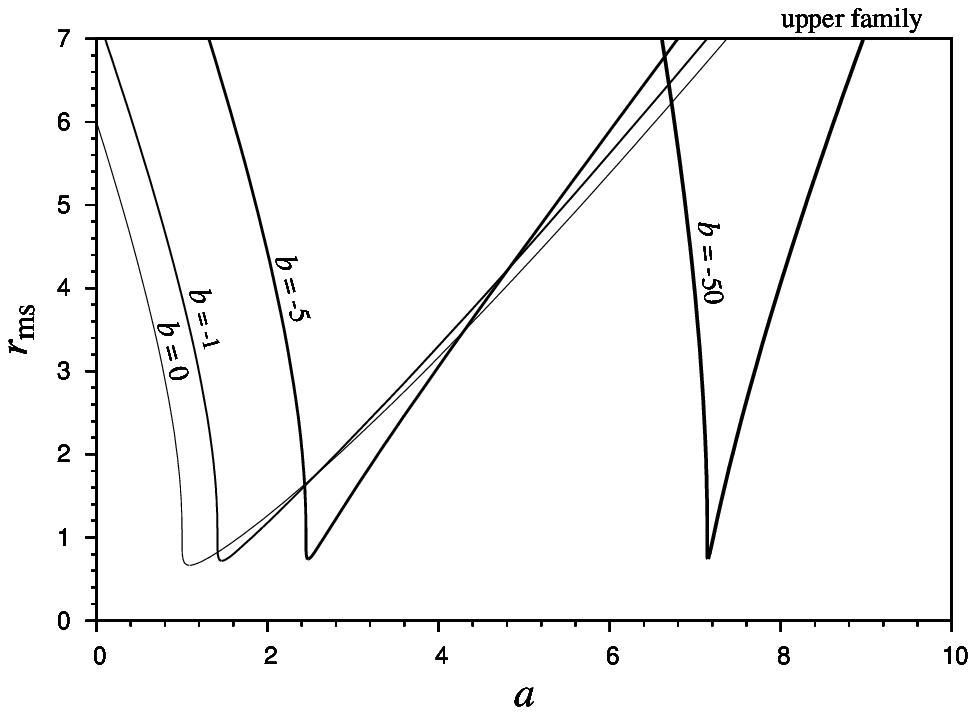}
\end{minipage}
\begin{minipage}{.5\linewidth}
\centering
\includegraphics[width=\linewidth]{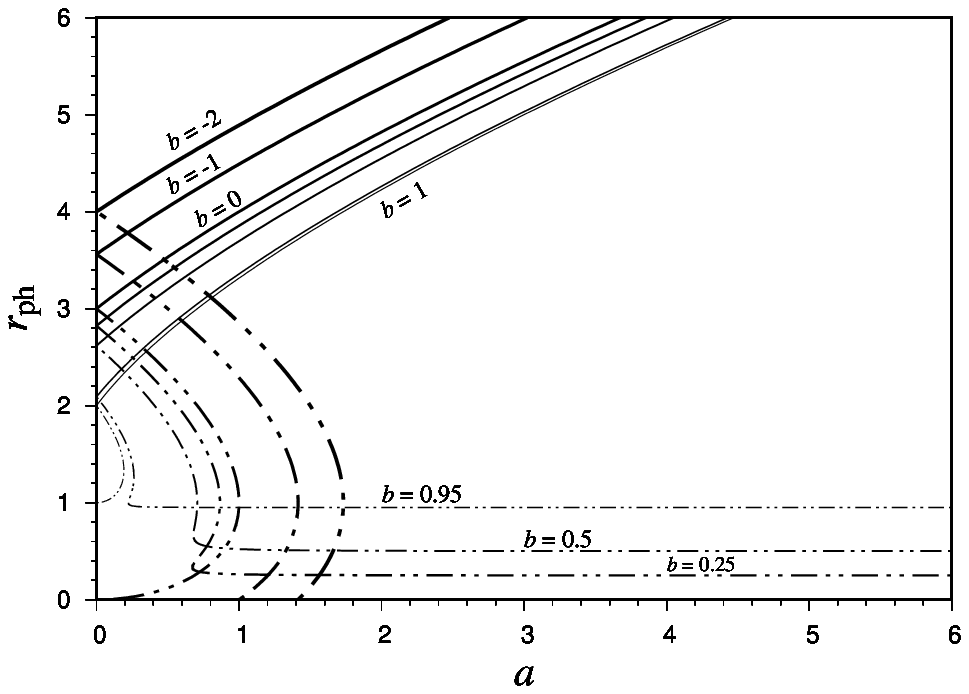}
\end{minipage}\hfill
\begin{minipage}{.5\linewidth}
\centering
\includegraphics[width=\linewidth]{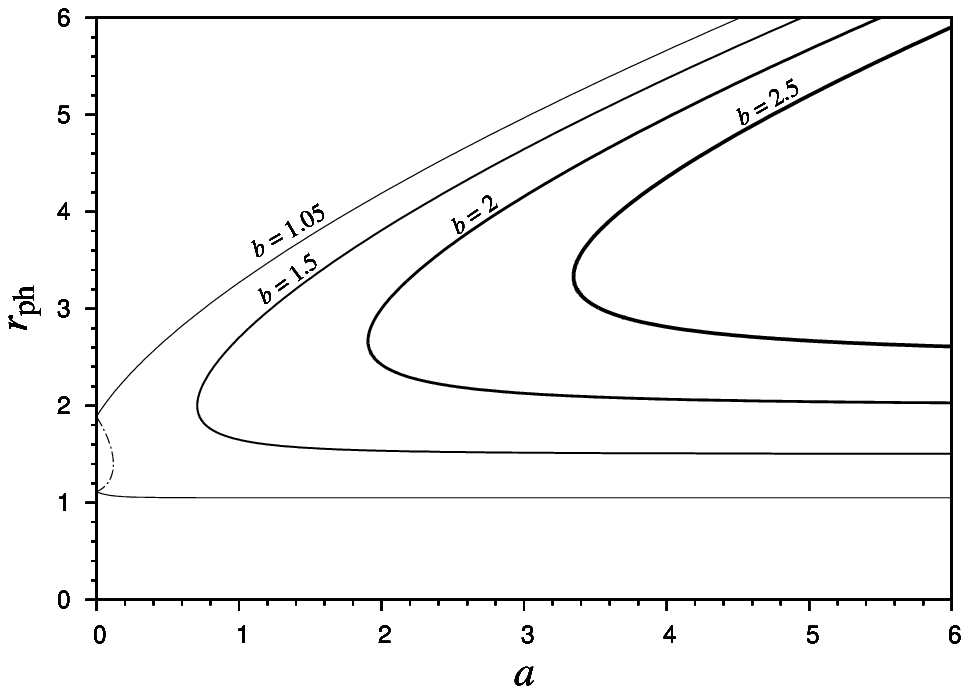}
\end{minipage}
\caption{\label{rmsrph}Marginally stable and photon circular geodesics. We give dependence of the radii of marginally stable or innermost stable and photon circular geodesics on the parameters $a,b$ of the braneworld KN spacetimes in the form of functions $r_{\mathrm{ms}}(a,b=\mathrm{const.})$ and $r_{\mathrm{ph}}(a,b=\mathrm{const.})$. In the upper row the radii of the upper family marginally stable orbits are given for positive tidal charges (left box) and negative tidal charges (right box). In the bottom row the radii of photon circular geodesics are given for both upper and lower family for values of tidal charge $b<1$ (left box) and $b>1$ (right box). Note that in the mining KN naked singularity spacetimes the marginally stable orbit of the upper family corresponds to the stable photon circular geodesic.}
\end{center}
\end{figure*}

\begin{figure*}[t]
\begin{center}
\begin{minipage}{.32\linewidth}
\centering
\includegraphics[width=\linewidth]{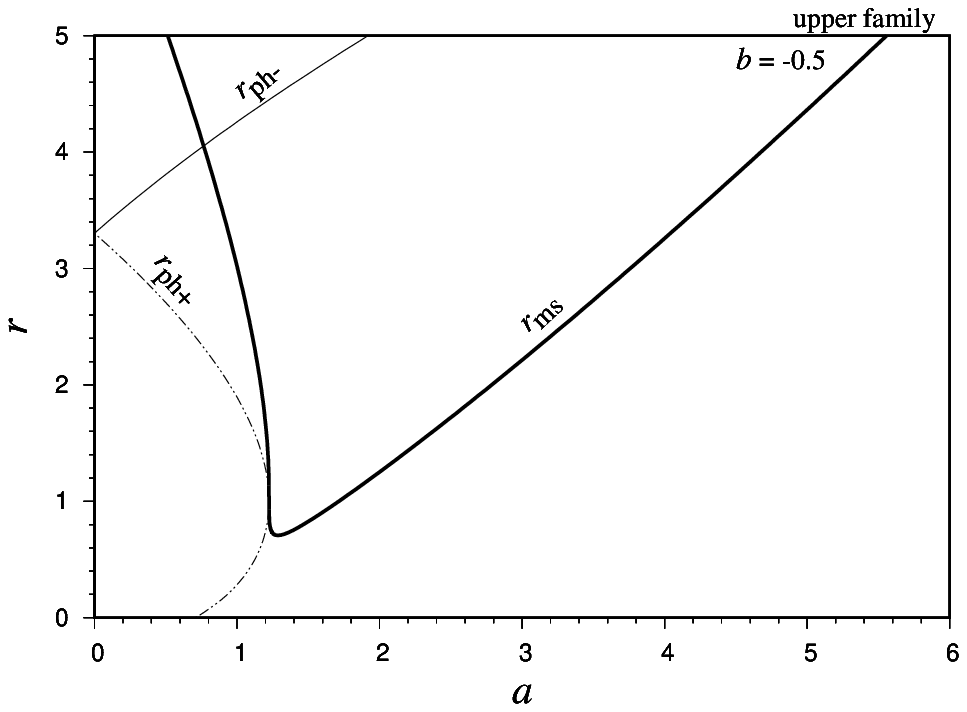}
\end{minipage}
\begin{minipage}{.32\linewidth}
\centering
\includegraphics[width=\linewidth]{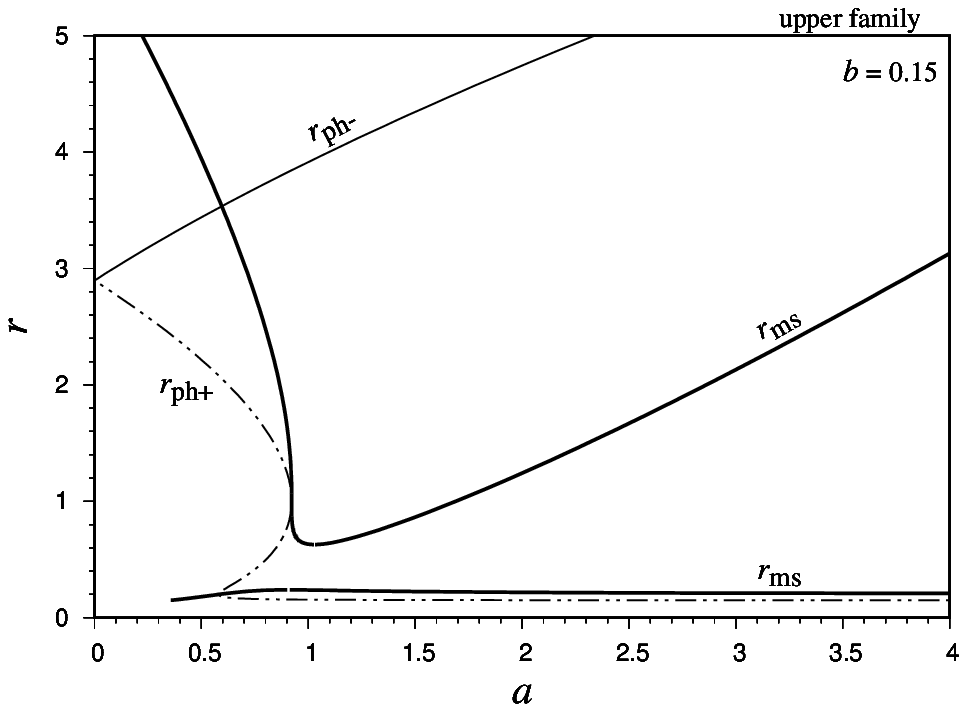}
\end{minipage}
\begin{minipage}{.32\linewidth}
\centering
\includegraphics[width=\linewidth]{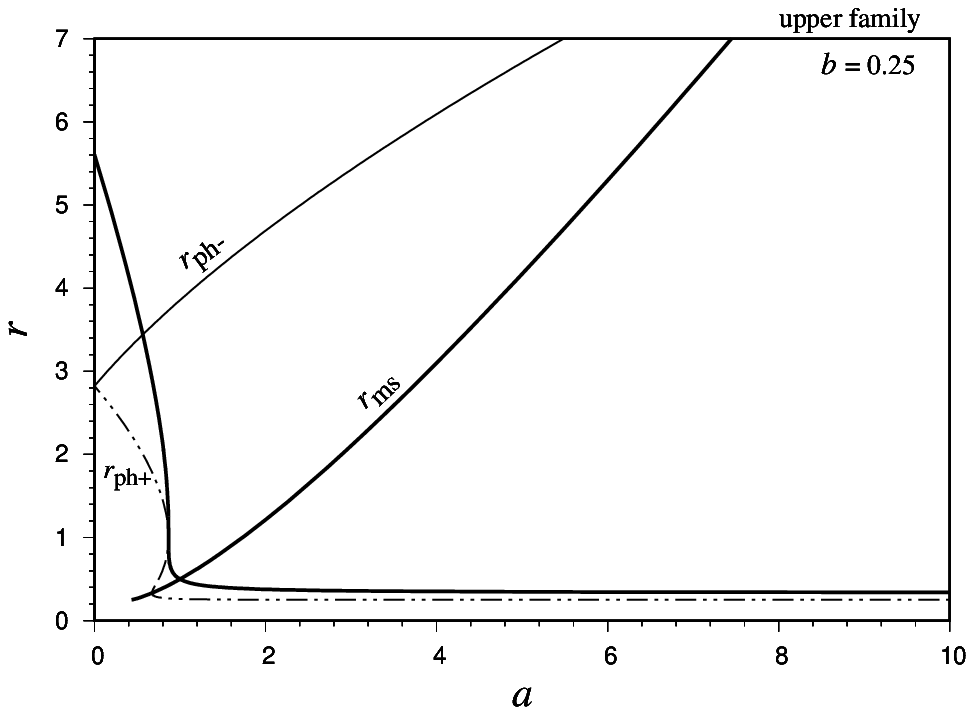}
\end{minipage}\hfill
\begin{minipage}{.32\linewidth}
\centering
\includegraphics[width=\linewidth]{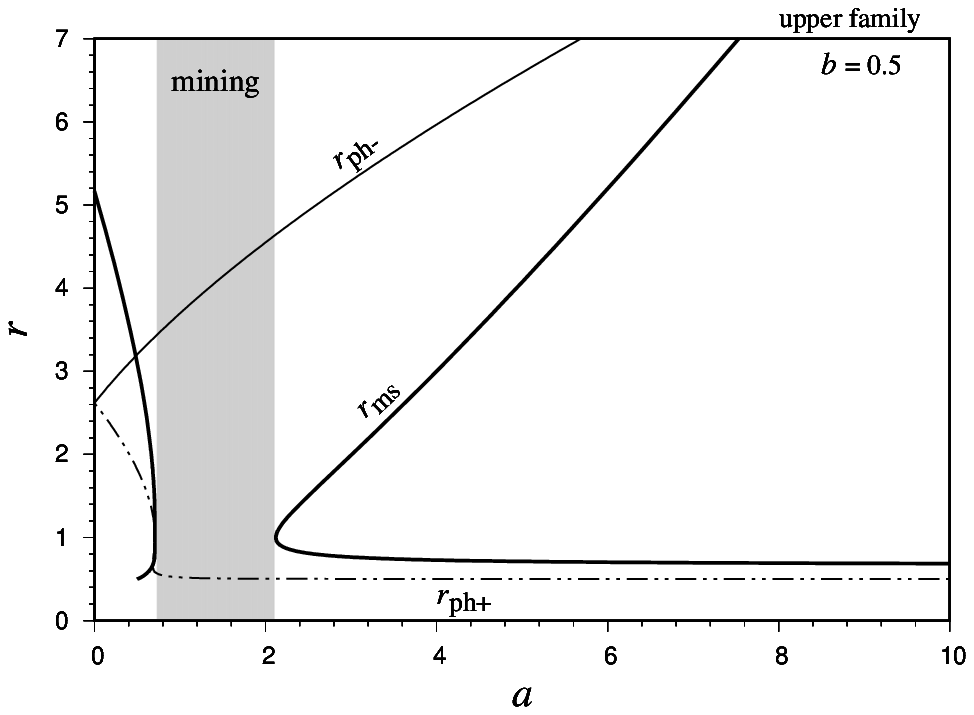}
\end{minipage}
\begin{minipage}{.32\linewidth}
\centering
\includegraphics[width=\linewidth]{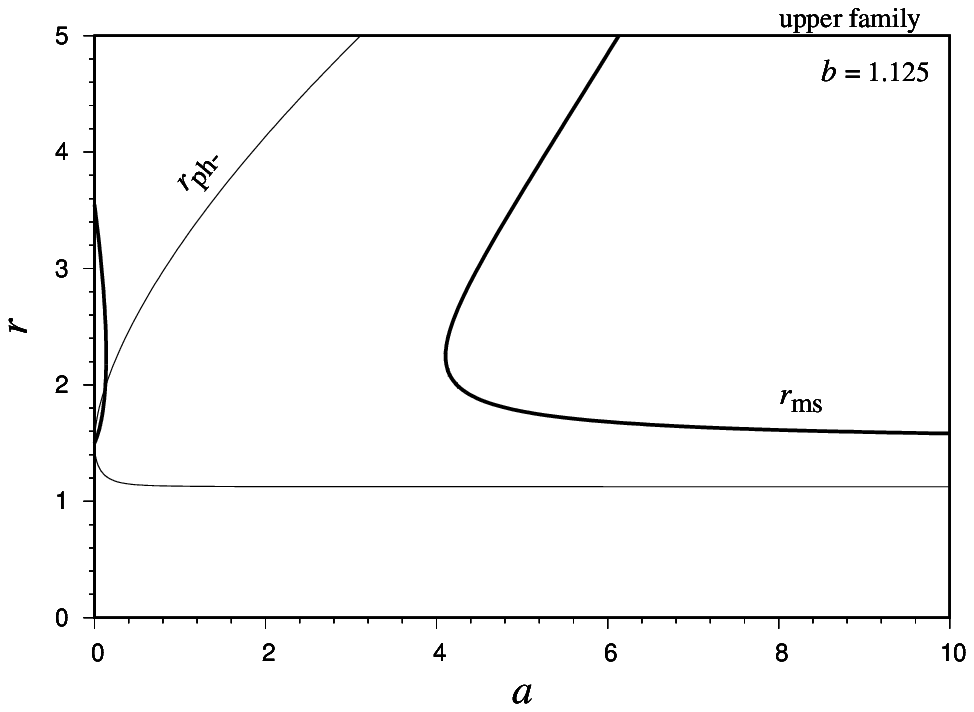}
\end{minipage}
\begin{minipage}{.32\linewidth}
\centering
\includegraphics[width=\linewidth]{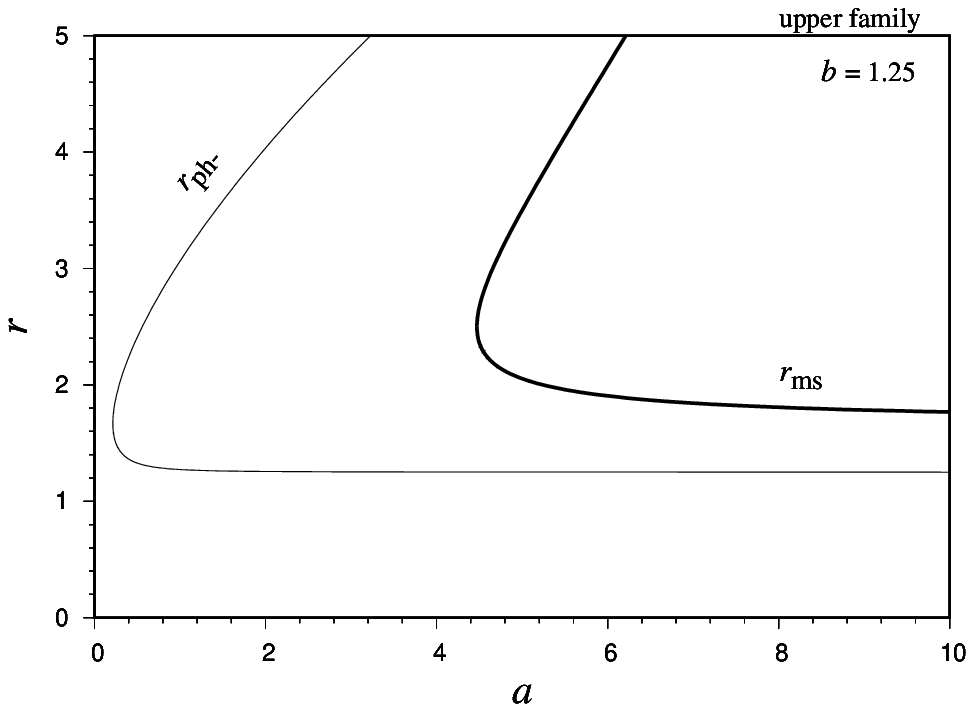}
\end{minipage}
\caption{\label{rms+rph} Relative locations of the marginally stable and photon circular geodesics in the braneworld KN spacetimes. We give the radial profiles of the marginally stable circular geodesics, $r_{\mathrm{ms}}(a,b=\mathrm{const.})$, and the photon circular geodesics, $r_{\mathrm{ph}}(a,b=\mathrm{const.})$, for typical values of the braneworld parameter $b$ reflecting qualitatively different behaviour of the geodesic structure of the braneworld KN spacetimes represented by the classification introduced in \cite{Bla-Stu:2016:PHYSR4:}. The case of the mining KN naked singularity spacetimes is reflected in the figure constructed for $b=0.5$ and is shaded there. Recall that in the mining KN naked singularity spacetimes the innermost stable circular geodesic of the upper family corresponds to the stable photon circular geodesic. A similar situations occurs also for other classes of the KN naked singularity spacetimes, but in these cases there exists also the standard marginally stable circular geodesic.}
\end{center}
\end{figure*}

Using the condition determining the marginally stable orbits $\partial^2 R/\partial r^2=0$ and inserting for specific energy and angular momentum $E,L$ the 
relations (\ref{E}) and (\ref{L}), we obtain for the braneworld KN spacetimes the relations determining radius of the marginally stable orbits \cite{Stu-Kot:2009:GRG:}\footnote{Formally the same results are relevant for the KN spacetimes  \cite{Ali-Gal:1981:}}
\begin{eqnarray}\label{rms}
r(6r - r^2 - 9b +3a^2) + 4b(b-a^2)\mp 8a(r-b)^{3/2}=0\nonumber\, .\\
\end{eqnarray}
The~solutions of Eq. (\ref{rms}) can be expressed in the form 
\begin{equation}
a = a_{\mathrm{ms}}(r,b) \equiv \mp\frac{4\left(r-b\right)^{3/2}\pm\sqrt{3 b r^2 - (2 + 4 b) r^3 + 3 r^4}}{4b-3r}\, ,
\end{equation} 
where the $\mp$ signs correspond to the upper and lower family of the circular geodesics while the $\pm$ signs correspond to the~two possible solutions of Eq. (\ref{rms}). The local extrema of the~function $a_{\mathrm{ms}}(r,b)$ are given by the relation 
\begin{equation}\label{extr}
  a = a_{\mathrm{ms(extr)}}(b) \equiv \mp\left(2 \sqrt{b}\pm\sqrt{b\left(4b-1\right)}\right) .   
\end{equation} 
In the braneworld KN spacetime parameter space, the function $a_{\mathrm{ms(extr)}}$ separates the KN spacetimes where marginally stable orbits exist for both the upper and lower family of circular geodesics, from those where they exist only for the lower family orbits, and finally those where the marginally stable circular geodesics does not exist neither for the upper and lower family circular geodesics -- for details see \cite{Bla-Stu:2016:PHYSR4:}. Dependence of the radii of the photon circular geodesics, and separately of the marginally stable circular geodesics, on the spacetime parameters $a$ and $b$ is illustrated in Fig. \ref{rmsrph}, while the \uvozovky{spin $a$} profiles of the radii $r_{\mathrm{ph}}(a,b=\mathrm{const.})$ and $r_{\mathrm{ms}}(a,b=\mathrm{const.})$ are compared in Fig. \ref{rms+rph} for typical values of the braneworld parameter $b$, representing thus the classification of the braneworld KN spacetimes introduced in \cite{Bla-Stu:2016:PHYSR4:}. Recall for completeness that the innermost limit on existence of the circular geodesics is given by the radius $r=b$. Notice that in the mining KN naked singularity spacetimes radius of the stable circular photon geodesic is almost independent of the spin parameter $a$, being located closely to the limiting radius $r=b$. Simultaneously, this radius corresponds also to the radius of the innermost stable circular geodesic. 

\subsection{Classification of the braneworld KN spacetimes}

Classification of the braneworld KN spacetimes according to the properties of the circular geodesics relevant for accretion in the quasi-geodesic (Keplerian) regime has been recently given in \cite{Bla-Stu:2016:PHYSR4:}. Note that the classification is mainly related to the regime of accretion that is at large distances corotating with respect to the rotating spacetime. 

Classification of the braneworld KN spacetimes due to the Keplerian accretion is in the spacetime parameter space $a - b$ governed by the functions $a_{\mathrm{ph-e}}(r=1,b)$, $a_{\mathrm{ph-ex}}(r=4b/3,b)$, and $a_{\mathrm{ms(extr)}}(b)$. Concerning the Keplerian accretion of matter corotating with the central naked singularity or black hole at large distances, three fundamentally different basic types of the KN spacetimes has been found. These simple cases correspond only to some of the introduced classes. In the other classes, a combination of the basic simple types occurs, as generally two sequences of the stable circular geodesics can exist in the KN naked singularity spacetimes (or even RN naked singularity spacetimes \cite{Stu-Hle:2002:ActaPhysSlov:}), being separated by a sequence of unstable circular geodesics or regions where existence of circular geodesics is forbidden - see \cite{Bla-Stu:2016:PHYSR4:} for details. The three basic types are the following.

\begin{figure*}[t]
\begin{center}
\begin{minipage}{.5\linewidth}
\centering
\includegraphics[width=\linewidth]{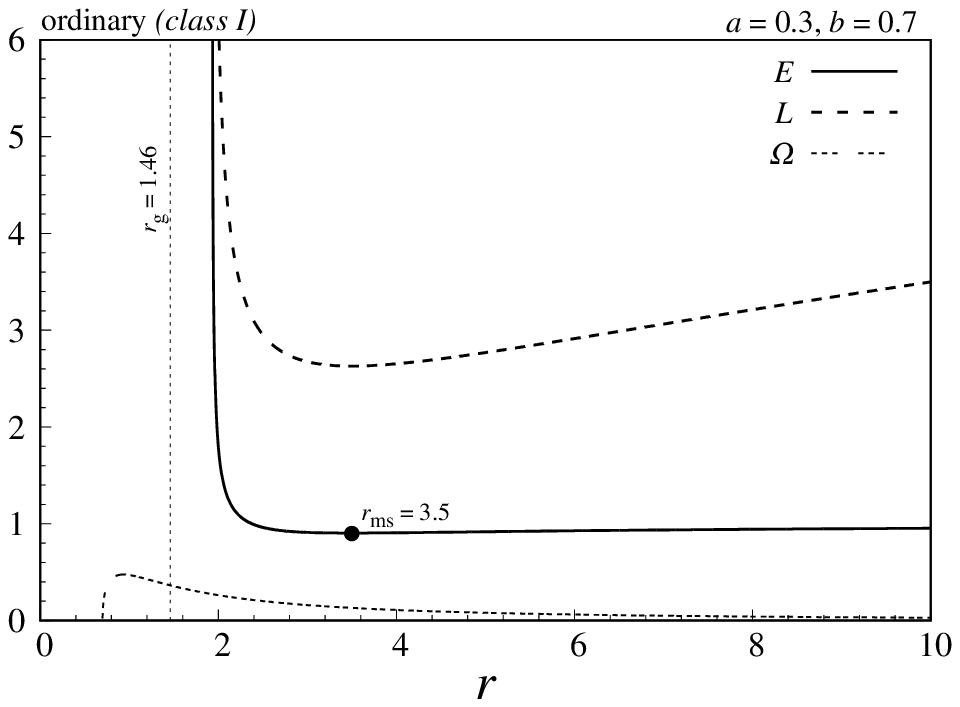}
\end{minipage}\hfill
\begin{minipage}{.5\linewidth}
\centering
\includegraphics[width=\linewidth]{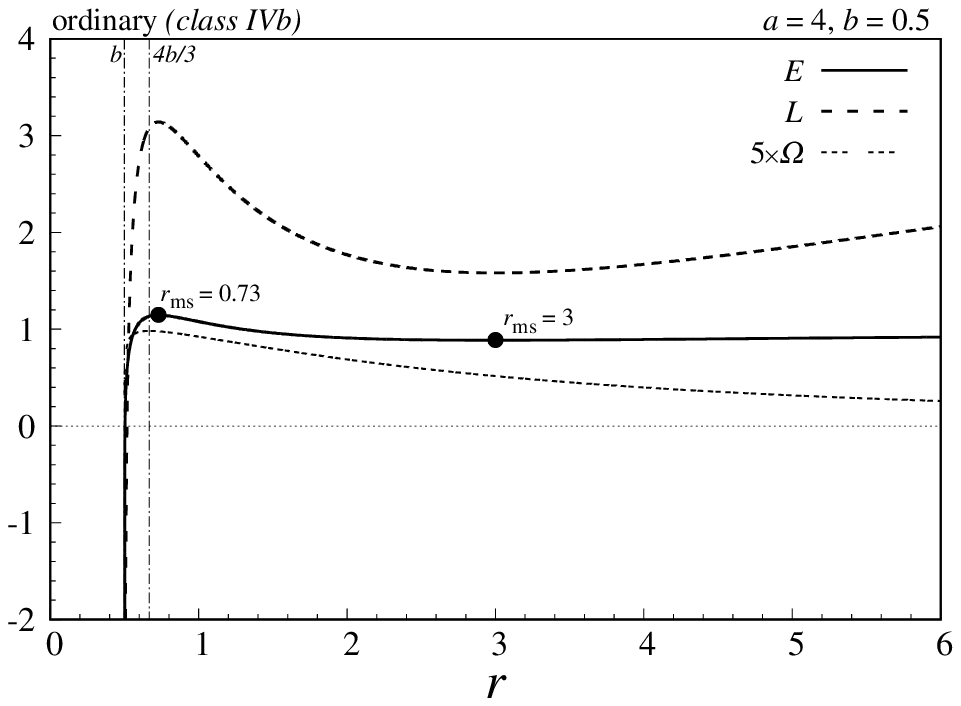}
\end{minipage}
\begin{minipage}{.5\linewidth}
\centering
\includegraphics[width=\linewidth]{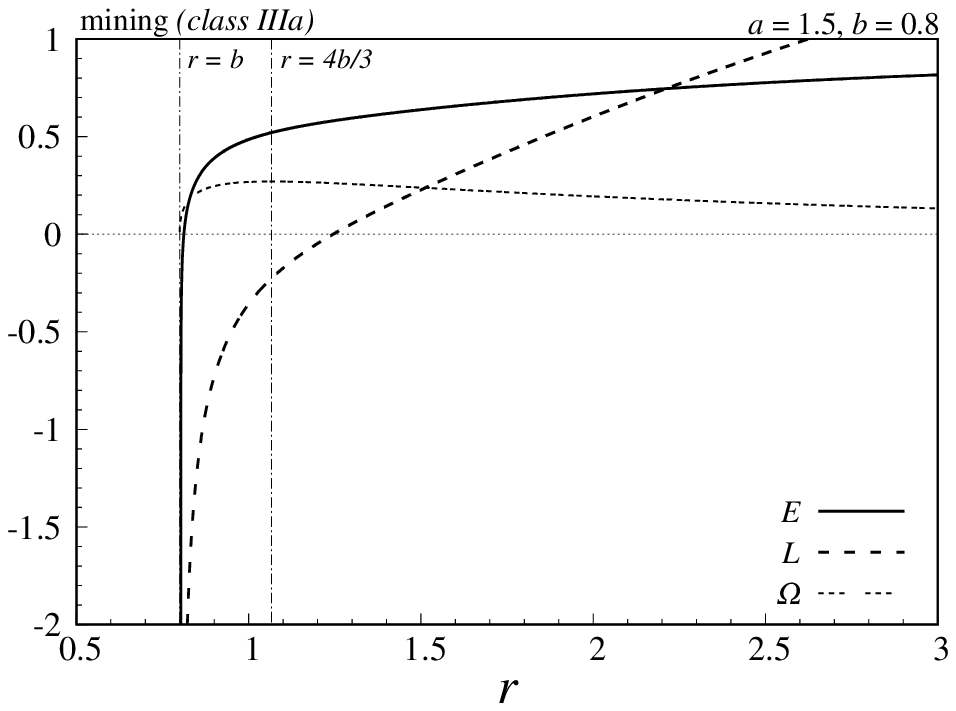}
\end{minipage}\hfill
\begin{minipage}{.5\linewidth}
\centering
\includegraphics[width=\linewidth]{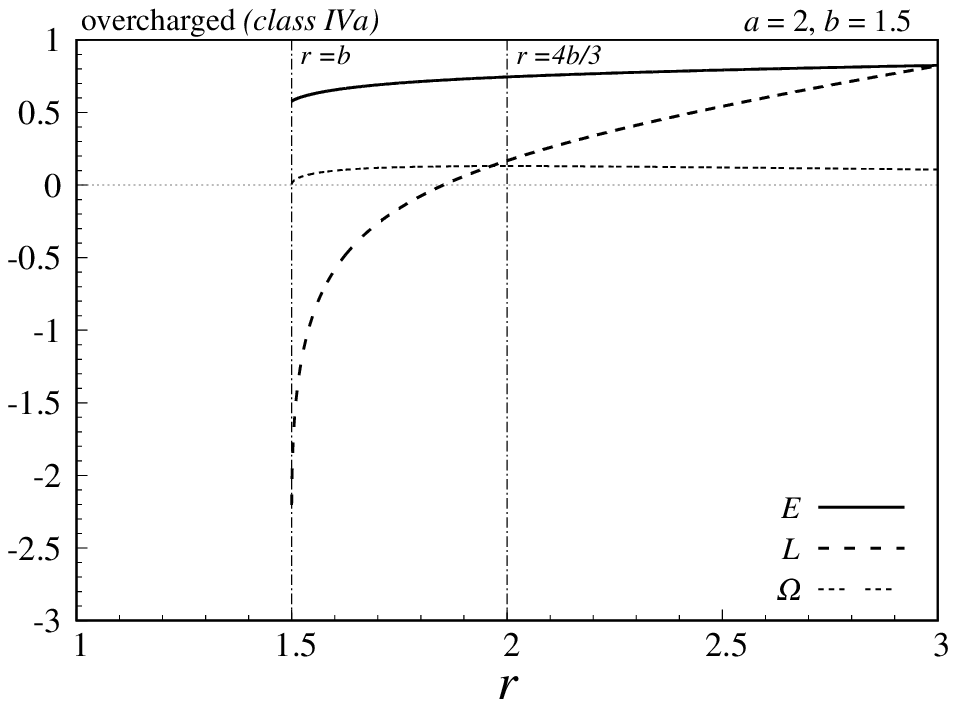}
\end{minipage}
\caption{\label{ELO} Radial profiles of the specific energy $E$, specific angular momentum $L$, and the angular velocity $\Omega$, given for spacetime parameters representing the ordinary, black hole ($a=0.3$ and $b=0.7$) and naked singularity ($a=4$ and $b=0.5$) spacetimes, the mining naked singularity spacetimes ($a=1.5$ and $b=0.8$), and the overcharged ($a=2$ and $b=1.5$) spacetimes. The black holes and naked singularities with $b<0$ can be considered to be of ordinary type. Notice that the ordinary naked singularity of class IVb, presented here, demonstrates in fact more complex behaviour, as it contains two sequences of stable circular geodesics. The sequence of unstable circular geodesics, located under the marginally stable circular geodesic giving the inner edge of the outer sequence of stable circular geodesics representing a Keplerian disk, ends at the outer marginally stable circular geodesic of the inner sequence of stable circular geodesics with inner edge at the stable photon circular geodesic, as in the mining spacetimes.}
\end{center}
\end{figure*}

\begin{figure}[t]
\begin{center}
\includegraphics[width=\linewidth]{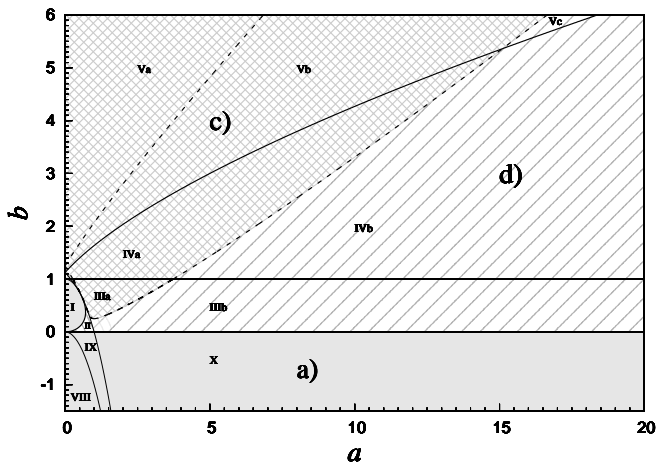}
\caption{\label{p4} Classification (a)-(e) of the braneworld KN spacetimes according to the properties of the radial profile of the angular velocity of the circular geodesics. Classes (a)-(e) are introduced according to the existence and position of the local extremum of the radial profile of the angular velocity $\Omega(r;a,b)$; the local maximum is located at $r=4b/3$. The classification is complementary to those related to the properties of energy $E$ and angular momentum $L$ radial profiles, introduced in \cite{Bla-Stu:2016:PHYSR4:}.}
\end{center}
\end{figure}

\begin{figure}[t]
\begin{center}
\includegraphics[width=\linewidth]{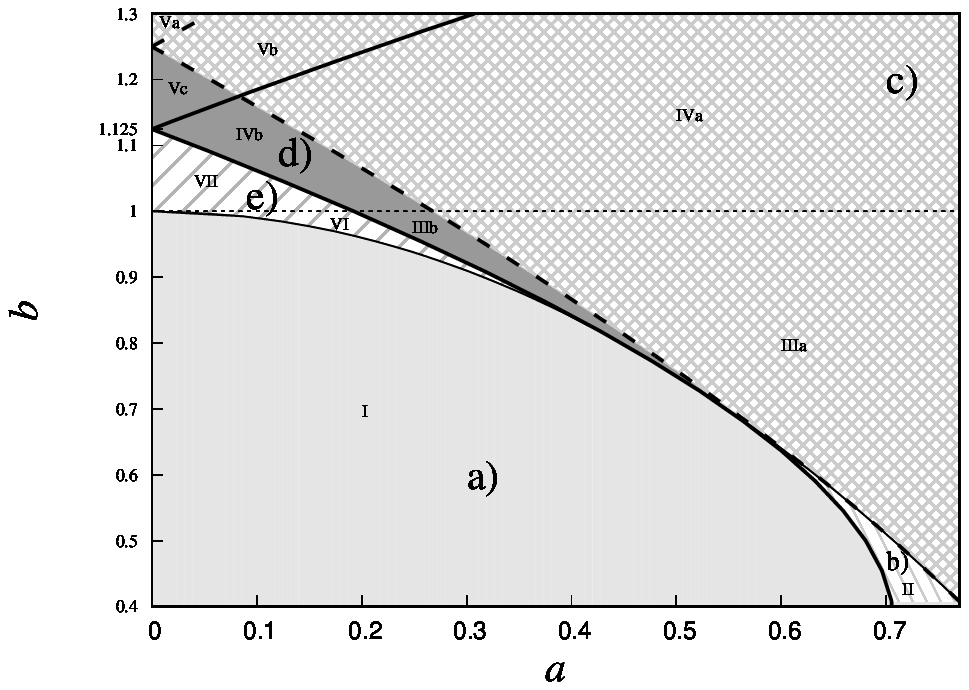}
\caption{\label{p5} Classification of the braneworld KN spacetimes according to the properties of the radial profile of the angular velocity of the circular geodesics. Detailed situation around the extreme black hole states.}
\end{center}
\end{figure}

a) Mining KN naked singularity spacetimes where the Keplerian accretion ends at the stable corotating photon circular geodesic were introduced in \cite{Bla-Stu:2016:PHYSR4:} and denoted as Class IIIa spacetimes. In this case the final stage of the accretion of matter corotating at large distances occurs in an unlimitedly deep gravitational well centered at the photon circular geodesic, and formally corresponds to unlimited efficiency of the accretion process.

The radial motion in the equatorial plane can be governed in the standard way due to the effective potentials defined by the relation 
\begin{equation}
g_{rr}\left(\frac{\mathrm{d}r}{\mathrm{d}\tau}\right)^2 = \left(E-V_{\mathrm{eff+}}\right)\left(E-V_{\mathrm{eff-}}\right)\, ,
\end{equation}
where $V_{\mathrm{eff+}}(r;a,b,L)$ ($V_{\mathrm{eff-}}(r;a,b,L)$) governs motion of particles in the positive-root states (negative-root states) \cite{Gravitation}. Here we consider the positive-root states governed by the effective potential $V_{\mathrm{eff+}}(r;a,b,L)$ that in the braneworld KN spacetimes takes the form \cite{Bla-Stu:2016:PHYSR4:}
\begin{widetext}
\begin{equation}
V_{\mathrm{eff+}}(r;a,b,L)= \frac{aL(2r-b) + r\sqrt{\Delta}\sqrt{L^2r^2 + r^4+a^2(r^2+2r-b)}}{r^4+a^2(r^2+2r-b)}\, . 
\end{equation}
\end{widetext}
For fixed values of the angular momentum $L$ and the spacetime parameters $a,b$, the local extrema of $V_{\mathrm{eff+}}$ (if they exists) determine the circular geodesic orbits. In the KN black hole spacetimes, the standard behaviour occurs - stable and unstable circular orbits (at minima and maxima of the effective potential) exist for $L > L_{\mathrm{ms+}}>0$ (corotating orbits), and $L<L_{\mathrm{ms-}}<0$ (counterrotating orbits). The situation is more complex for the KN naked singularities. In the Kerr naked singularity spacetimes with $a<1.3$, the marginal value of angular momentum of the stable plus family orbits, $L_{\mathrm{ms+}}$, becomes to be negative. In the mining KN naked singularity spacetimes, the effective potential has only one minimum for all $L>0$, but the sequence of these minima continues for decreasing $L<0$, with energy $E$ of these minima being negative and being decreasing without any limit $L\rightarrow -\infty$ and $r\rightarrow r_{\mathrm{ph+}}$, giving thus unlimited well of the effective potential. For $L<L_{\mathrm{ms-}}<0$, additional local extrema occur in the effective potential, corresponding to counterrotating unstable and stable circular orbits, energy of these circular orbits, both stable and ustable, is possitive, and in the limit $L\rightarrow -\infty$, there is $r\rightarrow r_{\mathrm{ph-}}>r_{\mathrm{ph+}}$, and $E\rightarrow \infty$ - for details see discussion of class IIIa in \cite{Bla-Stu:2016:PHYSR4:}. Nevertheless, it should be mentioned that some of the braneworld KN naked singularity spacetimes demonstrate behaviour of the effective potential that is formally even more extraordinary than in the case of the mining spacetimes - see, e.g., the class IIIb of the KN spacetimes discussed in \cite{Bla-Stu:2016:PHYSR4:}.   

Of course, such a \uvozovky{perpetuum mobile} regime of accretion must be limited by the validity conditions for the test particle motion that is involved in the assumption of the Keplerian accretion \cite{Bla-Stu:2016:PHYSR4:}. The typical radial profiles of the quantities characterizing the geodesic circular motion, $E$, $L$ and $\Omega$, are illustrated in Fig. \ref{ELO}. 

b) Overcharged KN naked singularity spacetimes where the final state of the accretion along the stable circular geodesics is located at the limiting radius $r=b$, given by the tidal charge. These KN naked singularity spacetimes have the tidal charge parameter $b>1$, and their accretion efficiency is independent of the spin parameter $a$ \cite{Bla-Stu:2016:PHYSR4:}. The typical radial profiles of $E$, $L$ and $\Omega$ are illustrated in Fig. \ref{ELO}. At the final state of the accretion, the specific energy 
\begin{equation}
                     E_{\mathrm{ISCO}} = \left(\frac{b-1}{b}\right)^{1/2}  
\end{equation}
and axial angular momentum and the angular velocity related to the distant static observers are given by 
\begin{equation}
                     L_{\mathrm{ISCO}} = -\frac{a}{\sqrt{b(b-1)}}, 
\end{equation}
\begin{equation}
                     \Omega_{\mathrm{ISCO}} = 0 .
\end{equation}
The accreting matter is thus concentrated at a ring at (and near) $r=b$, where it is at rest relative to distant observers, but it is counter-rotating relative to the LNRFs. For vanishing rotation parameter, $a=0$, the overcharged KN naked singularity spacetimes reduce to the Reissner-Nordstrom (RN) naked singularity spacetimes; properties of their geodesic structure were studied in \cite{Stu-Hle:2002:ActaPhysSlov:,Pug-Que-Ruf:2011:PHYSR4:}. The final state of the Keplerian accretion now has both  $L_{\mathrm{ISCO}} = 0, \Omega_{\mathrm{ISCO}} = 0$, and due to the spherical symmetry of the RN spacetimes, the final state of the accretion can be a sphere, similarly to the other strong gravity spacetimes (e.g. the regular spacetimes) without an event horizon \cite{Stu-Sche:2014:CLAQG:,Stu-Sche:2015:IJMPD:}. In the classification introduced in \cite{Bla-Stu:2016:PHYSR4:}, the simple situation corresponding to the continuously decreasing energy profile of accretion is represented by the class Va. In the other KN (RN) spacetimes exhibiting this kind of Keplerian accretion, only the inner sequence of circular geodesics ends at the final state $r=b$. The inner sequences can start at an outer marginally stable geodesic (located under the inner marginally stable orbit of the outer region of the Keplerian accretion), or at a stable photon circular geodesic \cite{Bla-Stu:2016:PHYSR4:,Stu-Hle:2002:ActaPhysSlov:}. 

c) Ordinary KN naked singularity and black hole spacetimes where the Keplerian accretion ends at the marginally stable circular geodesic. This is the standard case, relevant for the Kerr naked singularity and black hole spacetimes \cite{Bar-Pre-Teu:1972:ApJ:,Stu:1980:BAC:,Stu:Hle:Tru:2011:}. All the braneworld KN black hole spacetimes are of the ordinary type above the outer horizon. However, only the braneworld KN naked singularity spacetimes with negative tidal charges are of fully the ordinary type, having qualitatively the same character of the $E,L,\Omega$ radial profiles as Kerr naked singularities. We illustrate typical behaviour of the radial profiles of the specific energy $E$, specific axial angular momentum $L$, and angular velocity relative to distant observers $\Omega$ in Fig. \ref{ELO}. 

In the KN naked singularity spacetimes with positive tidal charge ($b>0$), the region of ordinary Keplerian accretion with $E,L,\Omega$ radial profiles similar to those of the Kerr naked singularity spacetimes is an outer one, and it is always combined with an inner region of Keplerian accretion having a non-standard character being, e.g., of the mining type or, inversely, representing a sequence of stable geodesics with $E$ and $L$ decreasing from stable photon circular geodesic \cite{Bla-Stu:2016:PHYSR4:}. Note that except the KN naked singularity spacetimes of Class IIIa and Va, demonstrating the \uvozovky{pure stable} regime of the (corotating) Keplerian accretion, all the other naked singularity spacetimes demonstrate a combined character of the Keplerian accretion where the outer region of the Keplerian accretion is of the ordinary type, ending at the marginally stable circular geodesic followed by region of unstable circular geodesics, while the inner Keplerian region is of non-standard character. In fact, we can consider the KN naked singularity spacetimes having the outer Keplerian region with inner boundary at a marginally stable circular geodesic as \uvozovky{effectively} ordinary KN naked singularities. 

\subsection{Gradient of the angular velocity of the Keplerian accretion}

In the present paper we extend our consideration of the Keplerian accretion for the notion of the gradient of the radial profile of the circular geodesic angular velocity related to distant observers $\Omega$ that is relevant for the mechanism governing the Keplerian accretion. It is well known that the so called magneto-rotational instability (MRI) mechanism \cite{Bal:Haw:1998:} can work only in regions where $\mathrm{d}\Omega /\mathrm{d}r < 0$. Therefore, we shall test in which regions related to the \uvozovky{corotating} Keplerian accretion in the three kinds of the braneworld KN naked singularity spacetimes, the condition $\mathrm{d}\Omega /\mathrm{d}r > 0$ requiring a different Keplerian accretion mechanism occurs. Such a different mechanism can be simply related to the gravitational (electromagnetic) radiation of the orbiting matter, or to some different viscosity mechanism; detailed discussion of the phenomena related to the switching of the angular velocity gradient, and possible behavior of matter in the region where $\mathrm{d}\Omega/ \mathrm{d}r \sim 0$, can be found in \cite{Stu-Sche:2014:CLAQG:,Stu-Sche:2015:IJMPD:}. Here, we only determine in which KN spacetimes the regions where $\mathrm{d} \Omega /\mathrm{d}r > 0$ occur. Using Eq. (\ref{Om}), we find that the condition  $\mathrm{d}\Omega /\mathrm{d}r = 0$ is satisfied just at the radii satisfying the relation 
\begin{equation}
                     r = \frac{4b}{3} . 
\end{equation}
The comparison of the radius $r=4b/3$ with the radii of the marginally stable circular geodesics, or the stable photon circular geodesics, is thus the clue for the extension of the classification of the braneworld KN spacetimes. We can introduce the following types of the braneworld KN spacetimes according to the behavior of the gradient of the angular velocity of the circular geodesics; we specify these regions also by using the classification introduced previously in \cite{Bla-Stu:2016:PHYSR4:}: 
 
a) Spacetimes with $\mathrm{d}\Omega/\mathrm{d}r < 0$ at all circular geodesics (Classes I, VIII, IX, X).

b) Black holes where $\mathrm{d}\Omega/\mathrm{d}r = 0$ occurs under the inner horizon (Class II). 

c) Naked singularities where $\mathrm{d}\Omega/\mathrm{d}r = 0$ occurs in the single continuous sequence of stable circular geodesics (Classes IIIa, IVa, Va and Vb). 

d) Naked singularities where $\mathrm{d}\Omega/\mathrm{d}r = 0$ occurs at the inner region of stable circular geodesics separated from the outer one by a region of unstable geodesics (Class IIIb, IVb and Vc). 

e) Naked singularities where $\mathrm{d}\Omega/\mathrm{d}r > 0$ at the inner region of stable circular geodesics starting at the stable photon circular geodesic, while $\mathrm{d}\Omega/\mathrm{d}r < 0$ at the outer region of the circular geodesics ending at the unstable photon circular geodesic (Class VI and VII). In this case there is no radius where $\mathrm{d}\Omega/\mathrm{d}r = 0$. 

We illustrate the extended classification of the braneworld KN spacetimes, based on the properties of the circular geodesic angular velocity radial profiles, in Fig. \ref{p4} with details given in Fig. \ref{p5}. Notice that naked singularity spacetimes where $\mathrm{d}\Omega/\mathrm{d}r = 0$ occurs at the region of unstable circular geodesics, separating two regions of stable circular geodesics, do not exist. 

We can conclude that in all the considered KN spacetimes, the MRI mechanism of the Keplerian accretion is possible only at the region of $r>4b/3\,$. Therefore, in the mining KN naked singularity spacetimes, the mining regime occuring in the vicinity of the radius of the stable photon circular geodesic is possible only due to some different mechanisms, similarly to the case of the final stages of accretion in the overcharged KN spacetimes ($b>1$) ending at $r=b$  - for details see \cite{Stu-Sche:2014:CLAQG:}. In the case of the ordinary braneworld KN spacetimes, or in the outer regions of the effectively ordinary KN naked singularity spacetimes having an outer region of standard Keplerian accretion ending at the marginally stable circular geodesic, the angular velocity is decreasing in the region of the stable circular geodesics, and the Keplerian accretion can be everywhere governed by the MRI mechanism. 

\subsection{Possible high-energy particle collisions on circular geodesics}

The circular geodesics in the deep gravitational well of the Kerr naked singularity spacetimes admit existence of astrophysically very interesting events demonstrating ultra-high energy that could be observationally relevant \cite{Stu-Sche:2012:CLAQG:,Stu-Sche:2013:CLAQG:}. In the braneworld KN naked singularity spacetimes the possibility of such ultra-high energetic processes is even much wider.  

We thus extensively study in the present paper the so called collisional Banados-Silk-West (BSW) processes \cite{Ban-Sil-Wes:2009:}, i.e., collisions of particles giving extremely large centre of mass (CM) energy, concentrating our attention to the braneworld KN naked singularity spacetimes, as the case of the braneworld KN black hole spacetimes was addressed in \cite{Sha-etal:2013:}. Such processes can occur in vicinity of the event horizon of the near-extreme rotating black hole spacetimes, if the motion constants of colliding particles are well tuned \cite{Zas:2014:,Zas:2015:}, or in special region $r \sim 1$ of the near-extreme Kerr naked singularity spacetimes under much more lesser restrictions on the motion constants of the colliding particles \cite{Stu-Sche:2012:CLAQG:,Stu-Sche:2013:CLAQG:}. Namely, in the near-extreme Kerr naked singularity spacetimes the particles freely falling from infinity along radial ($\theta = \mathrm{const.}$) trajectories and colliding at the radius $r=1$ with the related returning particles can yield efficiently extremely large CM energy \cite{Stu-Sche:2012:CLAQG:,Stu-Sche:2013:CLAQG:}. 

Here, we study the possibility to obtain ultra-high CM energy for particles incoming from large distances and colliding with particles orbiting the mining KN naked singularities in the \uvozovky{mining regime}, i.e., at circular orbits located extremely close to the stable photon circular geodesic radius. We then consider also the collisional processes at the specific radius $r=1$ in the braneworld KN naked singularity spacetimes of the ordinary type; notice that in the overcharged KN naked singularity spacetimes (having tidal charge parameter $b>1$) the special radius $r=1$ is located in the forbidden region \cite{Bla-Stu:2016:PHYSR4:}, and it is thus irrelevant for this specific kind of the collisional processes. Usually, it is assumed that the particles colliding at $r=1$ are incoming from infinity (or very large distances). Here, we extend the study of the BSW processes at $r=1$ for collisions of particles freely falling from small distances, concentrating attention to the simple case of the collisions of particles moving in equatorial plane with zero angular momentum. Then the CM energy on the colliding particles will be determined by their energy at the starting point. In the next section we thus give the analysis of the related test particle motion.  

\section{Geodesic motion with zero angular momentum and purely radial geodesics}

In order to determine family of geodesics that could reach regions where the ultra-high energy collisions can occur, i.e., the vicinity of the horizon in the black hole case, and the regions of $r \sim 1$ in the ordinary near-extreme naked singularity spacetimes, or the region of $r \sim r_{\mathrm{ph(s)}}$ in the mining naked singularity spacetimes, we consider a simple case of the equatorial motion with vanishing axial angular momentum ($L=0$). We also consider the special case of the \uvozovky{radial} geodesics starting at infinity \cite{Stu:1980:BAC:}. 

\subsection{Equatorial geodesics with zero angular momentum}

\begin{figure}[t]
\begin{center}
\centering
\includegraphics[width=0.9\linewidth]{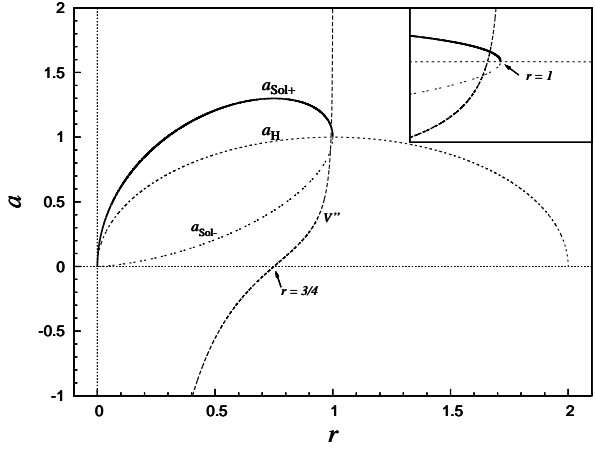}
\caption{\label{fg2} Equatorial geodesics with vanishing angular momentum. Characteristic functions for the Kerr spacetimes. }
\end{center}
\end{figure} 

\begin{figure}[t]
\begin{center}
\centering
\includegraphics[width=0.9\linewidth]{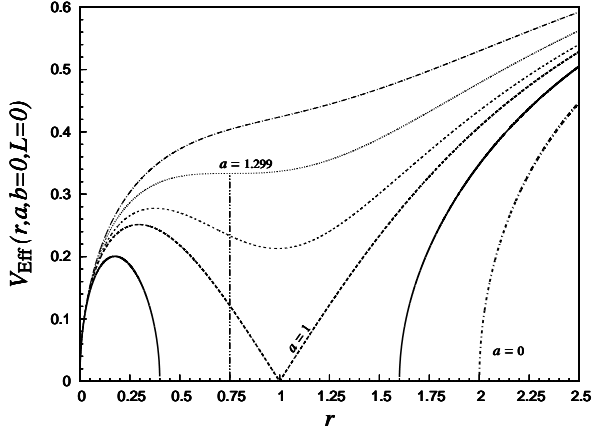}
\caption{\label{fg3} Equatorial geodesics with vanishing angular momentum. Effective potentials for the Kerr spacetimes are given for typical values of spin parameter $a$.}
\end{center}
\end{figure}

\begin{figure}[t]
\begin{center}
\centering
\includegraphics[width=0.9\linewidth]{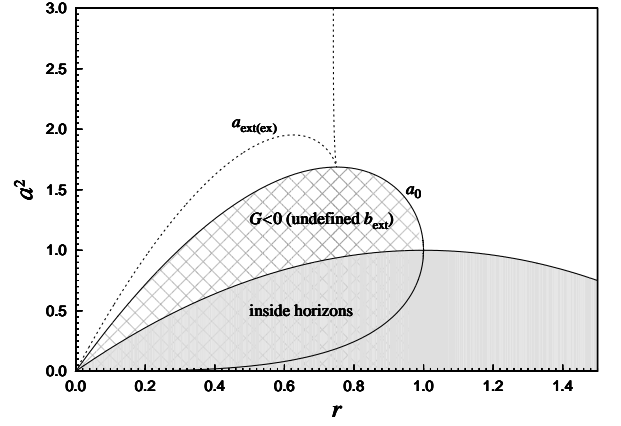}
\caption{\label{S2} Equatorial geodesics with vanishing angular momentum. Characteristic functions for the braneworld KN spacetimes -- functions $a_{\mathrm{ext(ex)}}$ and $a_{\mathrm{ext(r)}}$ governing behavior of the extrema function $b_{\mathrm{ext}}(r;a)$ and divergence function $b_{\mathrm{div}}(r;a)$.}
\end{center}
\end{figure}

\begin{figure*}[ht]
\begin{center}
\begin{minipage}{.5\linewidth}
\centering
\includegraphics[width=\linewidth]{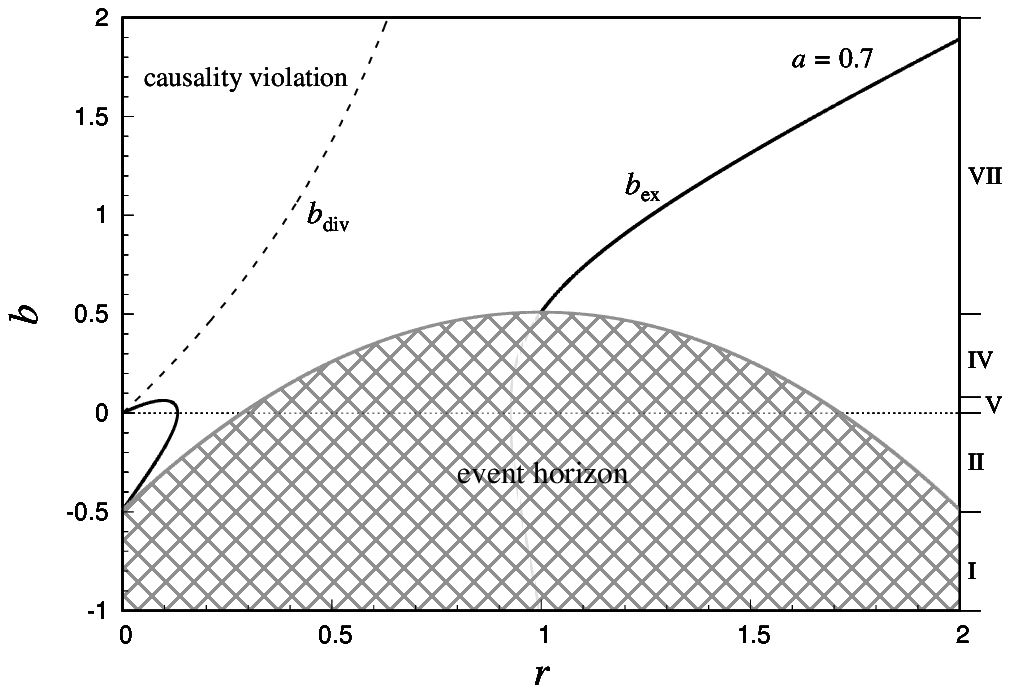}
\end{minipage}\hfill
\begin{minipage}{.5\linewidth}
\centering
\includegraphics[width=\linewidth]{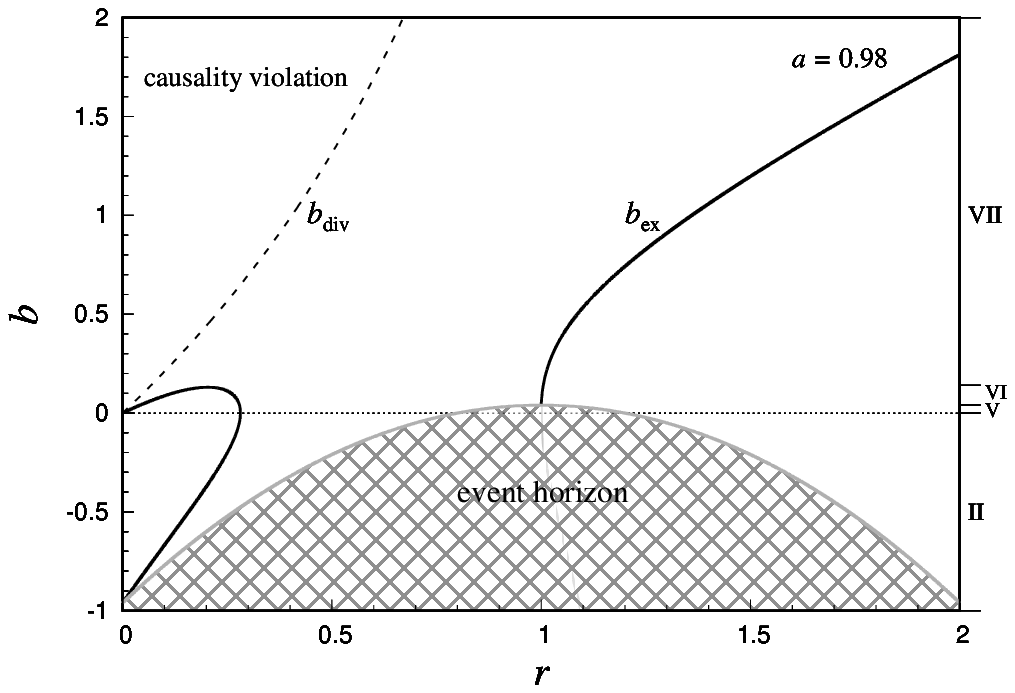}
\end{minipage}
\begin{minipage}{.5\linewidth}
\centering
\includegraphics[width=\linewidth]{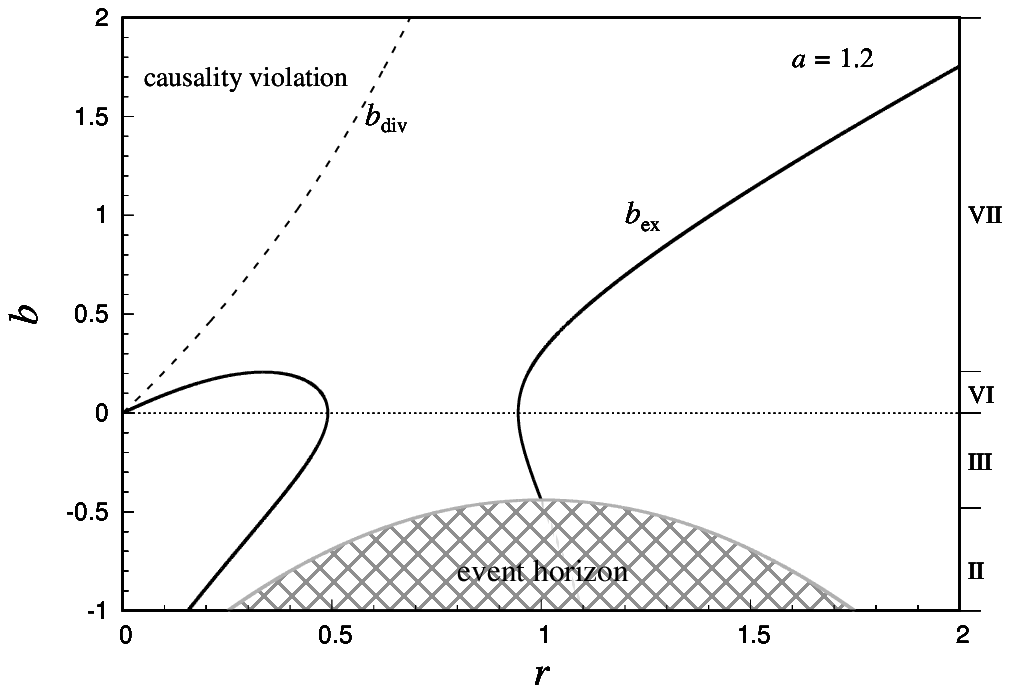}
\end{minipage}\hfill
\begin{minipage}{.5\linewidth}
\centering
\includegraphics[width=\linewidth]{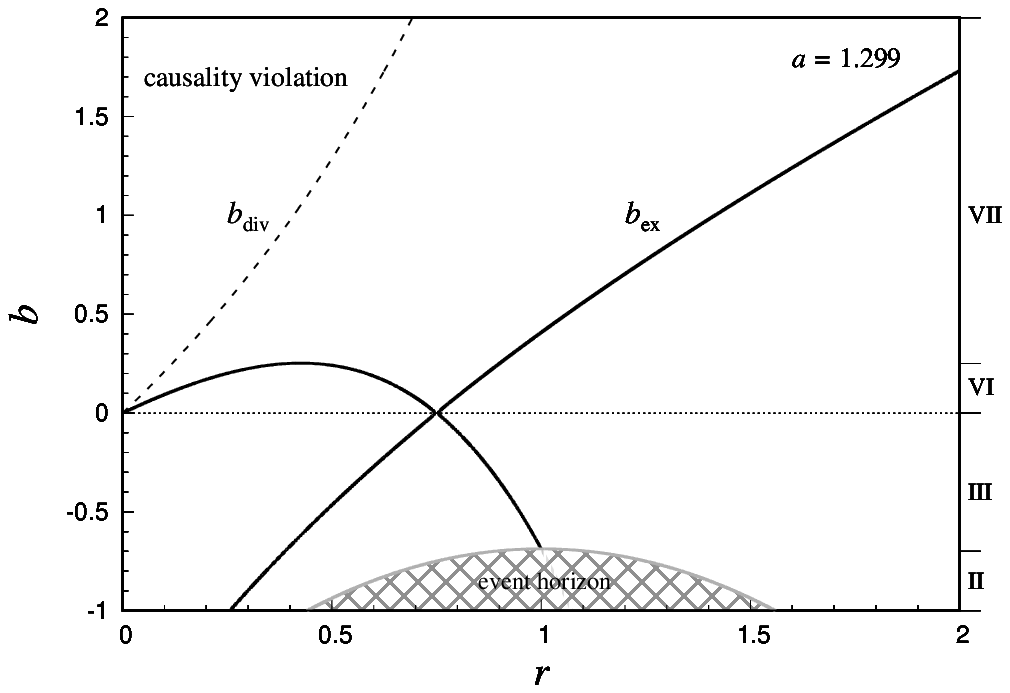}
\end{minipage}
\begin{minipage}{.5\linewidth}
\centering
\includegraphics[width=\linewidth]{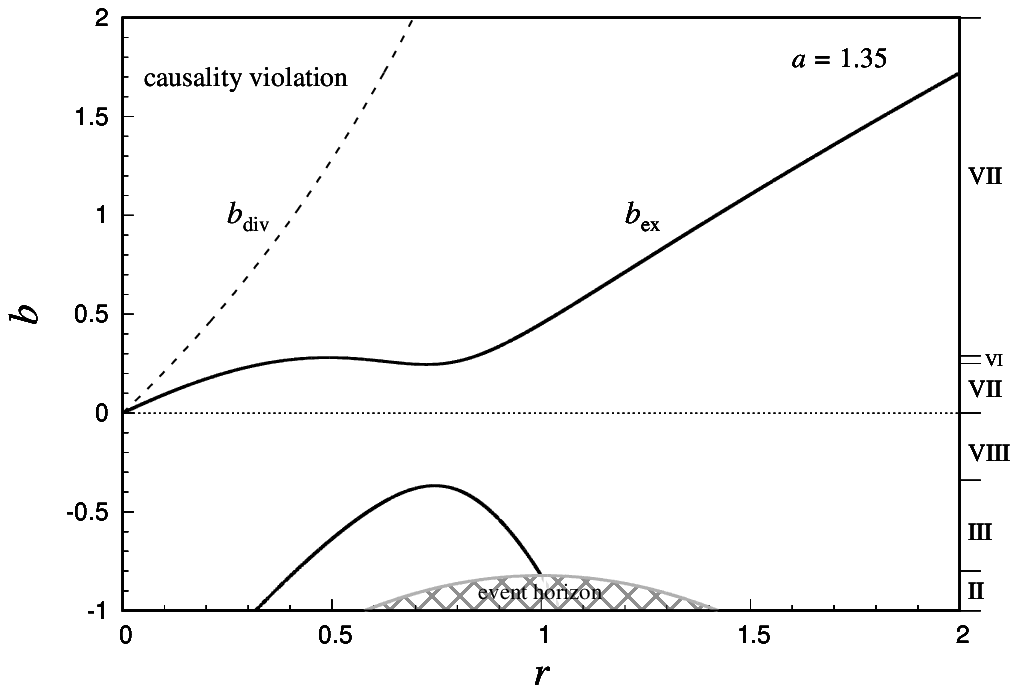}
\end{minipage}\hfill
\begin{minipage}{.5\linewidth}
\centering
\includegraphics[width=\linewidth]{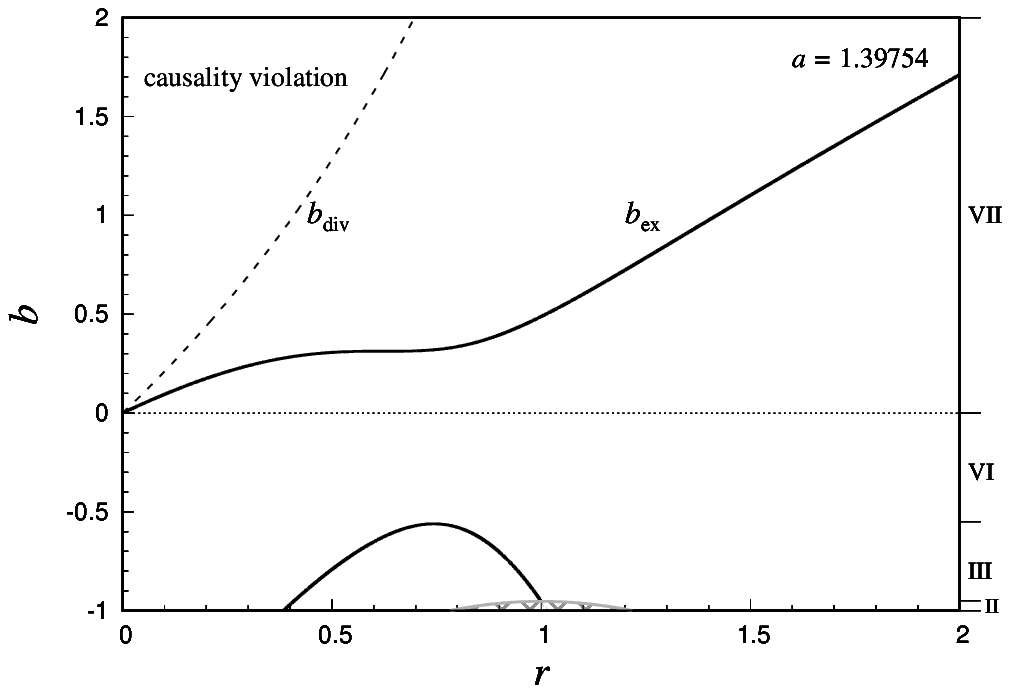}
\end{minipage}
\caption{\label{H} Equatorial geodesics with vanishing angular momentum. Characteristic functions governing the radial motion in the braneworld KN spacetimes -- $b_{\mathrm{ext}}(r;a)$ and $b_{\mathrm{div}}(r;a)$. We give the radial profiles of these functions for typical values of the spin parameter $a$ that govern the classification of the KN spacetimes related to the equatorial $L=0$ geodesics.}
\end{center}
\end{figure*}

\begin{figure}[t]
\begin{center}
\centering
\includegraphics[width=0.9\linewidth]{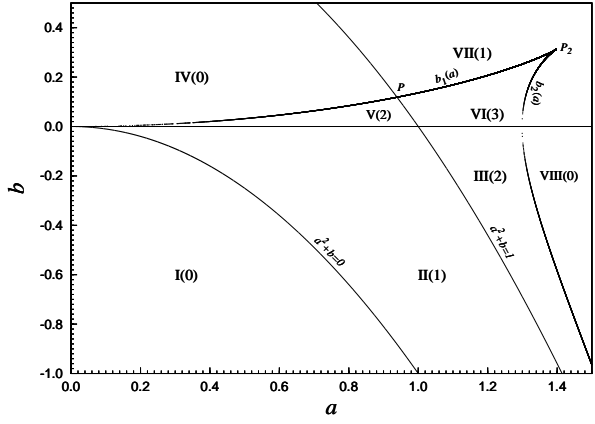}
\caption{\label{S3} Equatorial geodesics with vanishing angular momentum -- classification of the braneworld KN spacetimes, governed by the functions $b_{1}(a)$ and $b_{2}(a)$. Parameter space $(a-b)$ is separated into eight regions I-VIII. The number in brackets corresponds to the number of the extrema points of the effective potential for the motion with zero angular momentum. The critical points are $P=(0.938756\ ,0.118485)$, and $P_2=(1.39771\, ,0.312654)$. The function $b_2 (a)$ intersects the zero axis at $a=1.299\,$. }
\end{center}
\end{figure}

\begin{figure}[t]
\begin{center}
\centering
\includegraphics[width=0.9\linewidth]{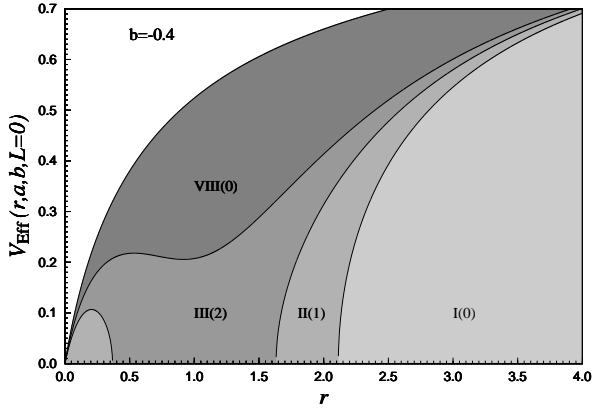}
\caption{\label{S5} Equatorial geodesics with vanishing angular momentum -- the effective potential $V_{\mathrm{Eff}}(r,a,b,L=0)$ given for braneworld KN spacetimes with negative tidal charge, in regions I(0),II(1),III(2) and VIII(0), for braneworld parameter $b=-0.4$ and corresponding values of the spin parameter $a$.}
\end{center}
\end{figure}

\begin{figure*}[ht]
\begin{center}
\begin{minipage}{.5\linewidth}
\centering
\includegraphics[width=\linewidth]{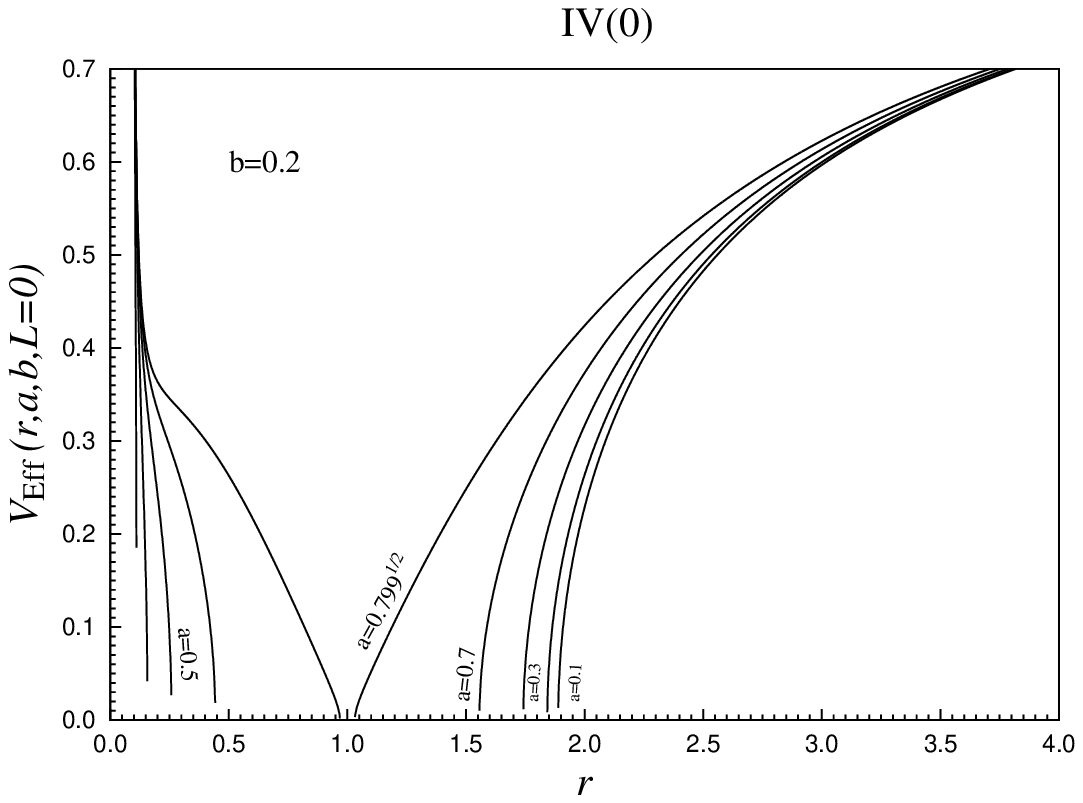}
\end{minipage}\hfill
\begin{minipage}{.5\linewidth}
\centering
\includegraphics[width=\linewidth]{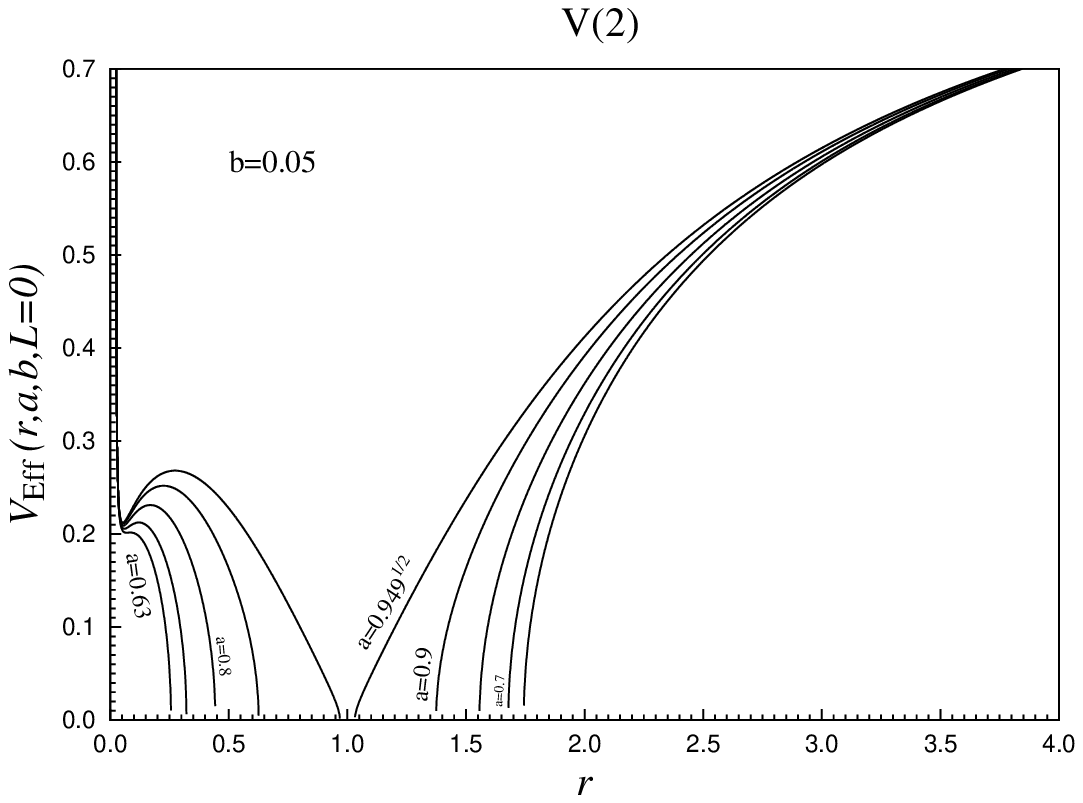}
\end{minipage}
\begin{minipage}{.5\linewidth}
\centering
\includegraphics[width=\linewidth]{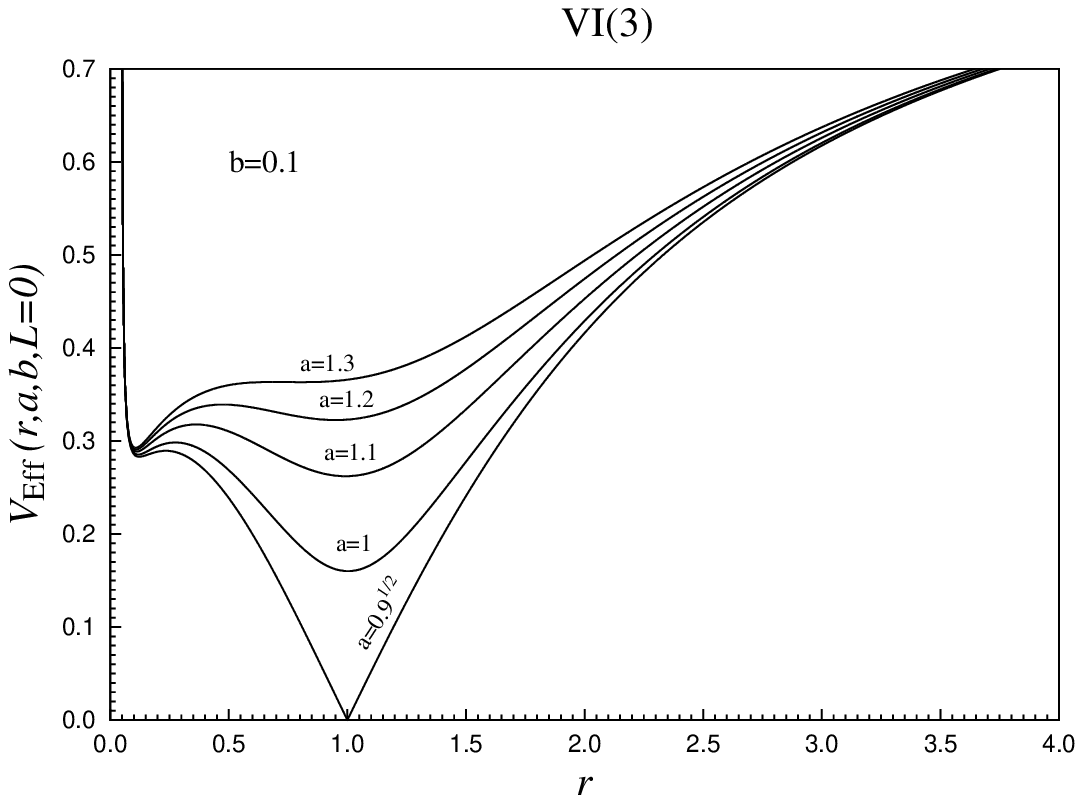}
\end{minipage}\hfill
\begin{minipage}{.5\linewidth}
\centering
\includegraphics[width=\linewidth]{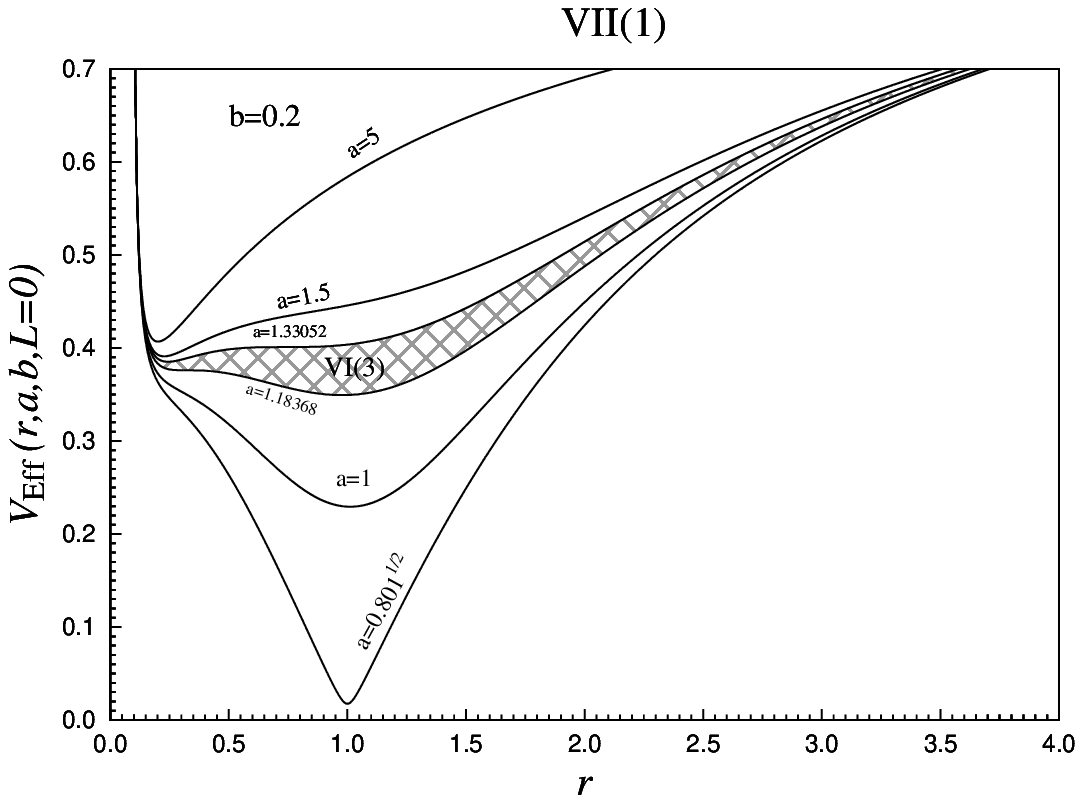}
\end{minipage}
\caption{\label{S6} Equatorial geodesics with vanishing angular momentum -- the effective potential  $V_{\mathrm{Eff}}(r,a,b,L=0)$ for braneworld (and standard) KN spacetimes with positive tidal charge, in regions IV(0),V(2),VI(3) and VII(1), are given for properly chosen values of $b$ and $a$.}
\end{center}
\end{figure*}

The equatorial motion with zero angular momentum requires automatically $Q=0$ and vanishing axial angular momentum, $L=0$. Then the motion is governed by the specific energy $E$ having the allowed values restricted by an effective potential giving turning points of the motion (where $\mathrm{d}r/\mathrm{d}\tau = 0$) due to the condition $E=V_{\mathrm{eff}}$. We shall thus discuss properties of the effective potential $V_{\mathrm{eff}}(r;a,b,Q=0,L=0) = V_{\mathrm{eff}}(r;a,b)$. 

\subsubsection{Kerr spacetimes}

It is convenient to remind first as a reference the case of the Kerr geometry when $b=0$. Then the effective potential takes the form 
\begin{equation}\label{ffggt}
V_{\mathrm{eff}}(r,a,b=0)=\frac{r \sqrt{\Delta}}{\sqrt{r^4+a^2(r^2-2r)}}\, .
\end{equation} 
The local extrema of the effective potential, given by the condition $V_{\mathrm{eff}}^{\prime}(r;a,b=0)=0\,$, where $\prime \equiv \mathrm{d}/\mathrm{d}r$, can be expressed by the relation  
\begin{equation}
a^2 = a^2_\mathrm{sol\pm}(r) \equiv 2r-r^2 \pm 2r \sqrt{(r-1)}\, . 
\end{equation} 
The local extrema of the effective potential determine the special class of equatorial circular geodesics with $L=0$. The character of these local extrema of the effective potential is determined by the sign of the second derivative $V_{\mathrm{eff}}^{\prime\prime}(r;a,b=0)\, $. As demonstrated in Fig. \ref{fg2}, for $r>3/4$ all the local extrema are minima corresponding to stable circular geodesics, while at $r<3/4$ maxima of the effective potential occur, corresponding to unstable circular geodesics. We can see that the function $a_{\mathrm{sol-}}(r)$ is always defined between the horizons, governed by the condition  
\begin{equation}
a=a_{\mathrm{H}}(r) \equiv \sqrt{2r-r^2}\, ,
\end{equation} 
and has no physical relevance. 
The inflexion point of the effective potential occurs at  
\begin{equation}
r=\frac{3}{4}\, ,\quad a=\frac{3\sqrt{3}}{4}\sim1.299\, .
\end{equation} 

The effective potential (\ref{ffggt}) has only six different types of behaviour -- see Fig. \ref{fg3}.  
The circular geodesics with $L=0$ exist only in the Kerr spacetimes with $a \leq 3\sqrt{3}/4 \sim 1.299$ \cite{Stu:1980:BAC:}. In the naked singularity spacetimes with $1 < a \leq 3\sqrt{3}/4$, there are both stable and unstable circular geodesics, the unstable being located under the stable one; in the spacetimes with $a > 3\sqrt{3}/4$, the effective potential is monotonically increasing with increasing radius. In the black hole spacetimes, unstable circular geodesic with $L=0$ can exist only, being located under the inner horizon. 

\subsubsection{Braneworld Kerr-Newman spacetimes}

For non-zero values of the tidal charge parameter $b$, the~effective potential takes the form 
\begin{equation}\label{efflo}
V_{\mathrm{eff}}(r;a,b)=\frac{r \sqrt{\Delta}}{\sqrt{r^4+a^2(r^2-2r-b)}}\, .
\end{equation}
Assuming the spin $a>0$, the effective potential vanishes at $r=0$ and at the inner and the outer horizons in the black hole spacetimes. Radius of divergence of the effective potential is governed by vanishing of denominator in the $V_{\mathrm{Eff}}(r;a,b)$ and can be determined by the relation 
\begin{equation}
  b = b_{\mathrm{div}}(r,a) \equiv \frac{r^4 + a^2(r^2-2r)}{a^2}. 
\end{equation}
Notice that the divergence radius of the effective potential corresponds to the boundary of the causality violation region in the equatorial plane \cite{Bla-Stu:2016:PHYSR4:}; the corresponding test particles have thus turning points of their radial motion before they can reach this physically forbidden region. 
The local extrema of the effective potential are for $a>0$ governed by the condition 
\begin{equation} 
 b = b_{\mathrm{ext\pm}}(r,a) \equiv \frac{-G(r;a) \pm \left(a^2+r^2\right)\sqrt{G(r;a)}}{2a^2} ,  
\end{equation}
where
\begin{equation}
     G(r;a) = r^4 + a^4 + 2a^2(r-2) , 
\end{equation}
while for the Reissner-Nordstron (RN) spacetimes with $a=0$, we obtain simple relation 
\begin{equation}
 b = b_{\mathrm{ext(RN)}}(r) \equiv r\, . 
\end{equation}

In the RN case the local extrema are located at $r=b$ and their character is given by the second derivative of the effective potential that takes the form 
\begin{equation}
\frac{\diff^2 V_{\mathrm{eff}}(r=b,a=0,b)}{\diff r^2}=\frac{1}{b^2\sqrt{(b-1)b}}\, .
\end{equation}
That is the reason why for $b\in(1;\infty)$ there is minimum of the effective potential at $r=b$, while for $b\in(0;1)$ the second derivative is not well defined and there is no local extremum of the effective potential outside the black hole horizons \cite{Stu-Hle:2002:ActaPhysSlov:,Pug-Que-Ruf:2011:PHYSR4:}. 

The behaviour of the effective potential $V_{\mathrm{eff}}(r,a,b)$ in the braneworld KN spacetimes is thus determined by the functions $b_{\mathrm{div}}(r;a)$ (discussed in detail in \cite{Bla-Stu:2016:PHYSR4:}) and $b_{\mathrm{extr}\pm}(r;a)$, where the condition $\Delta>0$ guaranteeing positions outside the dynamic regions of the spacetime has to be satisfied. For the braneworld KN black hole spacetimes, the equation $b_{\mathrm{extr}}=b\, ,$ has solutions only under the inner horizon. For naked singularities the situation is more complex. The limit of the reality of the function $b_{\mathrm{extr}\pm}(r;a)$ is governed by the function that is identical with the function governing its zero points. The limits of reality and the zero points are given by the relation 
\begin{equation}
  a^2 = a^2_{\mathrm{extr}(r)\pm}(r) = a^2_{\mathrm{extr}(z)}(r) \equiv r(2-r) \pm 2r\sqrt{1-r} . 
\end{equation}
The function $a^2_{\mathrm{extr}(r)\pm}(r)$ is represented in Fig. \ref{S2}. In order to fully understand the behaviour of the effective potential (\ref{efflo}), we have to determine the second derivative of the effective potential at the extrema points, \begin{equation}
\frac{\partial^2 V_{\mathrm{eff}}(r,a,b_{\mathrm{extr\pm}})}{\partial r^2}\, ,\end{equation}
that determine the character of the extremal points of the functions $b_{\mathrm{extr}\pm}(r;a)$. The local extrema of the function $b_{\mathrm{extr}\pm}(r;a)$ are given by the function $a^2_{\mathrm{extr(ex)}}(r)$ that is implicitly determined by the relation 
\begin{equation}
 4a^2(1-r) \pm 2r\sqrt{G(r;a)} \pm \frac{2(r^2+a^2)[(r-1)a^2+r^3]}{\sqrt{G(r;a)}} = 0 . 
\end{equation}
The resulting function $a^2_{\mathrm{extr(ex)}}(r)$ is determined numerically and is also illustrated in Fig. \ref{S2}. 

The zeros of the function $b_\mathrm{extr\pm}(r,a)$, given in Fig. \ref{S2}, determine a branching point at $a=1.299\, ,r_0=0.755$; i.e., the highest value of the spin parameter $a$ where the functions $b_{\mathrm{extr\pm}}(r;a)$ intersect the $b=0$ line. For $a>1.299$, there is $b_{\mathrm{extr+}}(r;a)>0$ and $b_{\mathrm{extr-}}(r;a)<0$, and both branches of the $b_\mathrm{extr\pm}(r,a)$ function have an extremal point. 

The characteristic functions of the effective potential, i.e., the functions $b_{\mathrm{div}}(r,a)$ and $b_{\mathrm{extr\pm}}(r,a)$, are illustrated for six typical situations in Fig. \ref{H} for six properly chosen values of the spin parameter $a$. We give also explicitly the character of the extremal points of the effective potential. In the black hole spacetimes, there is no extremum above the outer horizon, while under the inner horizon there can be none, one or two extrema if $b>0$, while only one local extremum exist for $b<0$. In the naked singularity spacetimes, one, two, or three local extrema exist if $b>0$, while none, one, or two local extrema exist, if $b<0$. 

Let $r_{\mathrm{sd\pm}}(a,b)$ be the positions of the inflexion points of the effective potential, i.e., the solution of the equation 
\begin{equation}
\frac{\partial^2 V_{\mathrm{Eff}}(r,a,b_{\mathrm{extr\pm}})}{\partial r^2}=0\, .
\end{equation}
Then we define the functions giving the limiting behavior of the effective potential in the form  
\begin{eqnarray}
b_1(a) \equiv b_{\mathrm{extr+}}(r=r_{\mathrm{sd+}}) 
\end{eqnarray}
\begin{eqnarray}
b_2(a) \equiv b_{\mathrm{extr-}}(r=r_{\mathrm{sd-}}) . 
\end{eqnarray}
The functions $b_1(a)$ and $b_2(a)$ govern the classification of the braneworld KN spacetimes according to the equatorial motion with vanishing angular momentum, giving the values of the spacetime parameters $a$ and $b$ for whom the effective potential changes its character. Namely, the effective potential changes the number and/or kind of its local extrema. These two functions are illustrated in Fig. \ref{S3} giving the classification of the braneworld KN spacetimes. 

In the classification we consider three criteria: number of the extrema of the effective potential, sign of the braneworld parameter $b$, and type of the spacetime (black hole or naked singularity). The spacetime parameter space $(a-b)$ can be then separated into eight areas (I-VIII) with differing number of solutions of the equation $b_{\mathrm{extr\pm}}(r,a)=b\, .$ The results can be seen in Fig. \ref{S3}. The arabic number in the brackets corresponds to the number of solutions for each region. The functions $b_1(a)$ and $b_2(a)$ were obtained by numerical calculations. They intersect at the point $P_2$ with the spacetime parameters taking the values $a=1.39771\, ,b=0.312654$; the function $b_2(a)$ intersects the zero axes at $a=1.299\, .$ The graph of the function $b_2(a)$ is always above the curve $a^2+b=1$ corresponding to the extreme black holes and separates black holes from naked singularities. Point $P$ in Fig. \ref{S3} where the function $b_1(a)$ intersects the curve $a^2+b=1$ is at $a=0.938756\, ,b=0.118485$.      

Now we can give for the established classes of the braneworld KN spacetimes typical sequences of the effective potential, taking fixed braneworld parameter $b$ and correspondingly varying spin parameter $a$. We separate the braneworld spacetime with negative and positive parameter $b$, and use only $a>0$ values as the equatorial motion with vanishing angular momentum is independent of the sign of the spin. We also give distribution of the local maxima (minima) of the effective potential giving unstable (stable) circular geodesics with $L=0$. 

Typical effective potentials corresponding to the classes I(0), II(1), III(2), VIII(0) of the braneworld KN spacetimes with negative values of brane parameter, $b<0$, are represented in Fig. \ref{S5}. The local extrema of the effective potential correspond to the function $b_{\mathrm{extr-}}(r;a)$. The behavior of the effective potential for the $b<0$ spacetimes is quite simple as suggested by Fig. \ref{S5}. In the black hole region I(0), the effective potential is simply descending to zero value at the horizon, having no local extremum. In the black hole region II(1), one local maximumm exists under the inner horizon. If we increase sufficiently value of the spin parameter $a$, we enter the naked singularity region III(2) where the effective potential demonstrates one local maximum and one local minimum. Further increase of the spin parameter $a$ will eventually cause that these two extrema coalesce into an inflexion point, and we then enter the naked singularity region VIII(0) where no extrema points of the effective potential exist. 

For the braneworld spacetimes with $b>0$, the behavior of the effective potential is more complex in comparison with the $b<0$ spacetimes, as shown in Fig. \ref{S6}. In the black hole region IV(0), there are no local extrema of the effective potential that is not defined between the horizons, being decreasing (increasing) above the outer (under the inner) black hole horizon. As the spin parameter $a$ increases and the brane parameter $b$ is kept at $b < 0.118485$, we enter the black hole region V(2). In this region there are two extrema points (one minimum and one maximum) under the inner horizon. Further increase of the parameter $a$ causes transition from the black holes region to the naked singularities region VI(3) where the effective potential demonstrates existence of two local minima and one local maximum. Finally, the region VII(1) corresponds to naked singularity spacetimes demonstrating one local minimum of the effective potential; notice that the corresponding case illustrated for $b=0.2$ is mixed with the case VI(3), as can be seen in the classification map reflected in Fig. \ref{S3}. 

\subsection{Radial fall from infinity}

As demonstrated in \cite{Bic-Stu:1976:MONNRS:,Stu:1980:BAC:,Stu-Bic-Bal:1999:GRG:}, there is a special class of test particle trajectories corresponding to the purely \uvozovky{radial} trajectories, i.e., trajectories keeping constant latitude $\theta=\mathrm{const.}$, but with varying azimuthal coordinate $\phi$ caused by the dragging of inertial frames. Such particles fall freely from infinity, having zero angular momentum, $L=0,Q=0$, and energy equal to the rest energy, $E=m$. Such test particles move purely radially relative to the family of the locally non-rotating frames introduced in \cite{Bar-Pre-Teu:1972:ApJ:}. 

For the purely radial geodesics with the motion constants $L=0\, , Q=0\, , E=m$, the function governing the radial motion takes in the braneworld KN spacetimes the following simple form 
\begin{equation}
R(r;a,b) = m^2(r^2+a^2)(2r-b) .
\end{equation}
We can see that the particles falling from infinity with $L=0, Q=0, E=m$ have turning point of their radial motion at the radius $r=b/2$. Clearly, in the equatorial plane the turning point is located outside the causality violating region, as follows from behaviour of the effective potential $V_{\mathrm{eff}}(r;a,b)$. Notice that the turning point of the equatorial radial geodesics is given by the solution of the equation $V_{\mathrm{eff}}(r;a,b)=1$. 

\section{Ultra-high energy particle collisions in the braneworld KN naked singularity spacetimes}

Kerr naked singularities (or Kerr superspinars) were extensively studied for a variety of astrophysical \cite{deF:1974:aap,Cal-Nob:1979:NUOC2:,Stu:1980:BAC:,Gib-Haw:1977:PRD:,Stu-Sche:2012:CLAQG:} and optical \cite{Stu:1981b:BAC,Stu:Hle:2000:,Stu-Sche:2010:CLAQG:} phenomena. Kerr superspinars, proposed in \cite{Gim-Hor:2009:PhysLetB:}, can be expected in Active Galactic Nuclei (AGN) where supermassive black holes are usually assumed \cite{Zio:2005:}, or in microquasars, i.e., Galaxy Black Hole Candidates (GBHC) observed in some X-ray binary systems \cite{Rem:McC:2006:ARAA:}. The primordial Kerr superspinars have to be converted into a near-extreme Kerr black hole due to accretion, but if they avoid a period of very efficient accretion from counterrotating Keplerian disc, they could well survive to the era of high redshift quasars \cite{Stu:Hle:Tru:2011:} when they could enter near-extreme states appropriate for the ultra-high energy collisions. The ultra-high energy collisions in the field of near-extreme Kerr naked singularities (superspinars) were studied in \cite{Pat-Jos:2011:,Pat-Jos:2010:,Pat-etal:2016:,Har:Kim:2014,Stu-Sche:2012:CLAQG:,Stu-Sche:2013:CLAQG:,Har-etal:2012:}

Here we shall study the ultra-high energy collisional processes in the braneworld spacetimes for two relevant cases: 

a) in the standard case of the near-extreme braneworld KN naked singularity spacetimes, when the ultra-high energy collissions occur near the radius $r=1$, recall that we have put $M=1$ 

b) in the special extraordinary class of the mining KN naked singularity spacetimes when one of the colliding particles is orbiting in the so called mining regime, being located very close to the radius of the stable photon circular geodesic. 

For comparison we give also the results for the collisions in the extreme black hole spacetimes. 

\subsection{Centre-of-mass energy of colliding particles}

The centre-of-mass (CM) energy of two colliding particles with 4-momenta $p^\alpha_1$, $p^\alpha_2$, rest masses $m_1$ $m_2$, and the total momentum $p^\alpha_{\mathrm{tot}}=p^\alpha_1+p^\alpha_2$, is given by the standard relation characterizing the so called Banados-Silk-West (BSW) process \cite{Ban-Sil-Wes:2009:}
\beq
         \tilde E^2_{CM}=-p^\alpha_{\mathrm{tot}} p_{\mathrm{tot}\, \alpha}= m^2_1+m^2_2 - 2 g_{\alpha\beta}p^\alpha_1 p^\beta_2 \nonumber . 
\eeq
The CM energy is a scalar independent of the coordinate system; the Carter equations imply its general form 
\bea
	\tilde E^2_{CM}&=&m_1^2+m_2^2+\frac{2 m_1 m_2}{\Sigma}\left[ \frac{P_{R1} P_{R2}-\epsilon_{1r}\sqrt{R_1}\epsilon_{2r}\sqrt{R_2}}{\Delta}\right.\nonumber\\
			&& \left. -\frac{P_{W1}P_{W2}}{\sin^2\theta}-\epsilon_{1\theta}\sqrt{W_1}\epsilon_{2\theta}\sqrt{W_2}\right]
\eea
where $\epsilon_{ir}$ takes value of $+$ ($-$) for outward (inward) radial motion, and $\epsilon_{i\theta}$ takes value of $+$ ($-$) for increasing (decreasing) latitude. 

In the standard BSW processes, we are looking for situations when for both the Kerr black hole or naked singularity spacetimes the collision occurs at radius where $\Delta(r;a) \to 0$, indicating thus possibility of extremely large values of the CM energy. In the black hole spacetimes, this possibility occurs near the black hole outer horizon, and it is shown that such processes are realistic for extreme or near-extreme Kerr black holes \cite{Ban-Sil-Wes:2009:,Har-etal:2012:,Zas:2014:,Zas:2015:}. This is also the case of the braneworld KN black holes \cite{Bobo:2013:}. In the naked singularity spacetimes, the standard mechanism assumes collisions of particles infalling from large distances (e.g., purely radially moving particles) that can be both inward and outward oriented at the special radius of collisions at $r=1$ where the ultra-high CM energy could occur in the near-extreme Kerr naked singularity spacetimes; the opposite orientation of the radial motion of the colliding particles is a necessary condition for the ultra-high CM energy \cite{Stu:Hle:Tru:2011:,Stu-Sche:2012:CLAQG:,Stu-Sche:2013:CLAQG:}. Note that the high CM energy can be obtained also due to collision of a particle orbiting at (or very close to) $r=1$ with radially infalling particle \cite{Stu-Sche:2012:CLAQG:}, but the efficiency is substantially reduced in comparison with the case of the radially opposite motion of the colliding particles. Here we shall consider the standard mechanism for the case of the near-extreme braneworld KN naked singularity spacetimes, testing also the dependence of the CM energy of collisions at $r=1$ on the energy of particles falling with vanishing angular momentum in the equatorial plane from rest at small distance to the collision radius. 

In the mining braneworld KN naked singularity spacetimes, a new mechanism of very efficient ultra-high energy collisions is possible for particle orbiting in the so called mining regime following the stable circular geodesics in the extremely deep gravitational well located very close to the stable circular photon geodesic; in this case we consider as the collisional partners both the particles incoming from large distances along the purely radial orbits, and the particles incoming from the marginally stable counterrotating circular geodesic. 

For simplicity we shall consider in the following that both colliding particles have the same rest energy. Therefore, we assume $m_1 = m_2 = m$. 

\begin{figure}[t]
\begin{center}
\centering
\includegraphics[width=1\linewidth]{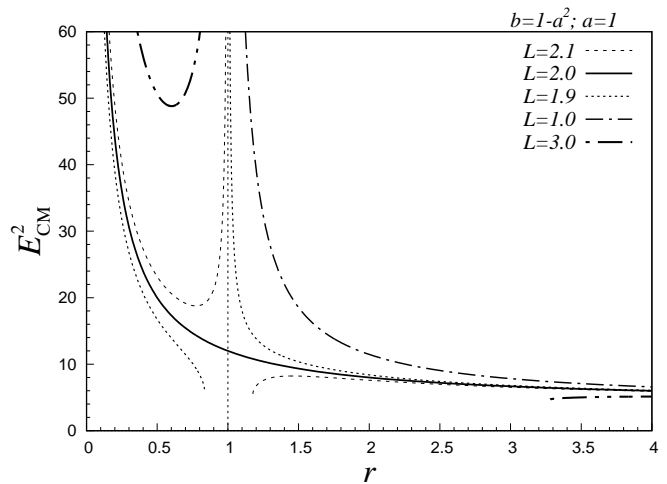}
\caption{\label{ecm0} Radial profile of the CM energy of the BSW collisions in the field of extreme Kerr black hole, when spin $a=1$ and tidal charge $b=0$. Angular momentum of the first particle is exactly $L=2$, in order to allow for the BSW effect, while for the secondary colliding particle $L<2$ in order to obtain diverging CM energy for $r \to 1$. Particles with $L > 2$ cannot reach the horizon.}
\end{center}
\end{figure}

\begin{figure}[t]
\begin{center}
\centering
\includegraphics[width=1\linewidth]{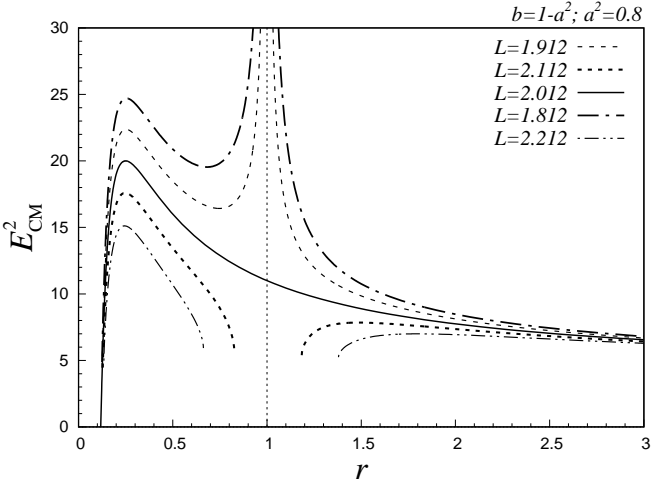}
\caption{\label{ecmb} Radial profile of the CM energy of the BSW collisions in the field of extreme KN black hole with spin $a=0.8$ and tidal charge $b=1-0.8^2$. Angular momentum of the first particle is exactly $L=2.012$, in order to allow for the BSW effect, while for the secondary colliding particle $L<2.012$ in order to obtain diverging CM energy for $r \to 1$. Particles with $L > 2.012$ cannot reach the horizon.}
\end{center}
\end{figure}

We first give the radial profiles of the CM energy for collisions near the horizon of the extreme Kerr black hole in Fig. \ref{ecm0}, or an extreme KN black hole in Fig. \ref{ecmb}. For completeness, we give the CM energy also for collisions under the horizon. We can see that the energy increases as the collision point is approaching the radius corresponding to the black hole horizon, for sufficiently low angular momentum of incoming particles. Generally, there is divergence of the CM energy for $r \to r_{\mathrm{h}}$ from both sides, with exception of the special limiting value of the angular momentum parameter when the radial profile is continuous (in the extreme Kerr spacetime the value is $l=2$). The divergence of the CM energy in the BSW effect in the extreme black hole spacetimes has been discussed in detail by \cite{Ban-Sil-Wes:2009:,Har-Kim:2011:}. 

\subsection{Collisions in the near-extreme braneworld KN naked singularity spacetimes}

First, we consider the standard ultra-high CM energy collisional processes occurring at (or near) the special radius $r=1$ in the near-extreme braneworld KN superspinning spacetimes with 
\begin{equation}
      a^2 + b = 1 + \delta\, ,\quad \delta \ll 1 . 
\end{equation} 
Then the condition $\Delta(r=1;a,b)=\delta \ll 1$ is satisfied, guaranteeing possibility of the ultra-high CM energy collisional processes. The ultra-high CM energy can be obtained only under the condition $\epsilon_{1r}\epsilon_{2r} = -1$, i.e., when the colliding particles move oppositely in the radial direction. \footnote{In the KN black hole spacetimes this condition cannot be satisfied for colliding particles incoming in the equatorial plane from infinity, as they must be inward moving in close vicinity of the outer black hole horizon; turning point of their radial motion has to be located above the photon circular geodesic. The outward directed colliding particles can be obtained only in the so called cascade collisional processes \cite{Har-Kim:2011:}.}

\subsubsection{Collisions of particles following the purely radial geodesics}

For the radially falling and returning particles with the motion constants $\tilde E=m$ and $L=Q=0$, the extremely high CM energy can be obtained, if they collide at the radius $r=1$ and at arbitrary latitude $\theta$, in the near-extreme KN naked singularity spacetimes. Since we assume the same rest energy of the colliding particles, $m_1=m_2=m$, the CM energy can be approximately expressed in the following form  
\beq
  \tilde E^2_{\mathrm{CM}} \sim \frac{4m^2}{1+a^2 \cos^{2}\theta}\left[\frac{1+2 a^2+a^4}{\delta}+a^2 \cos^2\theta-a^2\right].   
\eeq
Note that in this case the expression is in the basic approximation independent of the tidal charge parameter $b$ (the dependence on $b$ is hidden in parameter $\delta$), and follows the expression obtained for the Kerr naked singularity spacetimes in \cite{Stu-Sche:2012:CLAQG:} where detailed discussion is presented. Of course, in the more precise, higher-order, approximation, the tidal charge enters the $\tilde E^2_{CM}$ formula. 

\subsubsection{Collisions of particles moving with zero angular momentum in the equatorial plane}

Now we study the ultra-high CM energy collisional processes in the case of test particles moving in the equatorial plane ($Q=0$), with zero angular momentum $L=0$ and specific energy $\tilde E/m < 1$. We assume that such particles start their motion from rest at the radius given by the effective potential studied in the previous section. 

The collisional CM energy of the inward and outward moving particles with energy $\tilde E_1$ and $\tilde E_2$, which collide at an arbitrary allowed radius, takes the general form 
\begin{eqnarray}
  \tilde E^2_{\mathrm{CM}} &=& \frac{2m^2}{r^2 \Delta}\left[\frac{\tilde E_1 \tilde E_2}{m^2}\left(r^4+a^2 r^2 + 2a^2 r -a^2 b +\sqrt{A_1}\sqrt{A_2}\right)\right.\nonumber\\&+&\left. r^4-2r^3+\left(b+a^2\right)r^2\right] 
\end{eqnarray}
where 
\beq
A_i = \left(a^2+r^2\right)^2 - \Delta\left(a^2+\frac{r^2m^2}{\tilde E^2_i}\right)\,
\eeq
and the covariant specific energy $\tilde E_i/m$ of the colliding particles is given by the effective potential (\ref{efflo}) where we have to apply the substitution $r\rightarrow r_i$ for the initial position of the colliding particles, while $r$ in the previous equation denotes the radius of the collision event. 

An ultra-high CM energy can be obtained, if $\Delta\rightarrow 0$. Using again the near-extreme case $a^2+b=1+\delta$, we arrive to the basic approximation for the collisions at $r=1$ in the form  
\begin{equation}
\tilde E^2_{\mathrm{MC}}=\frac{4 \tilde E_1 \tilde E_2}{\delta}\left(1+a^2\right)^2\, .
\end{equation}
The modification due to the various starting positions of the colliding particles is given by the shift $m^2 \to \tilde E_1	\tilde E_2$. 

At the radius of the collision, the specific energy of the colliding particles must satisfy the relation 
\begin{equation}
  E^2_{1,2}\geq \frac{r^2 \Delta}{r^4 + a^2 \left(r^2+2r-b\right)}\equiv V^2_{\mathrm{eff}}(r,a,b,L=0)\, .
\end{equation}
Inspecting behavior of the effective potential $V^2_{\mathrm{eff}}(r,a,b,L=0)$, we conclude that the collisions in the KN naked singularity spacetimes with $b>0$ can occur for all particles  starting at rest, if they have specific energy in the interval 
\begin{equation}
     1 \geq E_i \geq E_{\mathrm{min(o)}} ,
\end{equation}
where $E_{min(o)}$ denotes the outer minimum of the effective potential $V^2_{\mathrm{eff}}(r,a,b,L=0)$. On the other hand, for the colliding particles in the braneworld KN naked singularity spacetimes with $b<0$, the specific energy has to satisfy the condition 
\begin{equation}
     E_{\mathrm{max}} \geq E_\mathrm{i} \geq E_{\mathrm{min}},
\end{equation}
where $E_{\mathrm{max}}$ ($E_{\mathrm{min}}$) denotes the maximum (minimum) of the effective potential. Clearly, the negatively charged braneworld spacetimes give only restricted possibilities for the ultra-high CM energy collisions. 

\subsubsection{Collisions of particles falling from rest from infinity in the equatorial plane with arbitrary axial angular momentum}

\begin{figure*}[ht]
\begin{center}
\begin{minipage}{.5\linewidth}
\centering
\includegraphics[width=\linewidth]{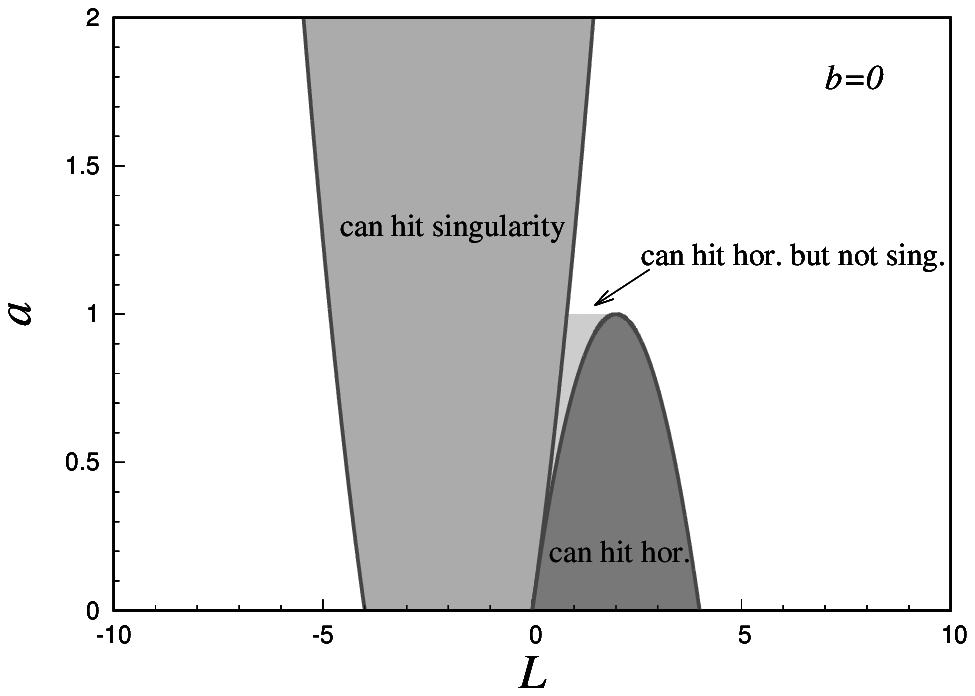}
\end{minipage}\hfill
\begin{minipage}{.5\linewidth}
\centering
\includegraphics[width=\linewidth]{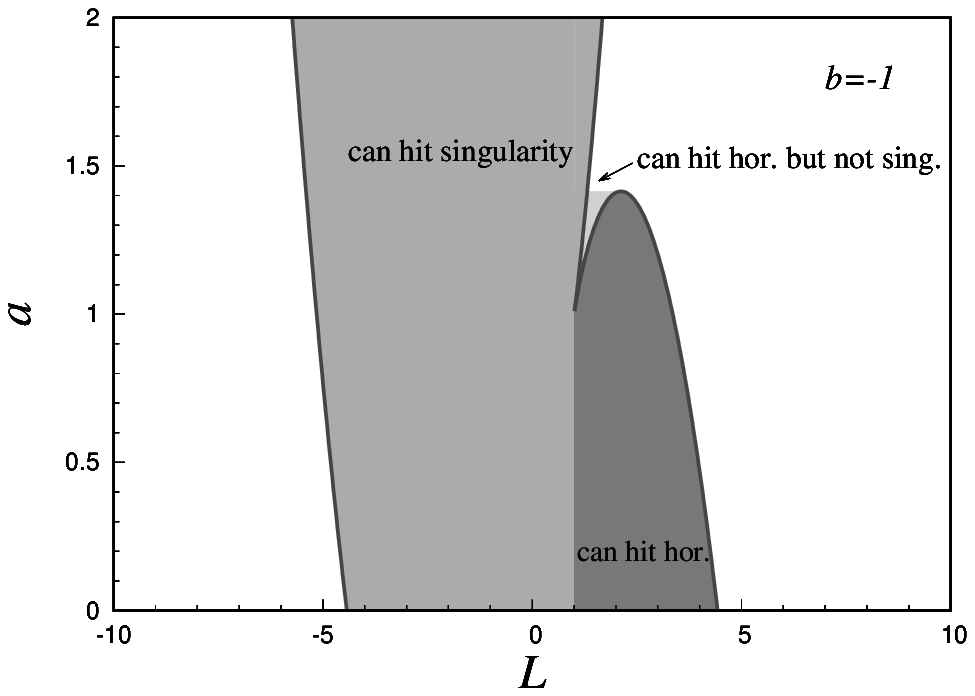}
\end{minipage}
\begin{minipage}{.5\linewidth}
\centering
\includegraphics[width=\linewidth]{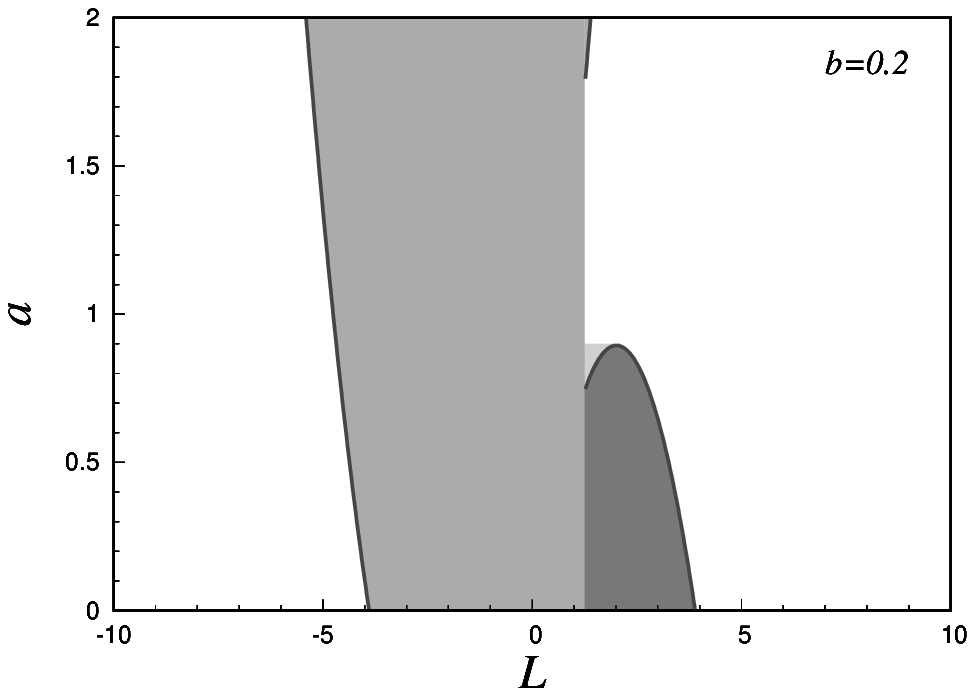}
\end{minipage}\hfill
\begin{minipage}{.5\linewidth}
\centering
\includegraphics[width=\linewidth]{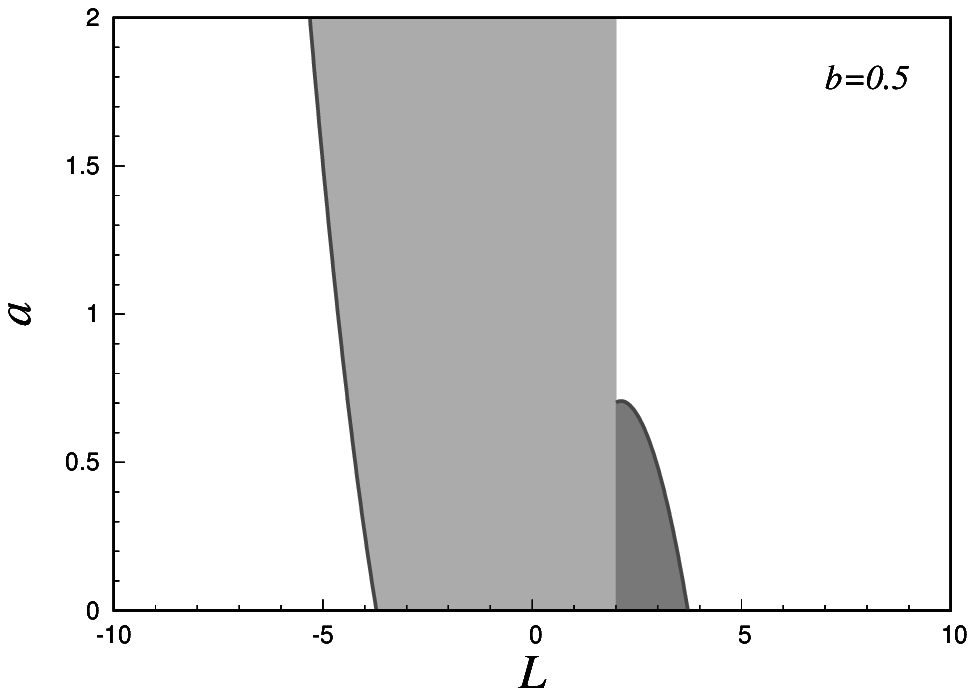}
\end{minipage}
\caption{\label{Al} Axial angular momentum $L$ of particles falling in the equatorial plane from rest at infinity ($\tilde E=m$) and reaching the radius $r=1$ where ultra-high CM collisional energy can be obtained. The regions are given in the plane $L - a$ for properly selected values of tidal charge parameter $b$. The light-shaded region correspond to the particles that can reach the singularity, the heavy-shaded regions correspond to particles that can reach $r=1$ radius.}
\end{center}
\end{figure*}

We consider two colliding particles falling from infinity with constants of motion $\tilde E_1=m\, , L_1\, , q_1=0$ and $\tilde E_2=m, L_2, q_2=0$ in the field of a near-extreme superspinning Kerr geometry with parameters satisfying the condition $a^2+b = 1 + \delta$. Assuming $\delta \ll 1$, we find for the particles with oppositely oriented radial motion the CM energy at the surface $r=1$ to be given in the first approximation by the formula 
\begin{equation}\label{Em21}
  \tilde E^2_{\mathrm{CM}} \sim \frac{4 m^2\left(1+a^2-a L_1\right)\left(1+a^2-a L_2\right)}{1+a^2\cos^2\theta}\frac{1}{\delta}\,  
\end{equation}
that is again explicitly independent of the tidal charge $b$; the dependence on $b$ is hidden in the parameter $\delta$. 
We assume $a>0$, $L_{1,2}>0$ and the conditions 
\begin{eqnarray}
L_1 &>& \frac{1+a^2}{a} \land L_2>\frac{1+a^2}{a} , \nonumber \\
L_1 &<& \frac{1+a^2}{a} \land L_2<\frac{1+a^2}{a}\, . 
\end{eqnarray}

We have to give also the conditions for the axial angular momentum, guaranteeing that the particles could reach the collision radius $r=1$. Recall that in the Kerr spacetime, the condition for $L$ allowing the radial fall reads \cite{Stu-Sche:2013:CLAQG:} 
\begin{equation}
-2\left(1+\sqrt{1+a}\right) \leq L \leq 2\left(1+\sqrt{1-a}\right)\, .
\end{equation} 
In the case of the braneworld KN spacetimes, the boundary of the allowed values of $L$ is implicitly given by the regular part of the relation 
\begin{equation}\label{ASA}
  a = a_{l(t)} \equiv L \pm \frac{\sqrt{\left(3L^2-X\right)\left(L^2+X\right)\left(4b+L^2+X\right)^2}}{8\left(L^2+X\right)}
\end{equation} 
where
\begin{equation}
X=\sqrt{L^2\left(L^2-8b\right)} . 
\end{equation}
This expression determines implicitly the boundary values $L_{\mathrm{b\pm}}(a)$, where $+$ sign corresponds to larger value of the specific angular momentum. In the parameter space $L - a$, the allowed regions of the axial momentum parameter $L$ are given for characteristic values of the tidal charge parameter $b$ in Fig. \ref{Al}. 

\subsection{Ultra-high energy collisions in the mining braneworld KN naked singularity spacetimes}

\begin{figure}[t]
\begin{center}
\centering
\includegraphics[width=1\linewidth]{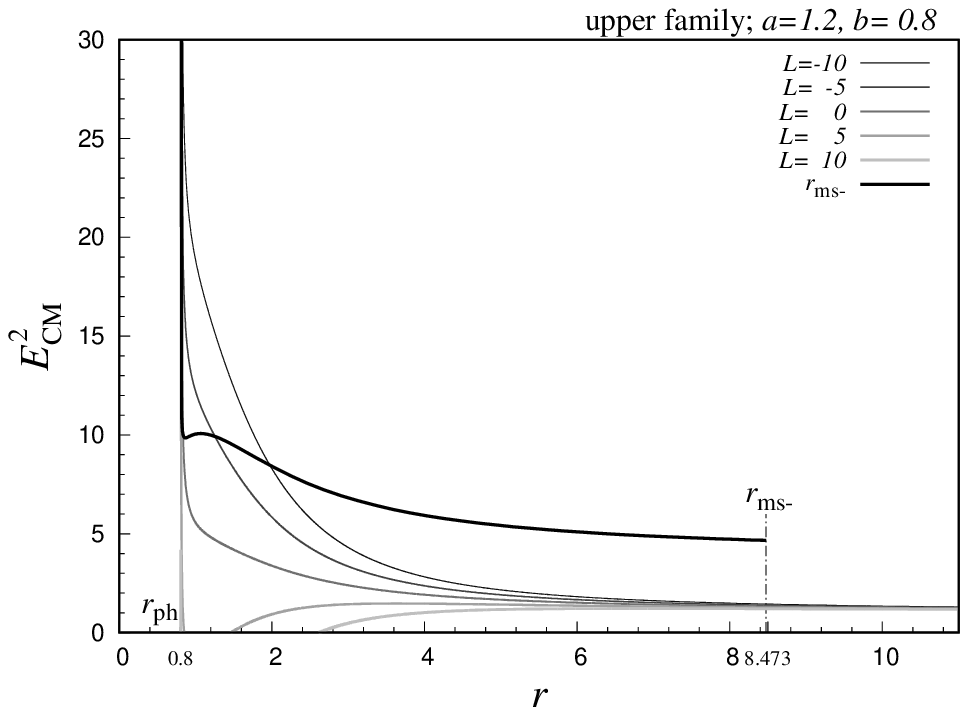}
\caption{\label{ecm} Radial profile of the CM energy of the BSW collisions in the field of mining KN naked singularity with spin $a=1.2$ and tidal charge $b=0.8$. The first particle is orbiting at a stable circular geodesic in the mining regime close to the stable photon circular geodesic, the second colliding particle fall from rest at infinity in the equatorial plane with axial angular momentum $L$. For all particles that can reach the position of the stable photon circular geodesic the CM energy diverges while the circular geodesic is approaching $r_{\mathrm{ph(s)}}$. For large values of $L$, the particles cannot reach the particles orbiting in the mining regime. By thick line represents the result of collisions of the orbiting particles with the particles leaving the counter-rotating (lower family) marginally stable geodesic -- then the CM energy differs for circular geodetics at large distance from $r_{\mathrm{ph(s)}}$, but nearly  coincides for $r\rightarrow r_\mathrm{ph(s)}\, $.}
\end{center}
\end{figure} 

Such high-energy phenomenon occurs at any braneworld KN naked singularity spacetimes, where the stable circular geodesics with negatively valued energy and angular momentum enter the so called mining regime near the stable circular geodesic representing a final state of the accretion process; notice that in the typical mining spacetime of class IIIa there are no near-extreme naked singularities \cite{Bla-Stu:2016:PHYSR4:} (with exception of those having $b \sim 1/2$). The ultra-high energy can be then reached due to an extremely low energy (extremely fast velocity) of the particle orbiting in the mining regime, for collisions with any particle -- incoming from large distances, or from vicinity of the particle orbiting in the mining regime. Moreover, the CM energy in this case is independent of the orientation of the radial motion of the colliding particle. We thus consider two special cases of the incoming particles. 

\subsubsection{Colliding particles falling from large distance}

We assume the first particle moving along the circular geodesics of the upper family, located extremely close to the stable photon circular geodesic, and having the constants of motion $(\tilde E/m)_c$ and $(\tilde \Phi/m)_c$. The second, incoming particle is assumed to be the radially freely falling particle with constants of motion $\tilde E=m$, $Q=0$, and $L$. We restrict the second particle to be moving in the equatorial plane, as the first particle is orbiting just in the equatorial plane. The CM energy can be then expressed by the relation 
\begin{equation}
\tilde E^2_{CM} = 2m^2 +\frac{2m^2\left(r^2\pm\left(a-L\right)\sqrt{r-b}\right)}{r\sqrt{r^2-3r+2b\pm 2a\sqrt{r-b}}}\, ,
\end{equation}
and we see immediately that the energy takes the extremely high values for the particle orbiting extremely close to the photon circular geodesic when $r^2-3r+2b + 2a\sqrt{r-b} \to 0$. This dependence is reflected in Fig. \ref{ecm}. Clearly, in the mining KN spacetimes, the ultra-high CM energy results due to the increasing (negatively valued) energy of the particle orbiting in the mining regime that results due to simple accretion process. Of course, a similar phenomenon could occur at any KN spacetime, if the incoming particle collides with a particle orbiting the central naked singularity (or a black hole) with extremely high (positively valued) energy. Such states are, however, unstable in the black hole spacetimes, or astrophysically unrealistic in the naked singularity spacetimes (for detailed discussion see \cite{Stu-Sche:2014:CLAQG:}). 

In order to understand the increasing of the CM energy in this kind of collisional processes, we make transformation of the CM energy formula. In dependence on the extremely small distance from the photon circular geodesic, we introduce the KN naked singularity spin in the form $a = a_{\mathrm{ph}}+\delta$ where the implicit expression of the photon circular geodesic radius reads 
\begin{equation}
a_{\mathrm{ph+}} = \frac{3r-r^2-2b}{2\sqrt{r-b}}\ .  
\end{equation} 
Using the first order Taylor expansion in $\delta \ll 1$, we arrive to the formula 
\begin{equation}
\tilde E^2_{CM} = 2m^2 + \frac{2m^2\left(r^2+3r-2b-2 L\sqrt{r-b}\right)}{\sqrt{2} r (r-b)^{1/4}}\frac{1}{\sqrt{\delta}} + O(\sqrt{\delta}).
\end{equation}
We have now obtained the same dependence on the parameter $\delta$, as the one that occurs for the collisions of incoming particles with particles orbiting at $r=1$ in the field of near-extreme Kerr naked singularities \cite{Stu-Sche:2012:CLAQG:}. Note that in our case the condition 
\begin{equation}
L \neq \frac{r^2+3r-2b}{2\sqrt{r-b}} 
\end{equation}
has to be satisfied. 

It should be stressed that there is a special subclass of the mining KN naked singularity spacetimes where the ultra-high energy collisions can occur also at the radius $r=1$, namely the exceptional near-extreme mining KN spacetimes that arise for $b \sim 1/2$ and $a \sim 1/\sqrt{2}$, if $a^2+b>1$. The other mining KN spacetimes cannot be near-extreme and the ultra-high energy collisions at $r=1$ cannot be obtained in such spacetimes. 

\subsubsection{Particles falling from the marginally stable lower-family circular geodesics}

In this case we consider the particle orbiting at the mining regime colliding with a particle that freely falls from the marginally stable geodesics of the second (retrograde) family of the circular geodesics. Then
the CM energy can be expressed in the form 

\begin{eqnarray}
\tilde E^2_{CM} = 2m^2&+&2m^2\left[\frac{r^2 b +\sqrt{r-b}\, r_{\mathrm{ms}}^2\left(a+\sqrt{r_{\mathrm{ms}}-b}\right)}{r\, r_{\mathrm{ms}} \sqrt{B_+(r)}\,  \sqrt{B_-(r_{\mathrm{ms}})}}\right.\nonumber \\
&+&\left. \frac{r^2\left[r_{\mathrm{ms}}(r_{\mathrm{ms}}-2)-a\sqrt{r_{\mathrm{ms}}-b}\, \right]}{r\, r_{\mathrm{ms}} \sqrt{B_+(r)}\,  \sqrt{B_-(r_{\mathrm{ms}})}}\right]\, ,
\end{eqnarray}
where
\begin{equation}
B(r)_{\pm} = r^2-3r+2b\pm 2a\sqrt{r-b}\, .
\end{equation}
where $r_{\mathrm{ms}}$ corresponds to the lower family orbits radius $r_{\mathrm{ms-}}$, implicitly determined by the relation \cite{Bla-Stu:2016:PHYSR4:}
\begin{equation}
r(6r - r^2 - 9b + 3a^2) + 4b(b - a^2) + 8a(r - b)^{3/2} = 0\, .
\end{equation}
We represent the dependence of this CM energy on the spacetime parameters in Fig. \ref{ecm}.  

It is interesting to note that very similar results were obtained in \cite{Gri:Pav:2013:} and subsequently in \cite{Zas:2013:}, where the authors showed that ultra-energy collisions could occur in the Kerr black hole ergosphere, between a particle with large negative values of angular momentum and another particle that are both incoming from large distances from the black hole. In the mining KN spacetimes, the particle with large negatively valued angular momentum follows the circular geodesic in the mining regime.

\section{Optical effects related to circular geodesics}

We study simple optical phenomena related to the circular geodesics in the braneworld KN naked singularity spacetimes, focusing to the frequency shift and appearance of the sky connected with the circular geodesics in the mining spacetimes, especially in the case when their radius is very close to the radius of the stable circular photon geodesic of the upper family. The frequency shift is calculated for the most fundamental principal null congruence (PNC) geodesics representing the radially moving photons that give the simplest and most representative results \cite{Bic-Stu:1976:BAC:}. The results obtained for the mining KN naked singularity spacetimes are compared to those related to the ordinary naked singularity spacetimes, or black hole spacetimes. 

\subsection{Frequency shift of the radiation from the circular geodesics}

The frequency shift $g$ represents the ratio of the observed photon energy $\gamma_\mathrm{o}$ to the emitted photon energy $\gamma_\mathrm{e}$ and can be expressed in the standard form
\beq
     g = \frac{\gamma_\mathrm{o}}{\gamma_\mathrm{e}} = \frac{U^{\nu}_{\mathrm{o}}(k_\mathrm{o})_{\nu}}{U^{\nu}_{\mathrm{e}}(k_\mathrm{e})_{\nu}}
\eeq
where $U^{\nu}_{\mathrm{o}}$ ($U^{\nu}_{\mathrm{e}}$) is the 4-velocity of the observer (emitter), and $(k_\mathrm{o})_{\nu}$ ($(k_\mathrm{e})_{\nu}$) is the covariant component of the photon 4-momentum at the observation (emission) event. 

Generally, the photon motion is characterized by the motion constants $\gamma,l,q$ and $m=0$, but the motion itself is governed by the impact parameters \cite{Bar:1973:BlaHol:,Bao-Stu:1992:ApJ:,Sche:Stu:2009b:,Schee:Stu:2015:} 
\beq
     \lambda \equiv \frac{l}{\gamma}, \eta \equiv \frac{q}{\gamma^2}. 
\eeq 
The 4-velocity of the static distant observers reads $U_{\mathrm{o}}=(1,0,0,0)$. For emitters following the circular geodesic orbits we have $U_{\mathrm{e}}=(U^{t}_{\mathrm{e}},0,0,U^{\phi}_{\mathrm{e}})$. The components of the emitter 4-velocity can be given in the form \footnote{Here we correct the misprint presented in \cite{Sche:Stu:2009:}.} 
\beq
     \left(U^{t}_{\mathrm{e}}\right)^{-2} = 1 - \frac{2}{r_\mathrm{e}}(1-a\Omega)^2 - (r_{\mathrm{e}}^{2} + a^2)\Omega^2 + 
     \frac{b}{r_{\mathrm{e}}^2}(1-a\Omega)^2
\eeq
and 
\beq
     U^{\phi}_{\mathrm{e}} = \Omega U^{t}_{\mathrm{e}}. 
\eeq 
where $\Omega = \mathrm{d}\phi/\mathrm{d}t$ denotes the angular velocity of the emitter relative to distant observers.
The frequency shift of a photon emitted from a circular geodesic located at radius $r_e$ and observed by static observers at infinity (large distances) reads 
\beq
     g = \frac{[1 - \frac{2}{r_\mathrm{e}}(1-a\Omega)^2 - (r_{\mathrm{e}}^{2} + a^2)\Omega^2 + 
     \frac{b}{r_{\mathrm{e}}^2}(1-a\Omega)^2]^{1/2}}{1-\lambda \Omega}\, .
\label{shiftco}                    
\eeq
Notice that for the photons emitted from the equatorial circular geodesics the frequency shift is independent of the impact parameter $\eta$, however, their trajectory depends on $\eta$. 

\subsubsection{Frequency shift of PNC photons}

The fundamental information on the frequency shift can be obtained using the so called PNC photons since these photons are purely radial in similar sense, as the freely radially falling observers, as they move along the $\theta=\mathrm{const.}$ trajectories, but with varying azimuthal coordinate; however, they are not purely radially moving relative to the family of LNRFs, but relative to the so called Carter frames \cite{Bar:1973:BlaHol:,Bic-Stu:1976:BAC:,Stu:1980:BAC:}. For the emitter moving along a circular geodesic, the PNC photon has to be radiated in the equatorial plane. The motion constants for the motion of the PNC photons in the equatorial plane read 
\beq
      l = a\gamma, q = -(l-a\gamma)^2 = 0 . 
\eeq
The frequency shift formula then simplifies to the form 
\beq
      g = \sqrt{\frac{1-\left(r_{\mathrm{e}}^2+a^2\right)\Omega^2}{\left(1-a\Omega\right)^2} +\frac{b-2r_{\mathrm{e}}}{r_{\mathrm{e}}^2}}\, .
\eeq

For the angular velocity $\Omega$ related to the circular geodesics, given by Eq. (\ref{Om}), we arrive to the simple relation 
\beq
g(r;a,\mathrm{e})=\frac{\sqrt{r_\mathrm{e}^2-3r_\mathrm{e}+2b\pm 2a\sqrt{r_\mathrm{e}-b}}}{r_\mathrm{e}}\, .
\eeq
Clearly, this relation is astrophysically relevant, if the emitter is located at a stable circular geodesic. (Emission of radiation from a source orbiting at an unstable circular geodesic causes departure of the source from the circular geodesic due to the radiation back-reaction effect.) The local extrema of the radial profiles of the PNC frequency shift of photons radiated from the circular geodesics, given by the condition $\mathrm{d}g/\mathrm{d}r=0$, is determined by 
\beq
\frac{\partial g(r)}{\partial r} = \frac{\left(4b-3r\right)\left(a-\sqrt{r-b}\right)}{2g(r)r^3\sqrt{r-b}}\, .
\eeq
Discussing the properties of the second derivative, $\mathrm{d}^2g/\mathrm{d}r^2$, we find that maxima of the profiles are located at 
\beq
 r = r_{\mathrm{g(max)}} \equiv \frac{4b}{3} , 
\eeq 
while minima of the radial profiles are located at 
\beq
 r = r_{\mathrm{g(min)}} \equiv a^2+b . 
\eeq
Notice that the radius $r_{\mathrm{g(max)}}$ coincides with the radius where $\mathrm{d}\Omega/\mathrm{d}r = 0$ and the angular velocity reaches maximal value. 

We can see that in the mining KN naked singularity spacetimes, for the photons radiated from geodesic circular orbits of the mining regime, there is clearly $g \to 0$ as the photon stable circular geodesic is approached. In the case of the overcharged KN naked singularity spacetimes ($b>1$), the frequency shift of PNC photons radiated from the inner-most circular geodesics at $r=b$ is independent of the spin parameter $a$, taking the value of 
\beq
 g(r=b;a,b) = \frac{\sqrt{b(b-1)}}{b} . 
\eeq

On the other hand, in the KN naked singularity spacetimes with negative tidal charges, there is $g \to \infty$, as the circular geodesics approach $r=0$, similarly to the case of pure Kerr naked singularities \cite{Stu:1980:BAC:}. The local extrema $r_{\mathrm{g(max)}}$ are irrelevant at the region $r>0$ in this case.

\begin{figure*}
\begin{minipage}{.5\linewidth}
\centering	
\includegraphics[width=\linewidth]{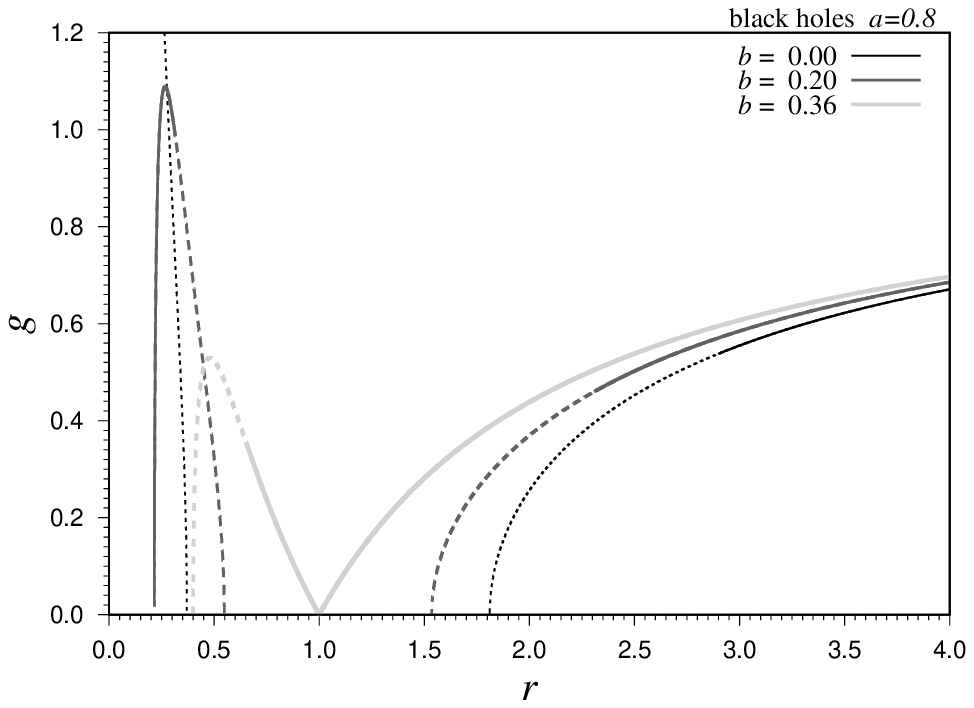}
\end{minipage}\hfill
\begin{minipage}{.5\linewidth}
\centering	
\includegraphics[width=\linewidth]{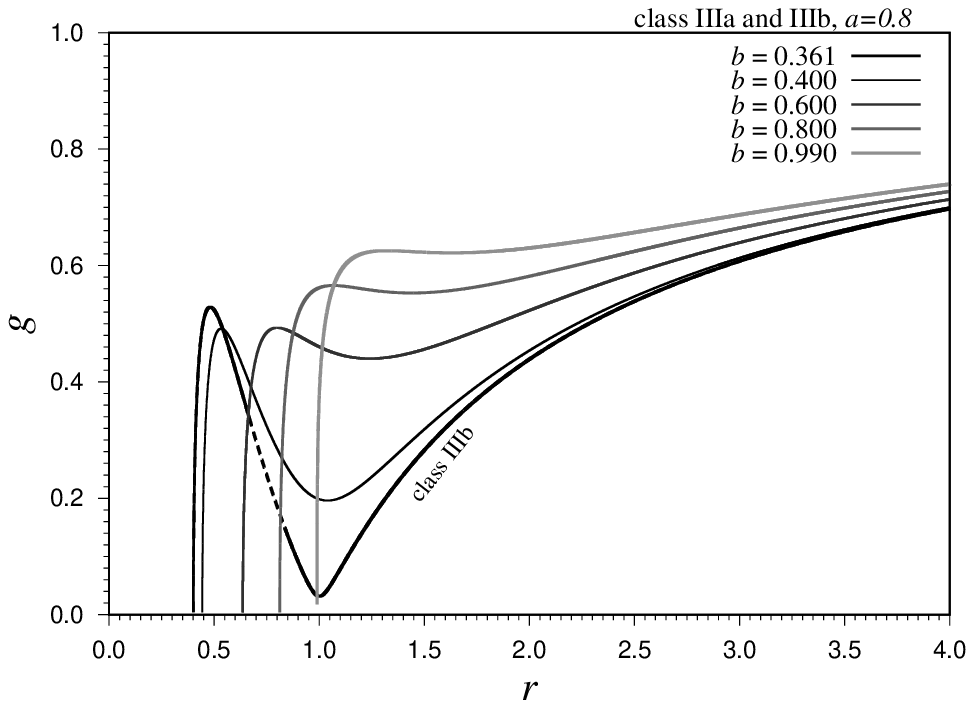}
\end{minipage}
\begin{minipage}{.5\linewidth}
\centering	
\includegraphics[width=\linewidth]{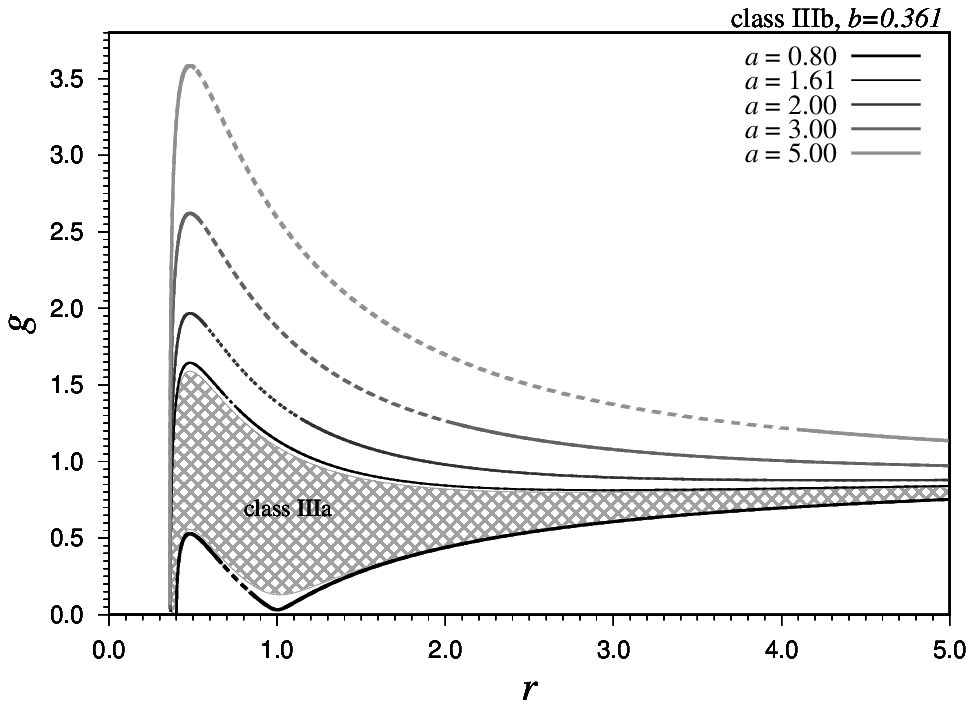}
\end{minipage}\hfill
\begin{minipage}{.5\linewidth}
\centering	
\includegraphics[width=\linewidth]{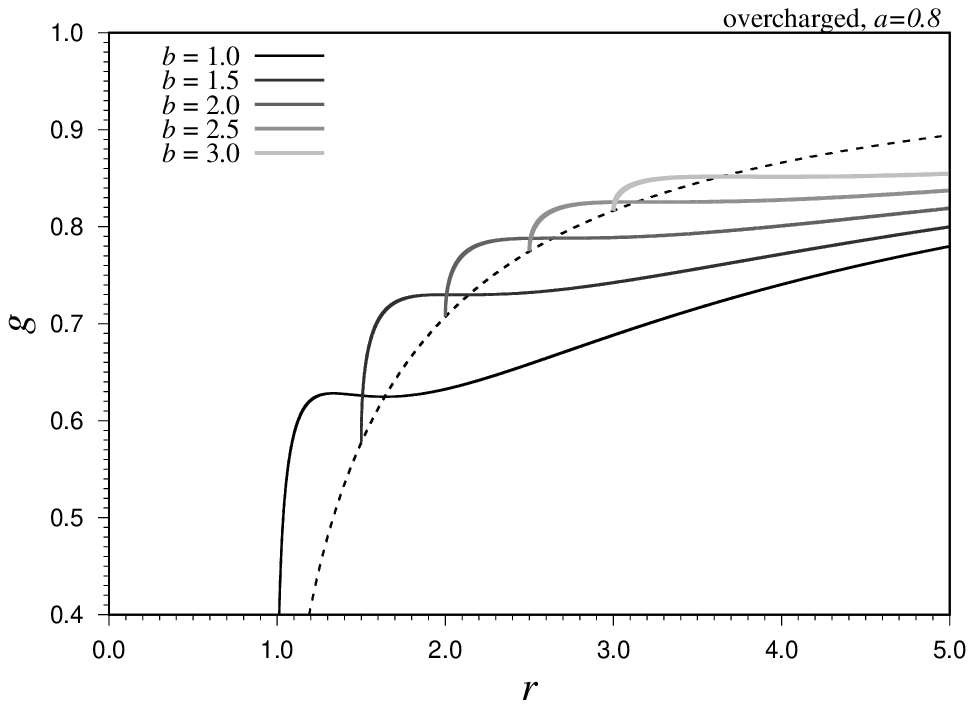}
\end{minipage} 
\caption{\label{PNC} The radial profiles of the frequency shift of the PNC photons radiated by sources following circular geodesics in the KN spacetimes with positive tidal charges. The solid lines correspond to stable geodesics, the dashed lines to unstable ones. We reflect the most interesting cases: black holes (upper left), mining naked singularities (upper right and lower left), overcharged naked singularities (lower right) where the dotted line gives the values of $g(r=b)$.} 
\end{figure*}

\begin{figure*}[t]
\begin{minipage}{.5\linewidth}
\centering	
\includegraphics[width=\linewidth]{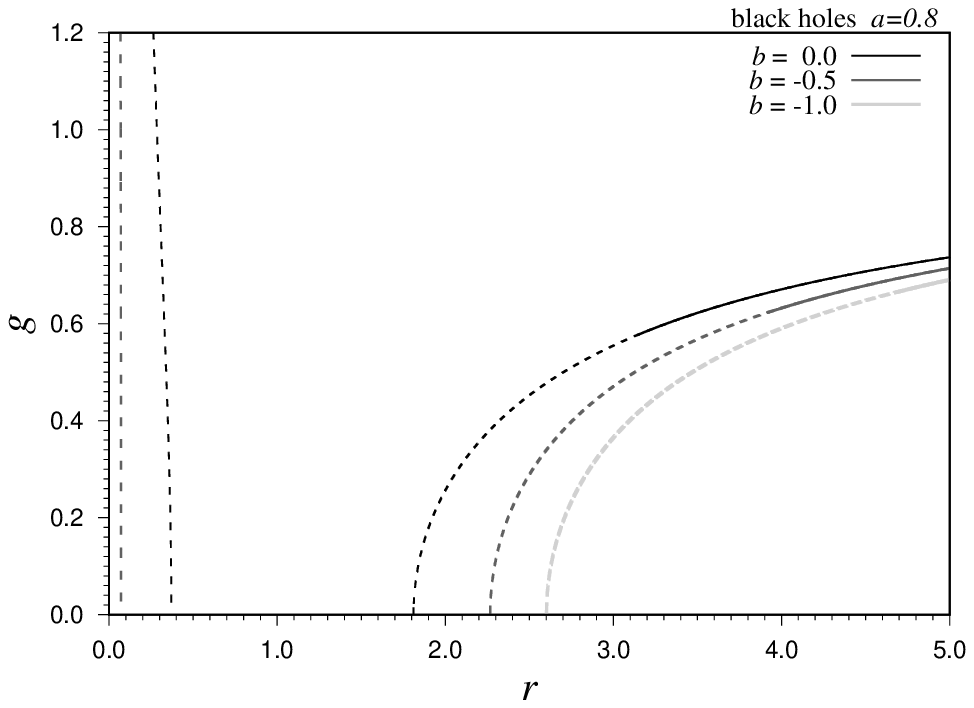}
\end{minipage}\hfill
\begin{minipage}{.5\linewidth}
\centering	
\includegraphics[width=\linewidth]{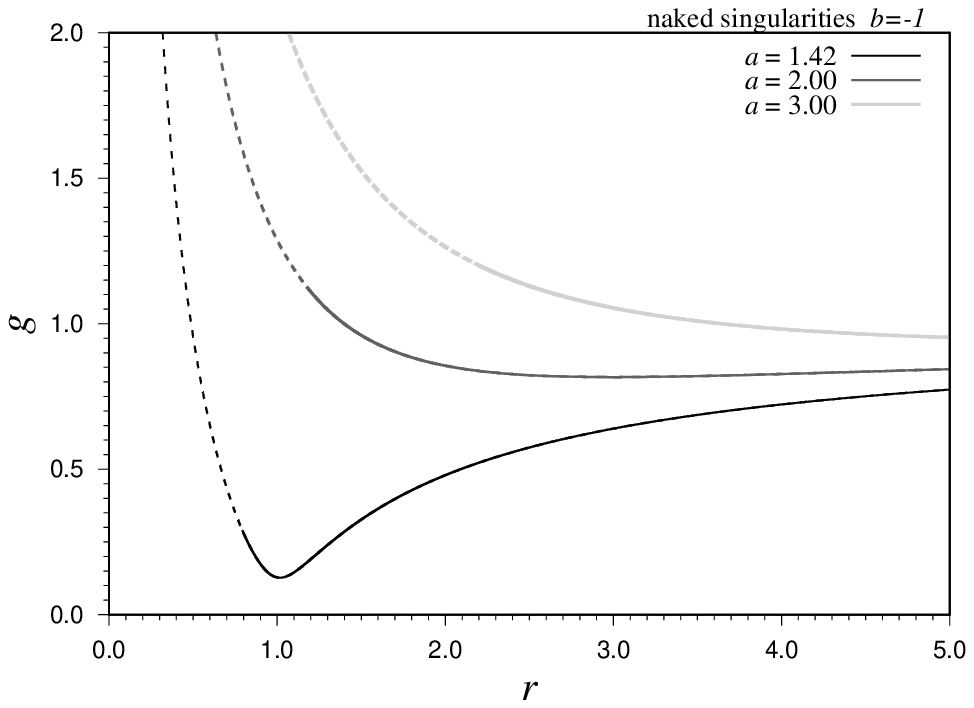}
\end{minipage}
\caption{\label{PNCneg} The radial profiles of the frequency shift of the PNC photons radiated by sources following circular geodesics in the KN spacetimes with negative tidal charges. The solid lines correspond to stable circular geodesics stable, the dashed lines to the unstable geodesics. The black holes (naked singularities) are represented by the left (right) figure.}
\end{figure*}

We represent in two ways the special optical signatures of the braneworld KN naked singularity spacetimes related to the representative congruence of radial null geodesics. First, we give the radial profiles of the frequency shift of the PNC photons radiated by sources following circular geodesics of appropriatelly selected KN spacetimes of various classes in Fig. \ref{PNC} for the KN spacetimes with $b>0$, and in Fig. \ref{PNCneg} for the KN spacetimes with $b<0$. We concentrate attention to the radial profiles $g(r;a,b)$ related to the black holes and the basical cases of the KN naked singularities, i.e., the mining and overcharged KN naked singularities with $b>0$ and ordinary type KN singularities with $b<0$. We can see that outside the black hole horizon, the radial profiles are always of the same character as in the Kerr black hole spacetimes, being increasing with radius $r$. Under the inner horizon of black holes with $b>0$ there is a maximum at the radius $r=4b/3$, while the profile is purely decreasing with increasing radius under the inner horizon of black holes with $b<0$. In the mining KN naked singularity spacetimes (class IIIa of the classification in \cite{Bla-Stu:2016:PHYSR4:}), the radial profile $g(r;a,b)$ demonstrates a minimum followed by a maximum, with decreasing $r$, approaching $g=0$ as the radius of the circular geodesic approaches the stable photon circular geodesic. It is explicitly demonstrated that a similar radial profile occurs for the KN naked singularity spacetimes of class IIIb, similar to the class IIIa, where some part of the radial profile located between the local minimum and maximum corresponds to the sequence of unstable circular geodesics. Similar radial profile of $g(r;a,b)$ occurs also in the case of the overcharged KN naked singularities, with different behavior at the innermost circular geodesics at $r=b$, where $g$ is non-zero. 
In the KN naked singularities or ordinary type (with $b<0$), the radial profile corresponds to the case of Kerr naked singularities, and we observe a local minimum of the profile at $r_{\mathrm{g(min)}}$, located at the region of stable circular geodesics. 

\begin{figure*}
\begin{minipage}{.5\linewidth}
\centering	
\includegraphics[width=\linewidth]{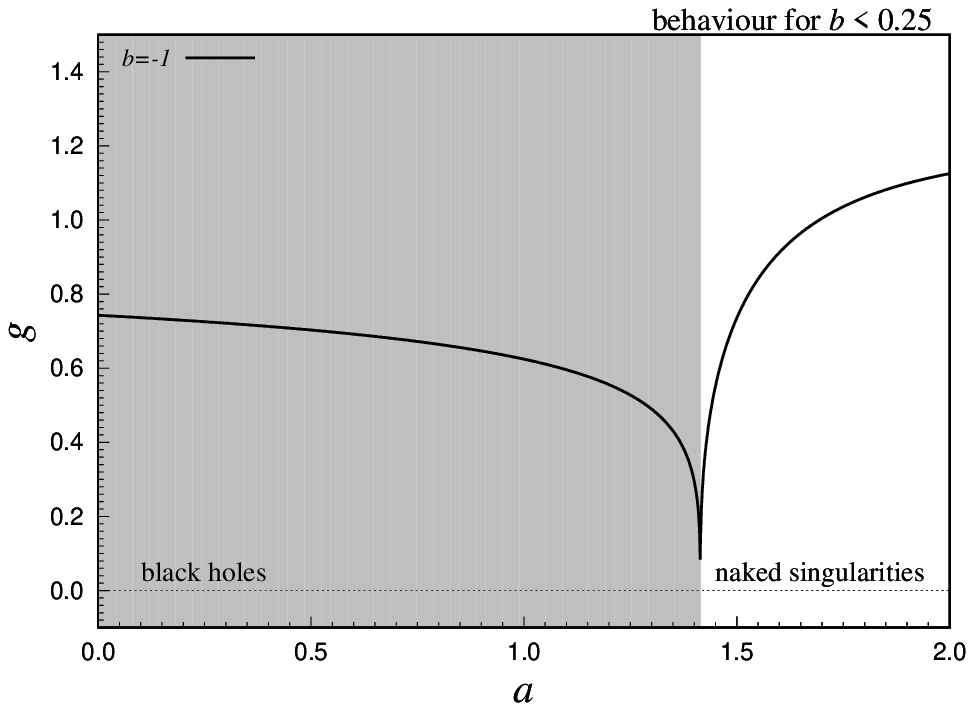}
\end{minipage}\hfill
\begin{minipage}{.5\linewidth}
\centering	
\includegraphics[width=\linewidth]{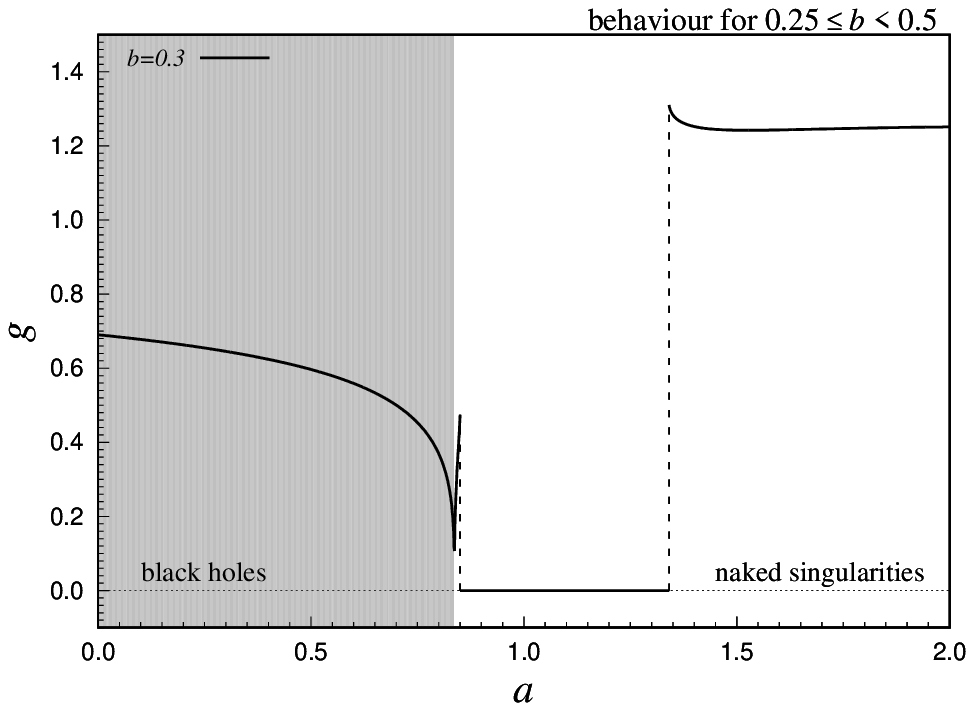}
\end{minipage}
\begin{minipage}{.5\linewidth}
\centering	
\includegraphics[width=\linewidth]{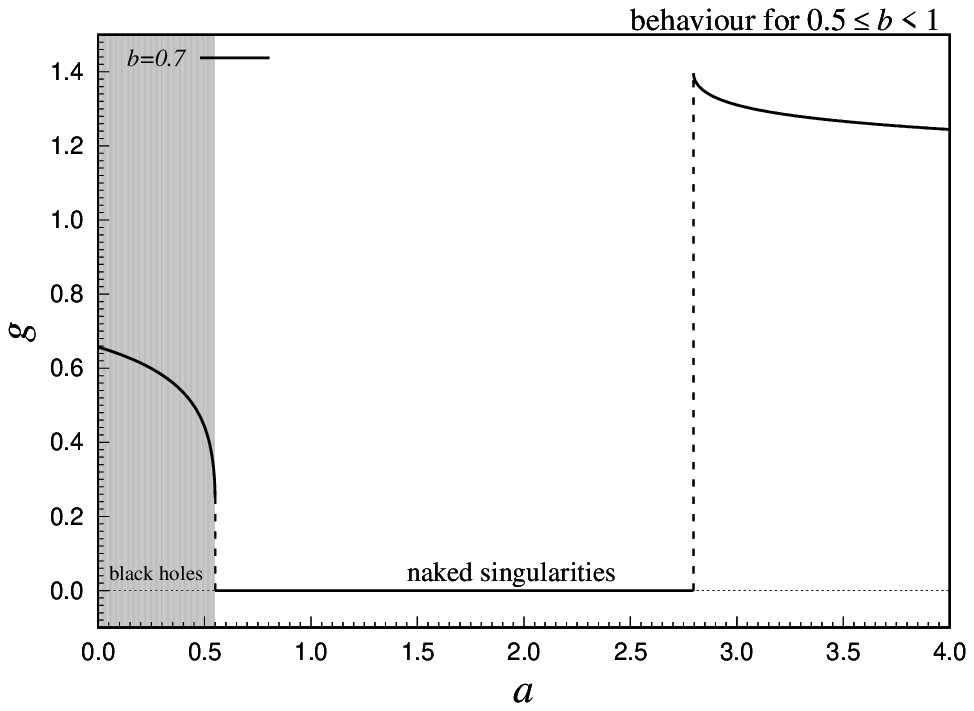}
\end{minipage}\hfill
\begin{minipage}{.5\linewidth}
\centering	
\includegraphics[width=\linewidth]{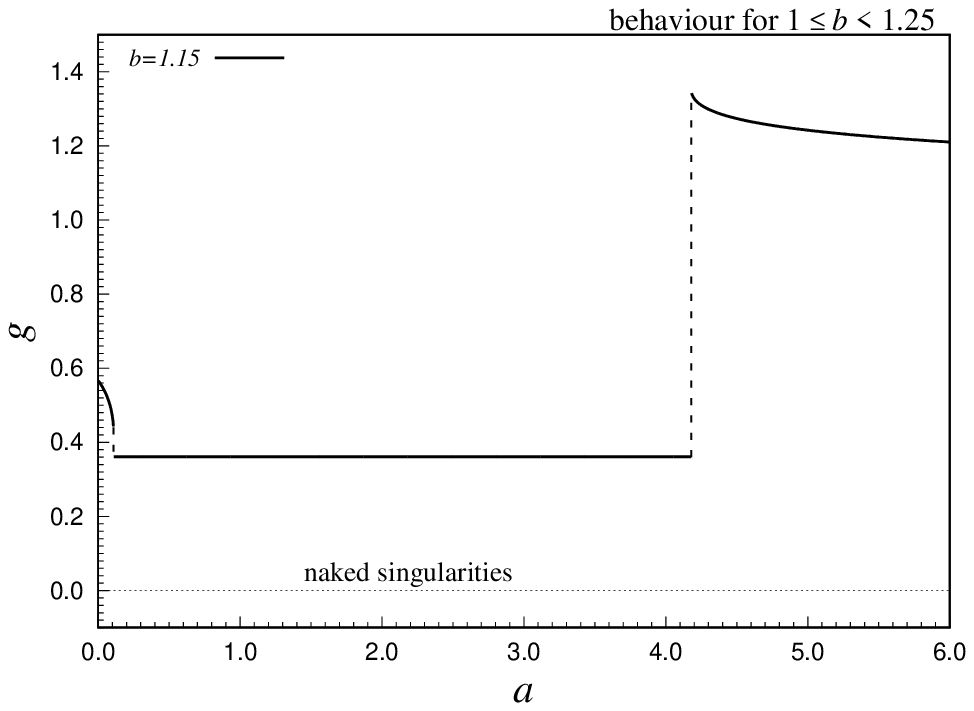}
\end{minipage}	
\caption{\label{PNC2} The dependence of the frequency shift $g(a,b)$ from the innermost stable circular geodesics in the braneworld Kerr spacetimes in dependence on the spin parameter $a$ for appropriately selected values of the tidal charge $b$. We give four characteristic profiles for typical values of the braneworld tidal charge. The left upper characterizes the cases of $b<0.25$, the right upper is relevant for $0.25<b<0.5$, the left lower holds for $0.5<b<1$, and the right lower reflects the spacetimes with $1<b<1.25$. The case of $b>1.25$ is similar to previous one, but the flat line starts at $a=0$.} 
\end{figure*}

Second, we give dependence of the frequency shift from the innermost stable circular geodesics in the braneworld Kerr spacetimes in dependence on the spin parameter $a$ for appropriately selected values of the tidal charge $b$ in Fig. \ref{PNC2}. Recall that the innermost stable circular geodesics, giving an endpoint of Keplerian accretion, can be of three types: marginally stable, stable photon circular geodesics, or innermost orbits at $r=b$. In the case of the effectively ordinary KN naked singularities (for the spacetimes with both $b<1$ and $b>1$), we consider as relevant the marginally stable orbit of the outer region of stable circular geodesics. 

We select four regions of the value of the braneworld parameter $b$ giving different behaviour of the profiles $g(a;b)$. In the case of $b<0.25$, including the negative values of $b$, the profiles reflect only the marginally stable geodesics, the minimum of the profile corresponds to the extreme KN black holes. In the case of $0.25<b<0.5$, the naked singularity region has two subregions related to the marginally stable orbits (for the effectively ordinary KN naked singularities), interrupted by the region related to the mining KN naked singularities where $g=0$. In the case of $0.5<b<1$, the region of the mining KN naked singularity spacetimes reaches the region of black holes, and only the outer region of effectively ordinary KN naked singularities remains. In the case of $1<b<1.25$ of overcharged KN naked singularities, for appropriately chosen values of the spin, when the Keplerian accretion ends at $r=b$, the radial profile $g(r;a,b)$ is independent of spin, being given by $\sqrt{b(b-1)/b}$, however, the effectively ordinary KN naked singularities exist for sufficiently small and sufficiently large values of the spin parameter $a$ and $g(r;a,b)$ is decreasing with increasing spin. In the case of $b>1.25$, the effectively ordinary overcharged KN naked singularity spacetimes exist only for sufficiently large values of spin $a$. Notice that the frequency shift of PNC photons radiated from the marginally stable orbits of the effectively ordinary KN naked singularity spacetimes with large values of spin $a$ takes relatively high values corresponding to blue-shift, $g(a;e) > 1$. 

Clearly, in the braneworld KN naked singularity spacetimes with $b>0$, the radial profiles of the frequency shift of the PNC photons radiated from the circular geodesics have qualitatively different character in comparison to the Kerr naked singularity case given in \cite{Stu:1980:BAC:}, while it is of the same character in the KN naked singularity spacetimes with $b<0$. 

\begin{figure}[t]
\begin{center}
\centering
\includegraphics[width=1\linewidth]{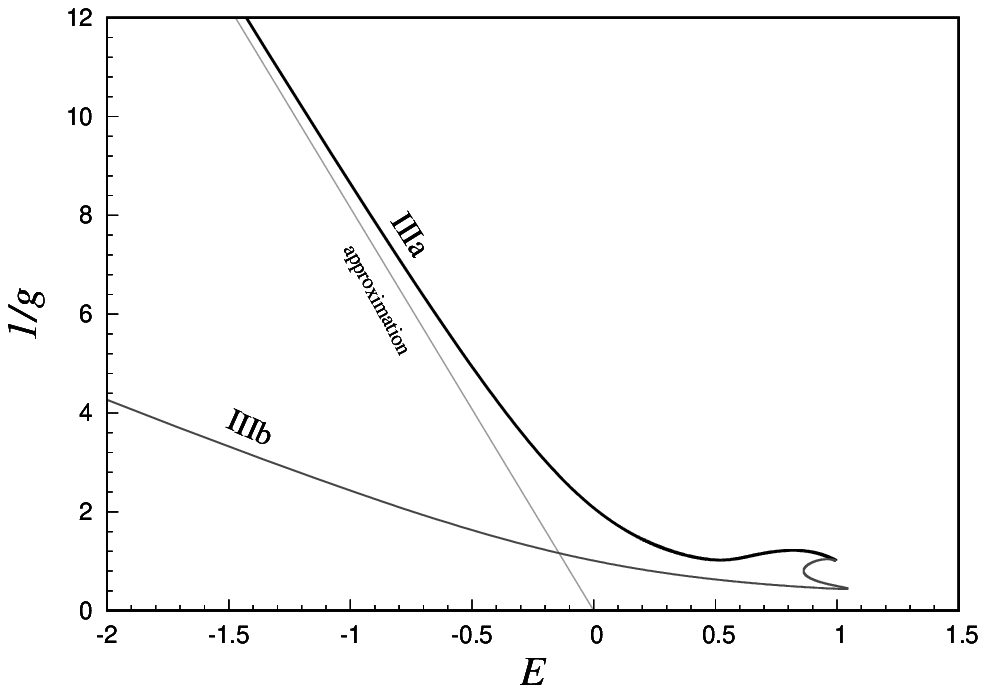}
\caption{\label{ge2} 
The frequency shift $1/g = 1+z$ of the PNC photons radiated by static observers at large distance and observed by geodesic observers orbiting the mining KN spacetimes of class IIIa, or the related effectively ordinary KN naked singularity spacetimes of class IIIb, is given in terms of the specific energy $E$ of the circular geodesic orbit. Such a way of characterizing the frequency shift is convenient especially for the observers orbiting in the mining regime in the mining KN naked singularity spacetimes of class IIIa ($a=1.5, b=0.8$) and the class IIIb ($a=3.5, b=0.5$), as the frequency shift can be expressed by very simple approximative relation in the limit of $E \to -\infty$. We give the exact energy (or corresponding radial) profiles, and compare them with the approximative line that is very precise for sufficiently low specific energy, and well approximates the exact relation for $E<-1$.}
\end{center}
\end{figure} 

\subsection{Frequency shift of CMB observed at the mining stable geodesics}

Now we study the frequency shift of photons incoming from large distance (e.g., the CMB photons) as observed by observers orbiting in the deep gravitational well of braneworld KN black holes and naked singularities. We focus our attention to the special case of the observers following the circular geodesics of the mining regime, with radius extremely close to the radius of the stable circular photon geodesics of the mining KN naked singularity spacetimes. The frequency shift is then governed by the general frequency formula used before, but with inverse meaning of the observer (following the circular geodesic) and emitter (for simplicity static source at large distance) -- this formally means $U_{\mathrm{o}} \to U_{\mathrm{e}}$ and $U_{\mathrm{e}} \to U_{\mathrm{o}}$. Thus we could simply use the inversion of the formula in Eq. (\ref{shiftco}) with the frequency shift denoted as $1/g$. However, for the purpose of the study in the mining regime, it is convenient to apply an alternate formula where we use directly the specific energy $E$ and specific angular momentum $L$ of observer following a circular geodesic in the mining regime, as in this regime extremely small changes of radius imply extremely large changes of $E$ and $L$ that are thus appropriate quantities for description of the physical phenomena in the mining regime. The frequency shift $(1+z)_{\mathrm{MNS}}\equiv1/{g_{\mathrm{MNS}}}$ is then given by the relation 
\begin{equation}
(1+z)_{\mathrm{MNS}}=-g^{tt} E +g^{t\phi} L + \left(g^{t\phi} E - g^{\phi\phi} L\right)\frac{l}{\gamma}\, ,
\end{equation}
where $\gamma$ and $l$ denote the energy and axial angular momentum of the arriving photon. 

In the case of the observers in the braneworld mining KN naked singularity spacetimes, following the circular geodesics in the mining regime with specific energy $E<0\, $, the radius of the circular geodetic can be well approximated as $r\approx r_{\mathrm{ph}}$, where $r_{\mathrm{ph}}$ represents the radius of the stable photon circular geodesic that is related to the spacetime parameters $a$ and $b$ as 
\begin{equation}
a = a_{\mathrm{ph\pm}} =\pm \frac{3r_{\mathrm{ph}}-r_{\mathrm{ph}}^2-2b}{2\sqrt{r_{\mathrm{ph}}-b}}\, .
\end{equation}  

Since $r\approx r_{\mathrm{ph}}\,$, in the mining regime the following relations hold 
\begin{equation}
E = \frac{\gamma}{l}L\, ,
\end{equation}
and
\begin{equation}
\frac{l}{\gamma} = \pm\frac{3r_{\mathrm{ph}}+r_{\mathrm{ph}}^2-2b}{2\sqrt{r_{\mathrm{ph}}-b}}\, .
\end{equation}
Using these relations, we arrive for the frequency shift of incoming photons observed by observers following circular geodesics in the mining regime to the resulting approximative formula
\begin{equation}
1+z_{\mathrm{MNS}}=\frac{2Er_{\mathrm{ph}}(a,b)}{r_{\mathrm{ph}}(a,b)-1}\, .
\end{equation}
Note that in the mining regime, there is $r_{\mathrm{ph}}(a,b)<1$ for all the values of energy $E<0$, so that we always obtain a frequency blueshift. In Fig. \ref{ge2}, we demonstrate the radial profile of the frequency shift related to the observers orbiting a typical braneworld mining KN naked singularity of Class IIIa with $a=1.5,b=0.8$. The frequency shift is constructed for both the approximative and exact formulae, and it is demonstrated explicitly that the approximative formula works quite well for orbits in the mining regime, with sufficiently large values of magnitude of the specific energy of the circular geodesic. For comparison we show the frequency shift radial profiles for the braneworld KN spacetimes of the class IIIb (effectively ordinary type, where we consider the inner mining region). 

For the CMB radiation, the total frequency shift is governed by the simple formula 
\begin{equation}
1+z_{\mathrm{tot}} = (1+z_{\mathrm{MNS}}) \times (1+z_{\mathrm{CMB}}) , 
\end{equation}
where $1+z_{\mathrm{CMB}}$ denotes the standard cosmological frequency shift of the CMB radiation at the corresponding cosmic time, related to local stationary observers at sufficiently large distance from the KN naked singularity (or black hole) where the spacetime can be considered nearly flat. Clearly, for an arbitrarily large cosmic redshift of the CMB radiation, one can obtain any sufficiently high total blueshift of the CMB radiation, if the circular orbit of the observer is located sufficiently close to the stable photon circular geodesic of the mining KN naked singularity. Its characteristic specific energy can be determined by the relation 
\begin{equation}
    E \sim  \frac{r_{\mathrm{ph}}(a,b)-1}{2r_{\mathrm{ph}}(a,b)} \frac{1+z_{\mathrm{tot}}}{1+z_{\mathrm{CMB}}} .  
\end{equation}
We have to stress that such a mechanism of enhancement of the CMB radiation, representing an enormous energy supply to the orbiting observer, can work, if the magnitude of the energy of the circular orbit is sufficiently smaller than the mass parameter $M$ of the braneworld mining KN spacetime \cite{Bla-Stu:2016:PHYSR4:}. 

\section{Appearance of the sky to observers orbiting on the circular geodesics}

\begin{figure*}[t]
\begin{minipage}{.5\linewidth}
\centering	
\includegraphics[width=\linewidth]{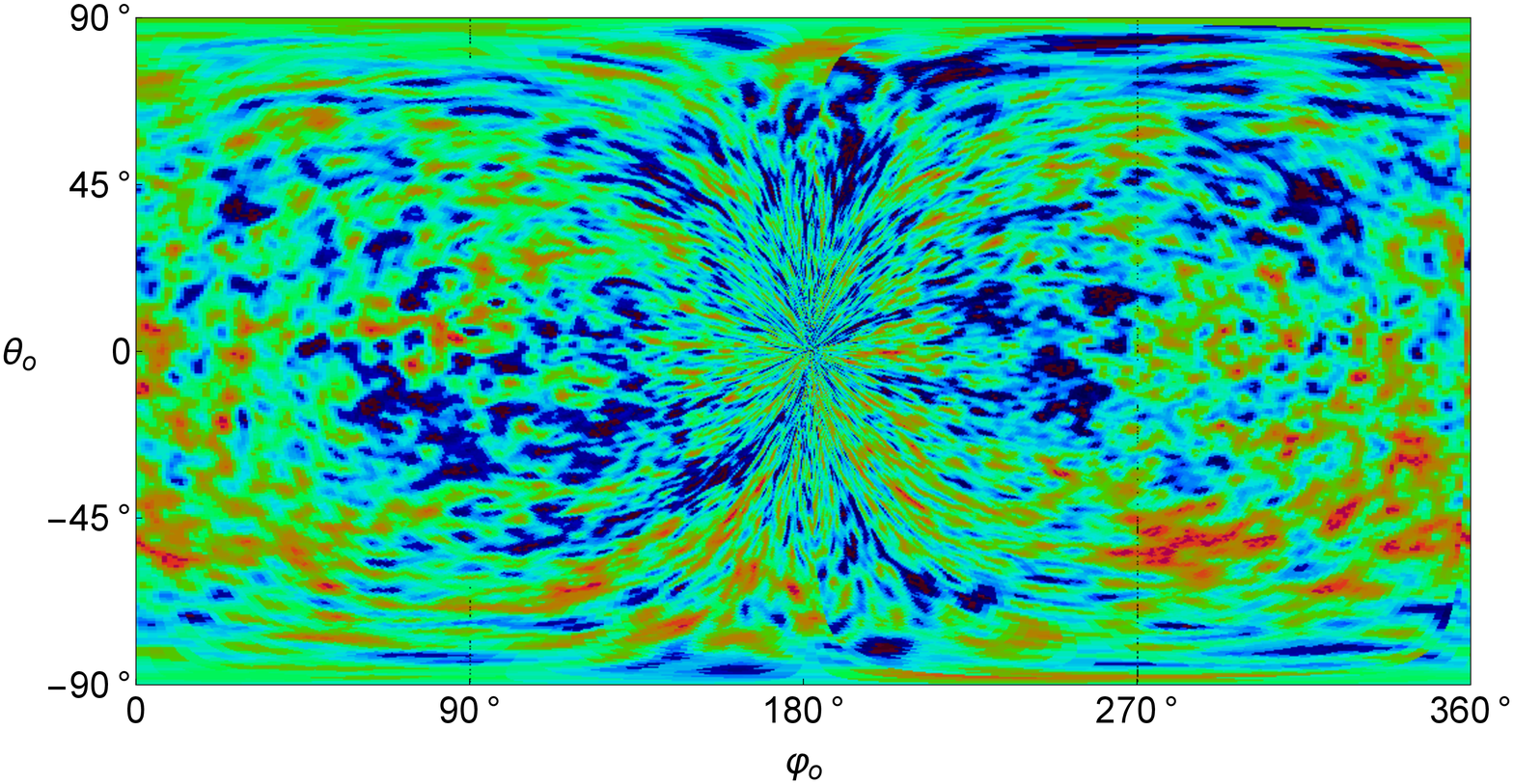}
\end{minipage}\hfill
\begin{minipage}{.5\linewidth}
\centering	
\includegraphics[width=\linewidth]{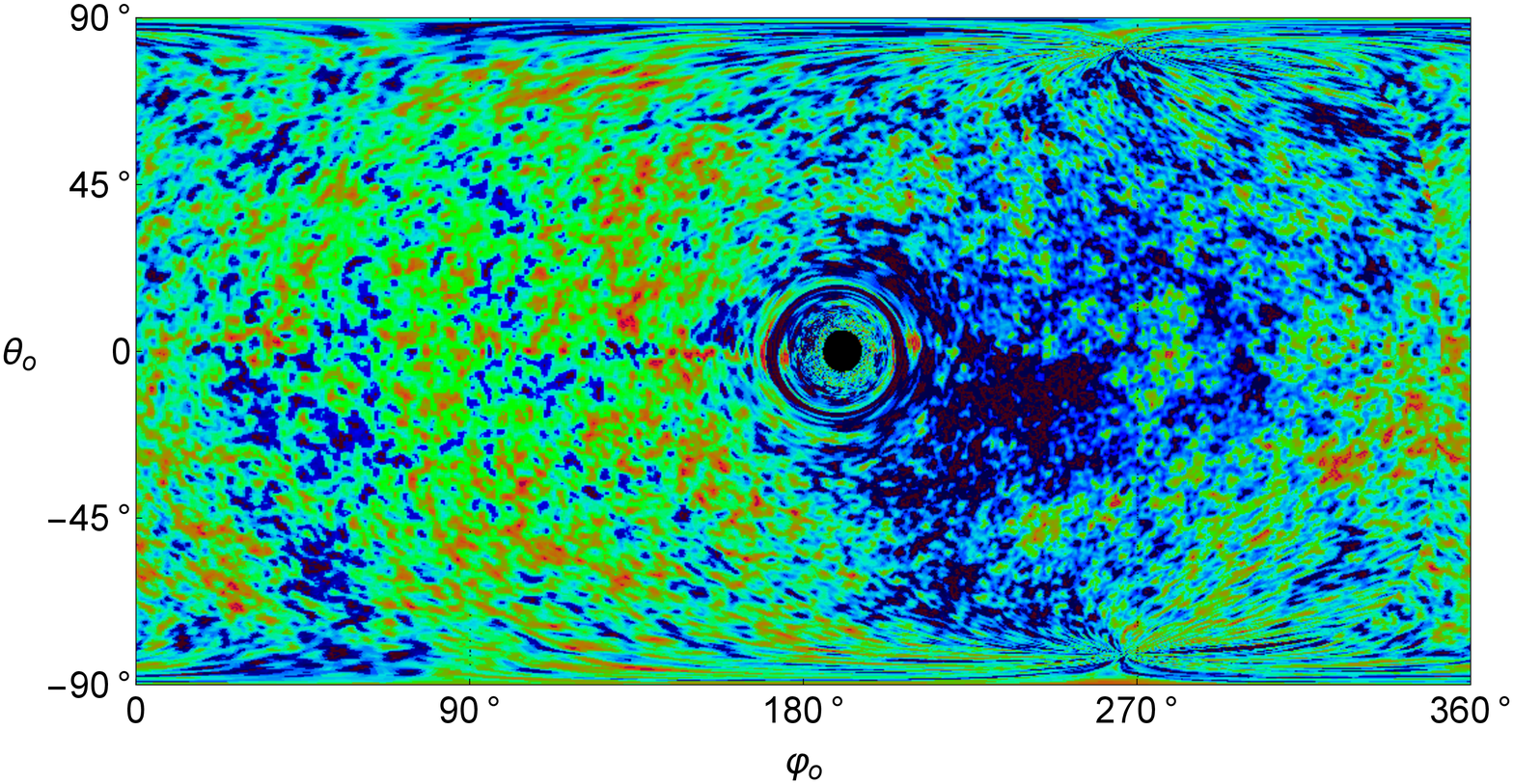}
\end{minipage}
\begin{minipage}{.5\linewidth}
\centering	
\includegraphics[width=\linewidth]{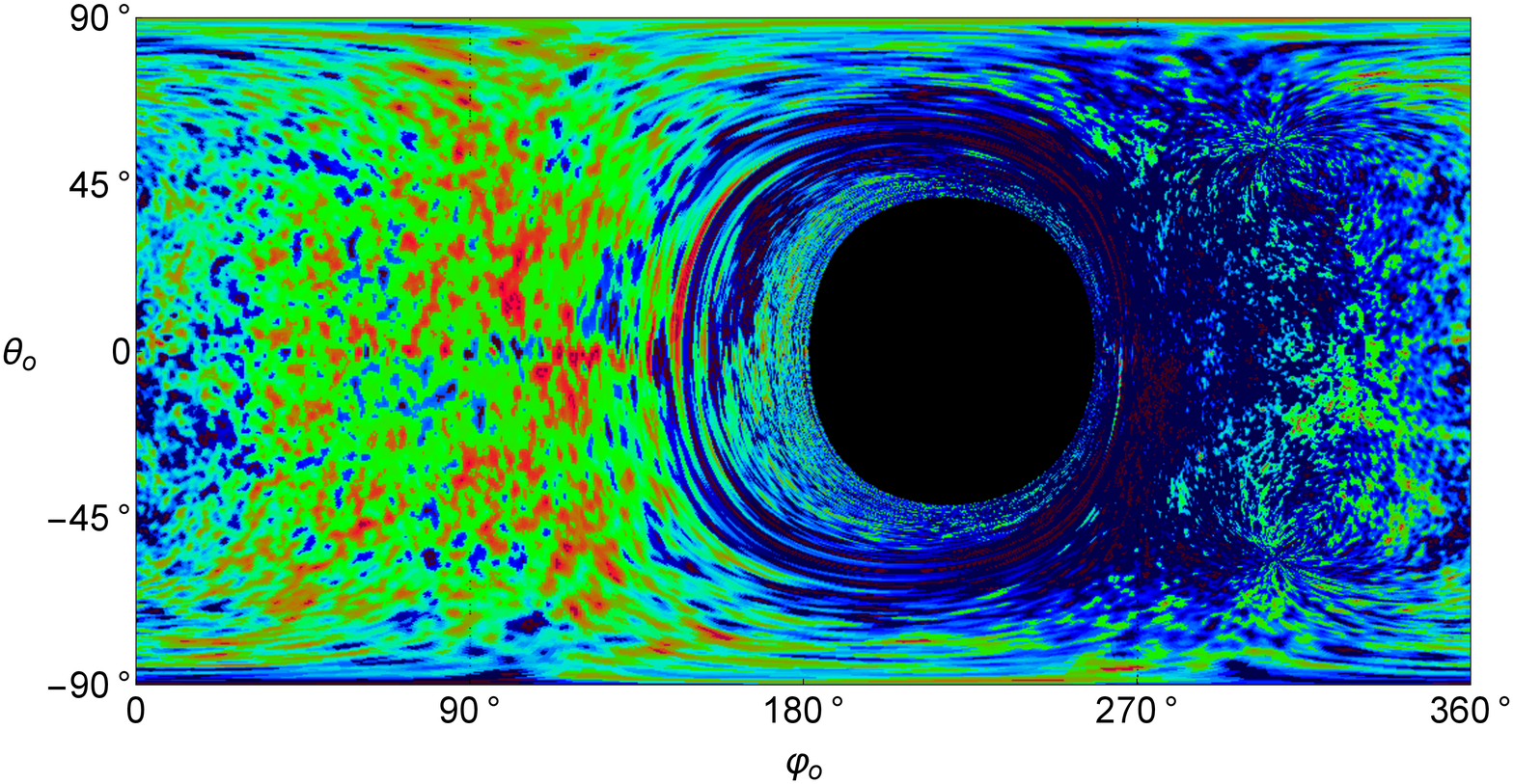}
\end{minipage}\hfill
\begin{minipage}{.5\linewidth}
\centering	
\includegraphics[width=\linewidth]{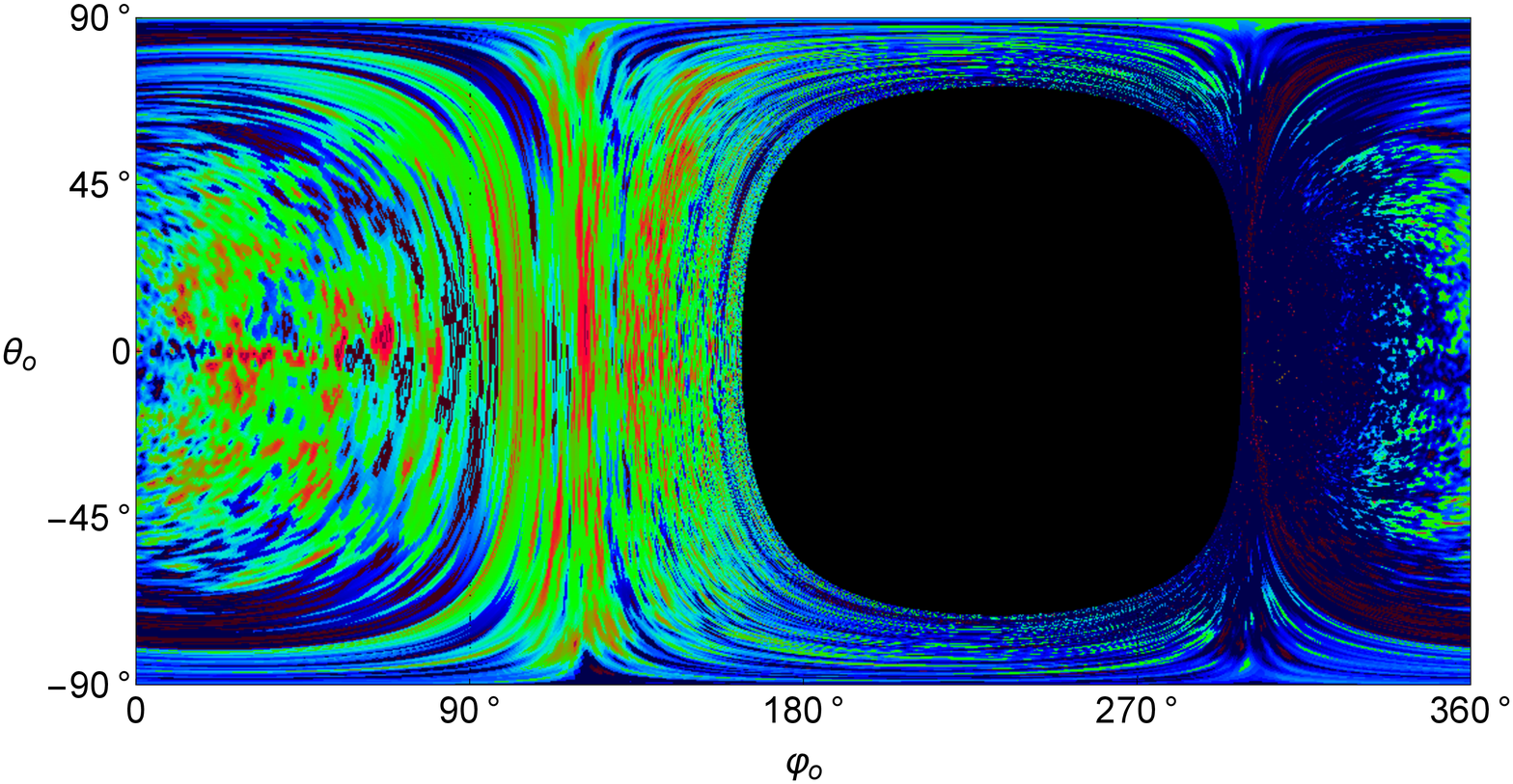}
\end{minipage}
	\caption{\label{skyBH}Sky as observed in the CGFs in the KN black hole spacetime with parameters $a=0.9$, $b=1/10$. The radial coordinate of the circular geodesics of the CGFs takes values $r_0=10^3$, $50.0$, $5.0$, and $2.014$ (from left to right). }
\end{figure*}

\begin{figure*}[t]
\begin{minipage}{.5\linewidth}
\centering	
\includegraphics[width=\linewidth]{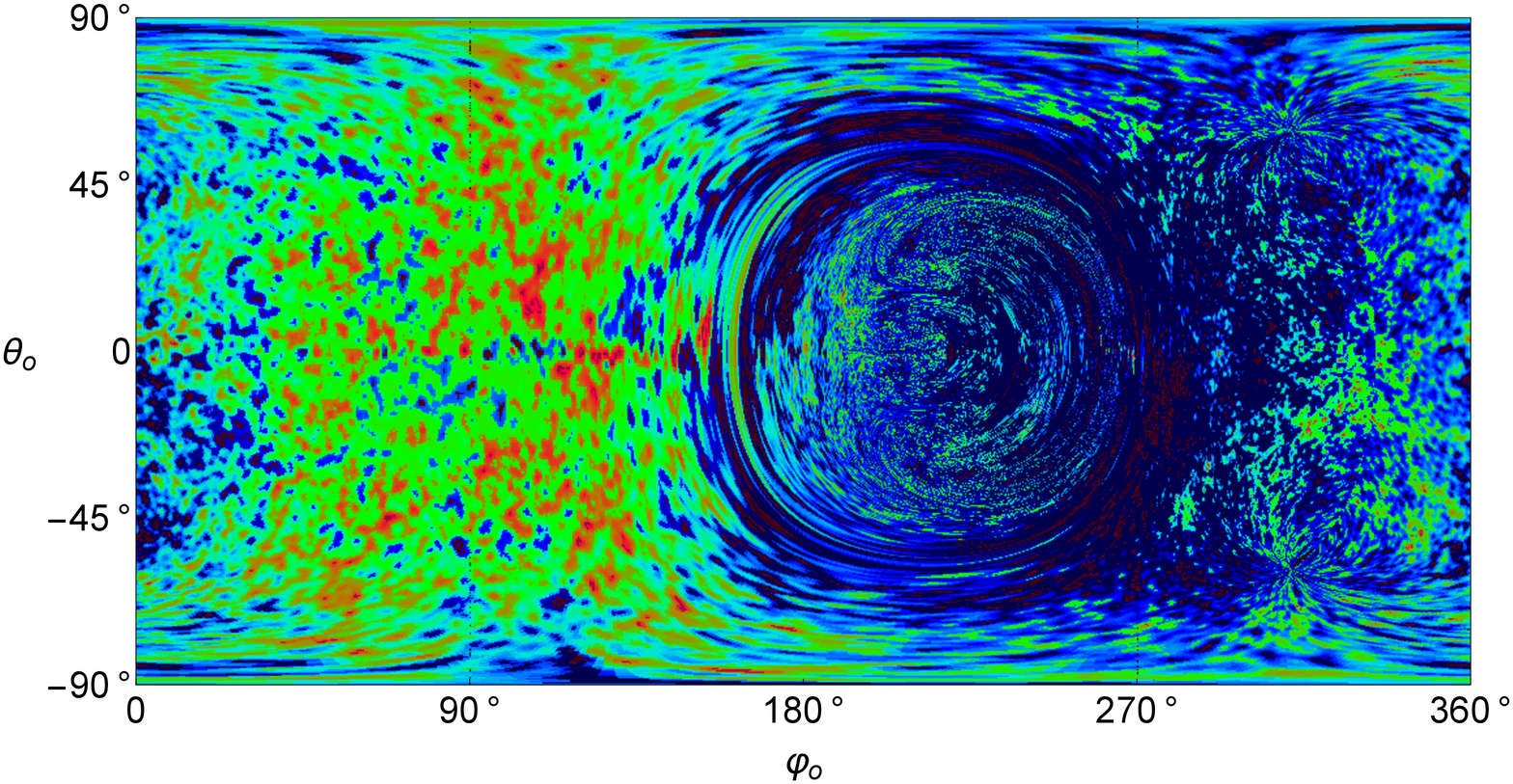}
\end{minipage}\hfill
\begin{minipage}{.5\linewidth}
\centering	
\includegraphics[width=\linewidth]{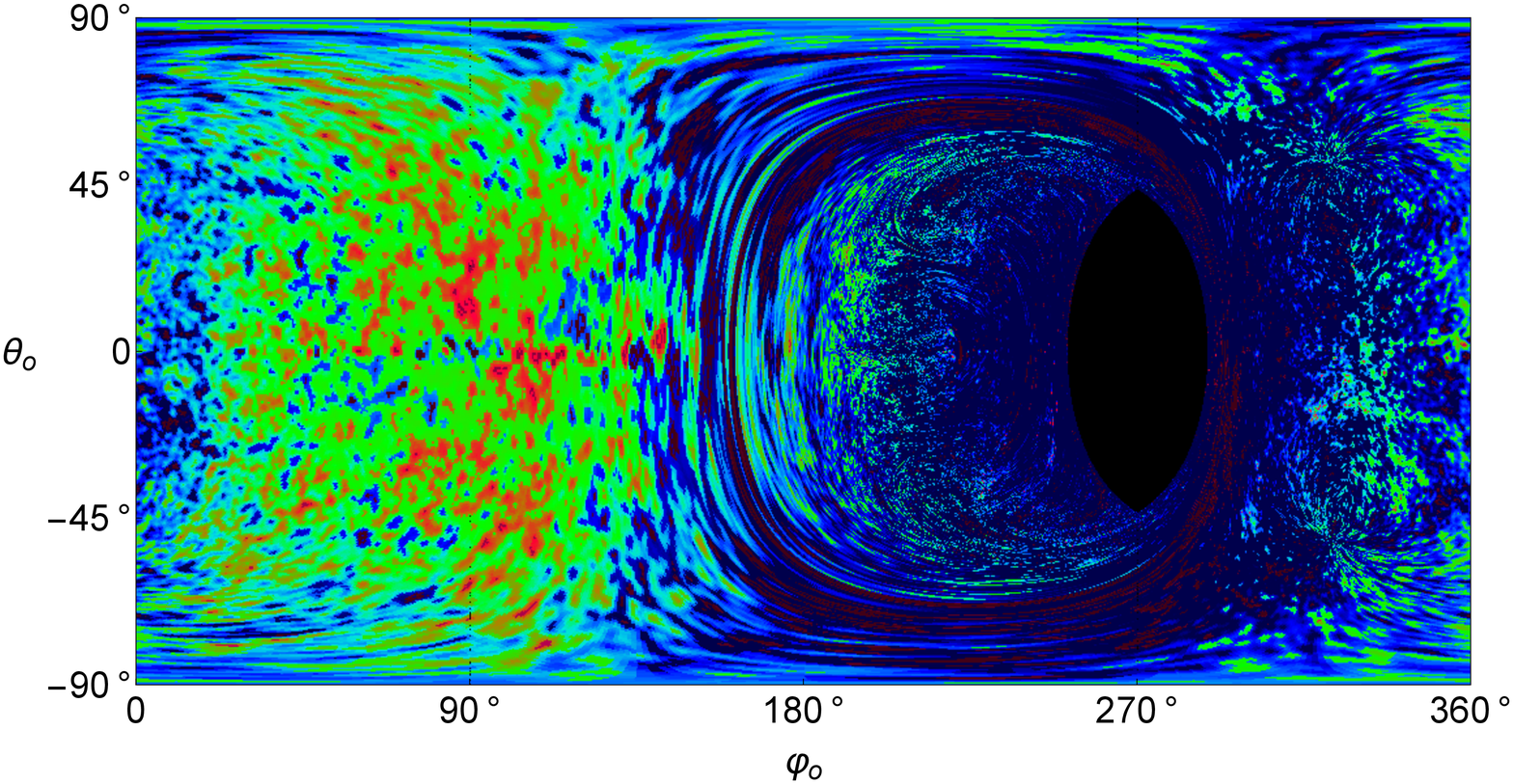}
\end{minipage}
\begin{minipage}{.5\linewidth}
\centering	
\includegraphics[width=\linewidth]{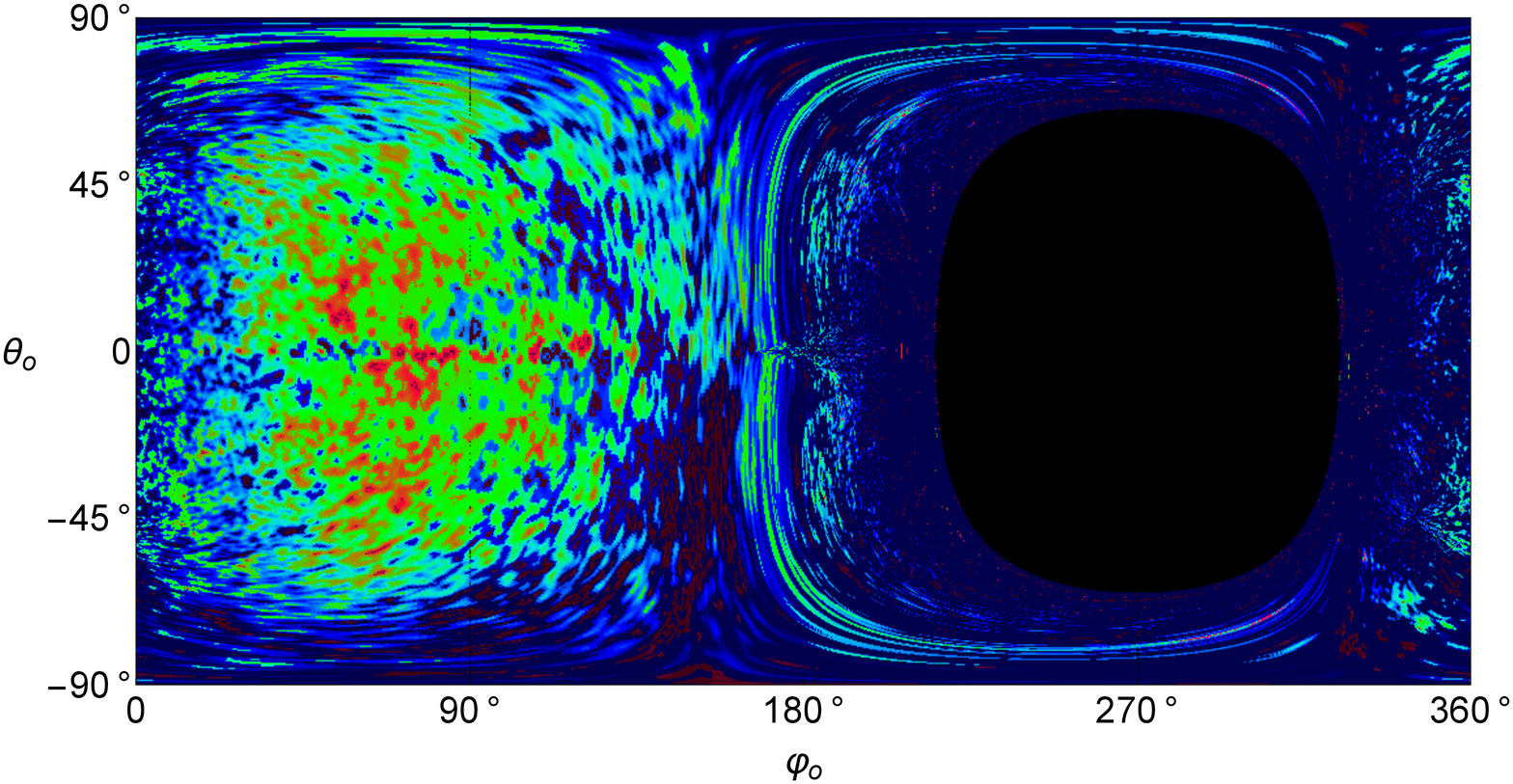}
\end{minipage}\hfill
\begin{minipage}{.5\linewidth}
\centering	
\includegraphics[width=\linewidth]{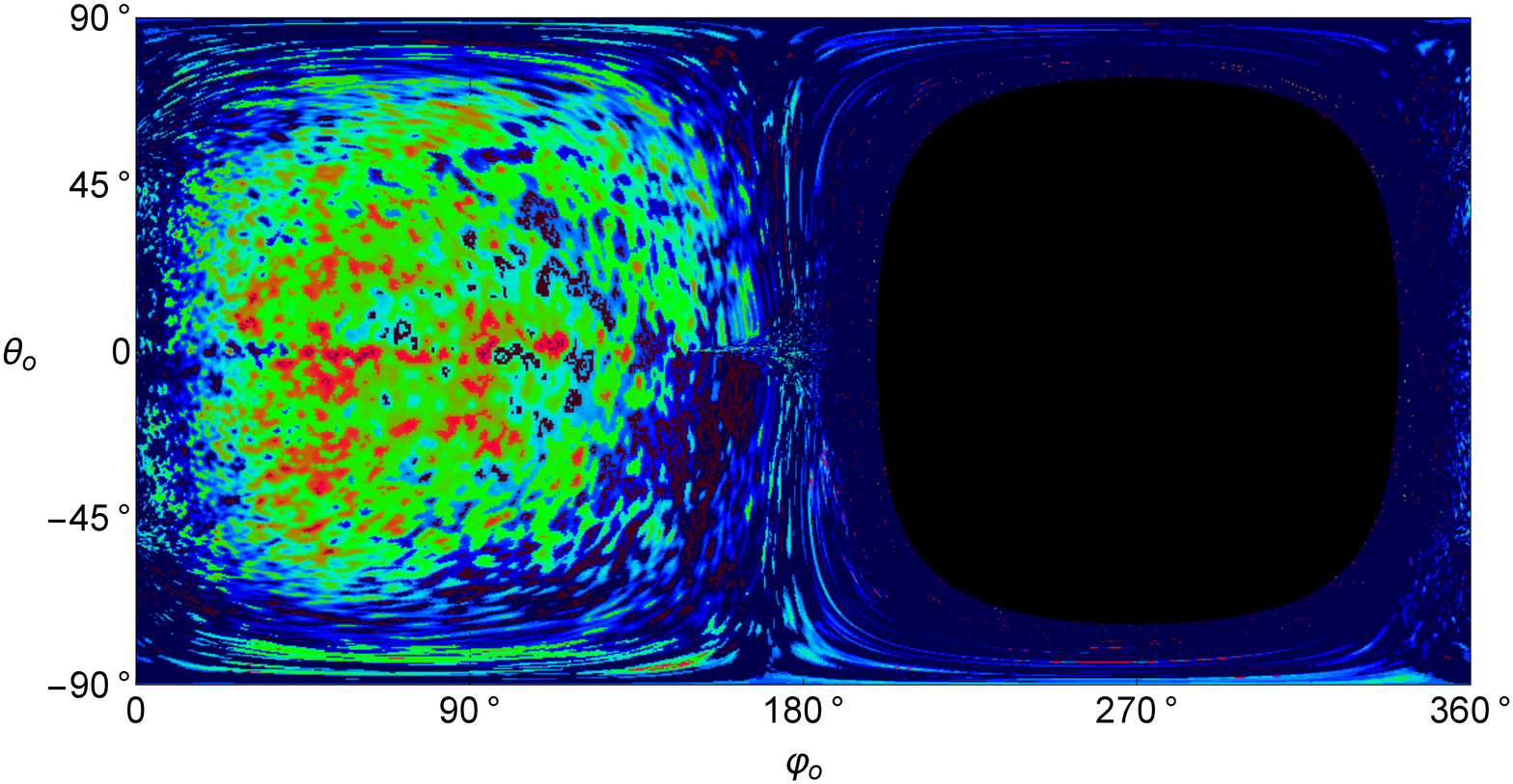}
\end{minipage}	
\caption{\label{skyNS} Sky as observed in the CGFs in the ordinary KN naked singularity spacetime with parameters $a=1.48803$, $b=1/10$. The radial coordinate of the circular geodesics of the CGFs takes values $r_0=5.0$, $3.0$, $1.5$, and $0.8643$ (from left-top to right-bottom). }
\end{figure*}

\begin{figure*}[t]
\begin{minipage}{.5\linewidth}
\centering	
\includegraphics[width=\linewidth]{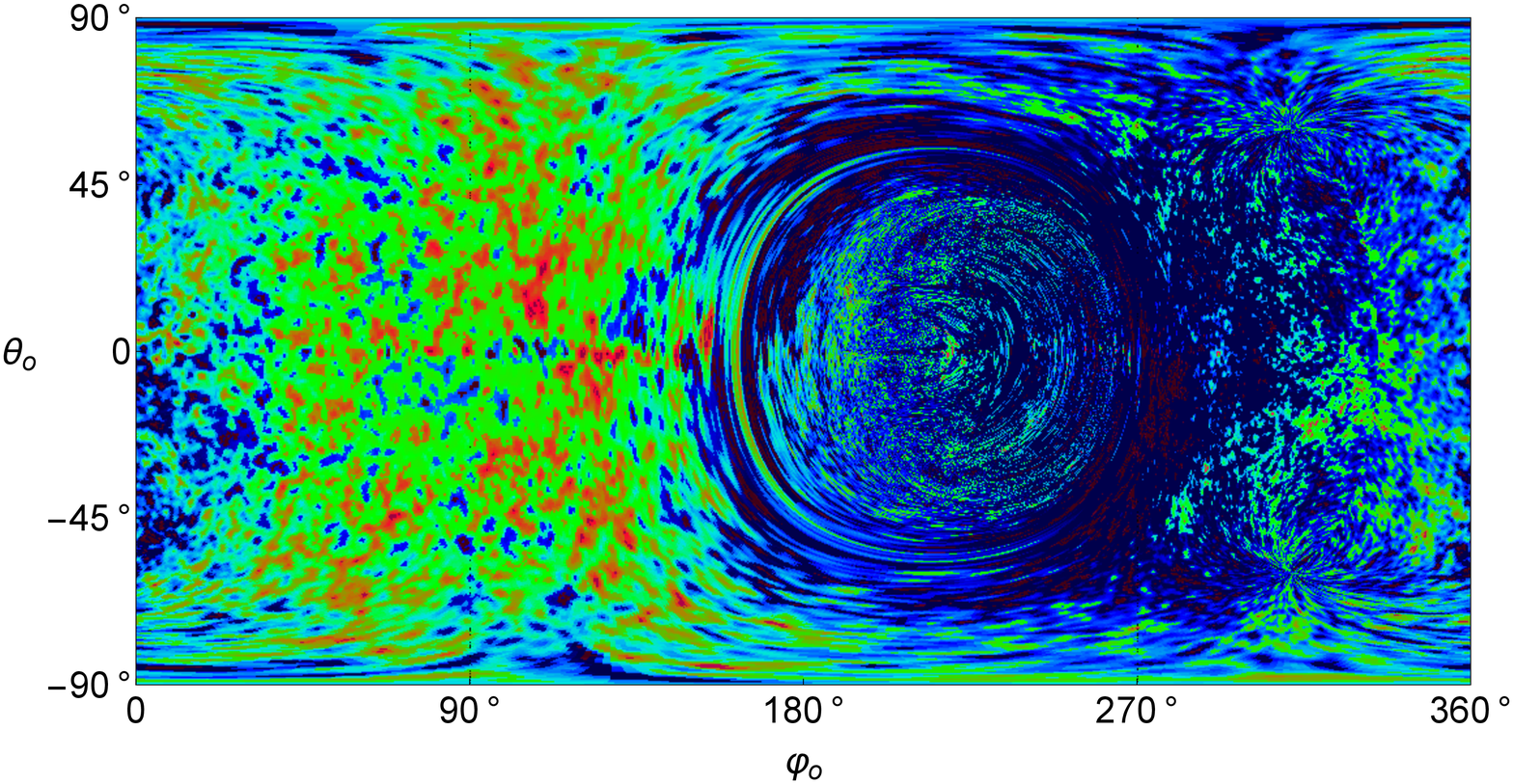}
\end{minipage}\hfill
\begin{minipage}{.5\linewidth}
\centering	
\includegraphics[width=\linewidth]{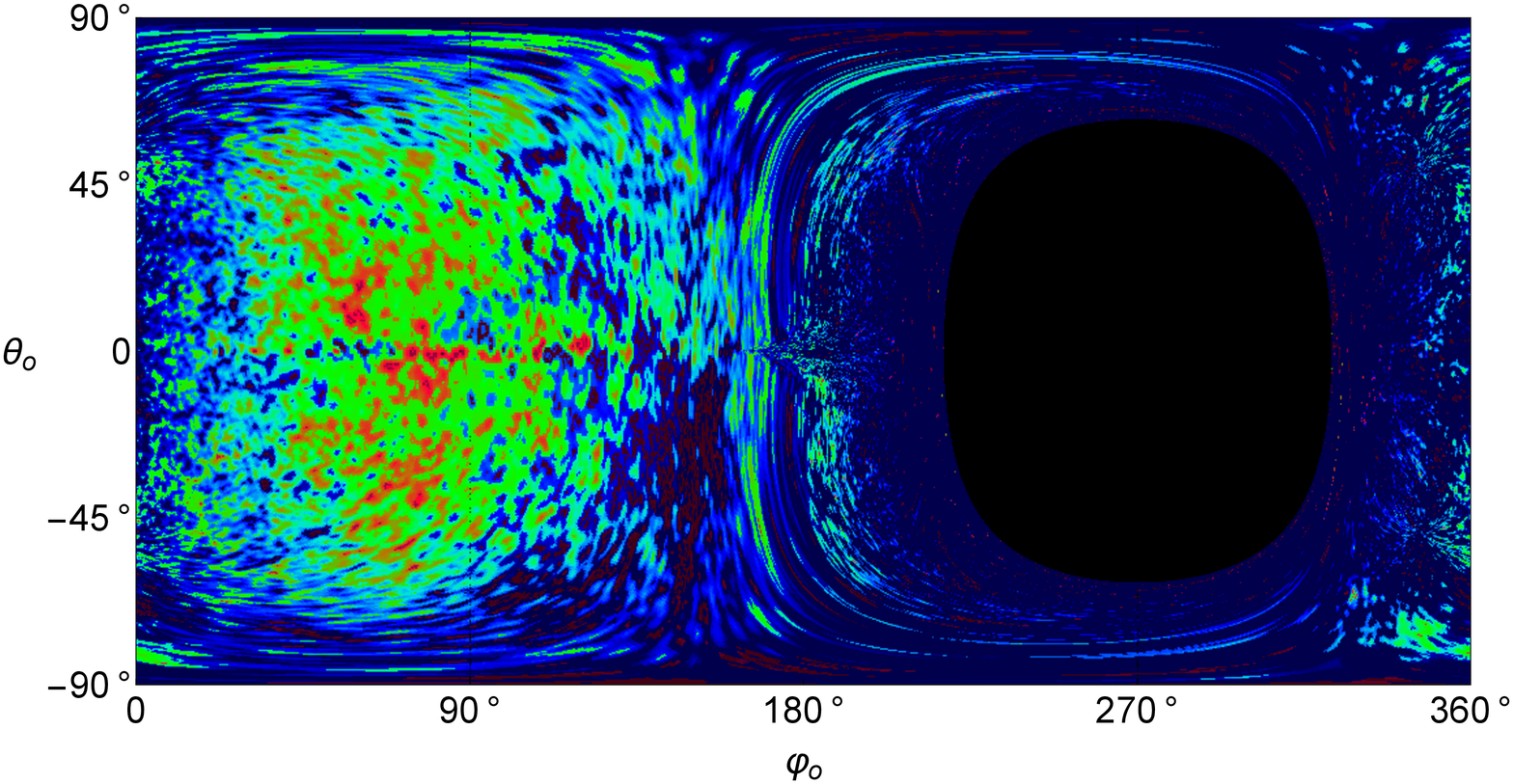}
\end{minipage}
\begin{minipage}{.5\linewidth}
\centering	
\includegraphics[width=\linewidth]{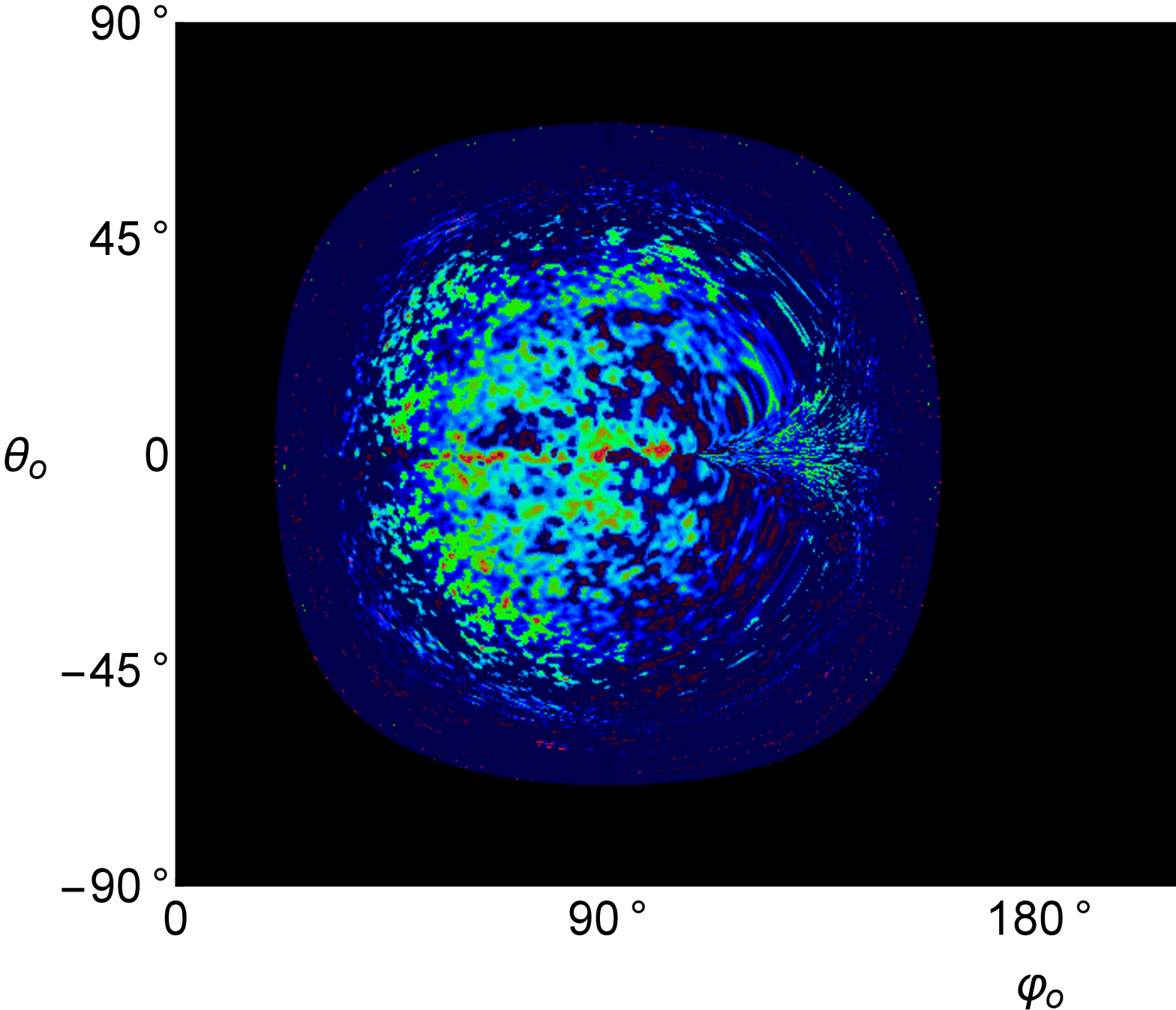}
\end{minipage}\hfill
\begin{minipage}{.5\linewidth}
\centering	
\includegraphics[width=\linewidth]{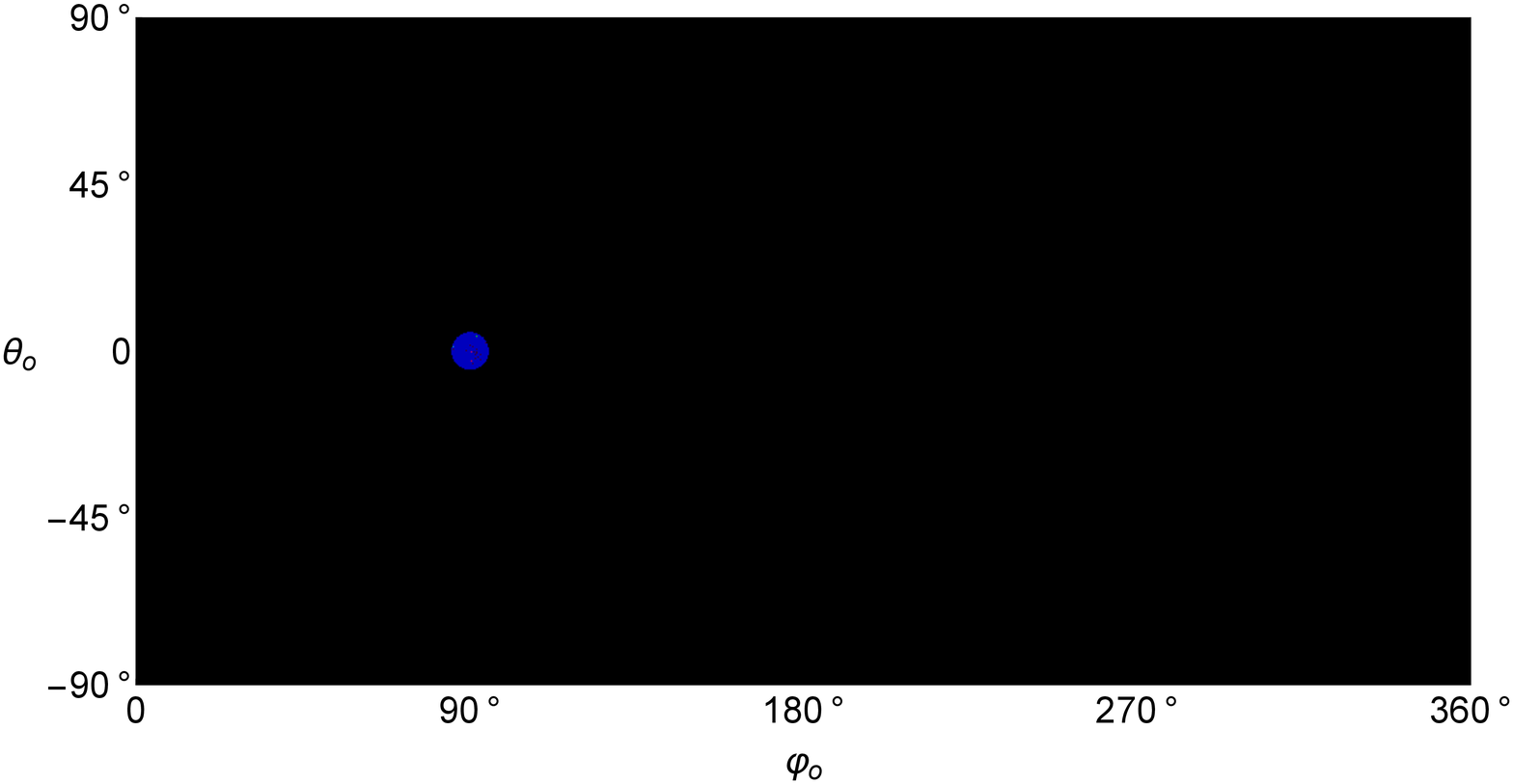}
\end{minipage}	
	\caption{\label{skyMNS} Sky as observed in the CGFs in the mining KN naked singularity spacetime with parameters $a=1.48803$, $b=1/3$. The radial coordinate of the circular geodesics of the CGFs takes values $r_0=5.0$, $1.5$, $0.35$, and $0.3397$ (from left-top to right-bottom).}
\end{figure*}

In order to calculate the appearance of the sky as given for the observers following the circular geodesics, we have to make transformation from the LNRFs to the frame tetrad related to the circular geodesic observers. For this purpose we have to calculate first the velocity of the geodesic observers relative to the LNRFs and realize the Lorentz transformation to the Circular Geodesic Frames (CGF). Then we can make calculations in the CGFs giving the deformations of the observed sky in dependence on the KN spacetime parameters $a$ and $b$ and the radius of the circular geodesic. 

\subsection{Velocity of the circular geodesics relative to LNRFs}

The circular geodesic frames have just single non-zero, axial, velocity component relative to the LNRFs. We can express the axial $\phi$-component of the velocity of the circular motion with 4-velocity $U^{\mu}=(U^t,0,0,U^{\phi})$ relative to the LNRFs in the form   
\begin{equation}
\mathcal{V}^{(\phi)} \equiv \frac{U^{(\phi)}}{U^{(t)}} = \frac{U^\mu {\bf e}^{(\phi)}_{\mu}}{U^\nu {\bf e}^{(t)}_{\mu}}\, , 
\end{equation} 
using the LNRF tetrad. 
The axial velocity of the orbiting observers, related the LNRFs, takes the form 
\begin{equation}
\mathcal{V}^{(\phi)}_{\mathrm{LNRF}}= \frac{g_{\phi\phi}\Omega+g_{t\phi}}{\sqrt{g_{t\phi}^2-g_{tt}g_{\phi\phi}}}
\end{equation}
where 
\begin{equation}
\Omega \equiv \frac{U^\phi}{U^t}
\end{equation}
represents the angular velocity of the circular orbits of the observers relative to distant static observers. 

Using for the angular velocity the Keplerian relation relevant for the circular geodesics, we arrive to  
\begin{equation}
\mathcal{V}^{(\phi)}_{\mathrm{LNRF}\pm}=\pm \frac{\left(r^2+a^2\right)\sqrt{r-b}\mp a(2r-b)}{\sqrt{\Delta}\left(r^2\pm a\sqrt{r-b}\right)}\, . \label{exact}
\end{equation}
Concentrating attention to the case of the mining KN spacetimes and the motion in the mining regime, we obtain a simple zero-approximation relation 
\begin{equation}
\mathcal{V}^{(\phi)}_{\mathrm{LNRF,m}} \sim \frac{{|r_{\mathrm{ph(s)}}-1|}}{r_{\mathrm{ph(s)}}-1}\, , 
\end{equation}
where $r_{\mathrm{ph(s)}} < 1$ denotes the radius of the stable photon geodesic -- as expected, the LNRF velocity of such geodesic circular observers has to be negative (the observers are counter-rotating relative to LNRFs) and close to velocity of light, as these observers are located close to the stable circular geodesic; in the zero-approximation given above, the velocity is equal to the velocity of light. We thus can expect strong focusing of light rays along the direction of the orbital motion. 

\begin{figure}[ht]
\begin{center}
\includegraphics[width=\linewidth]{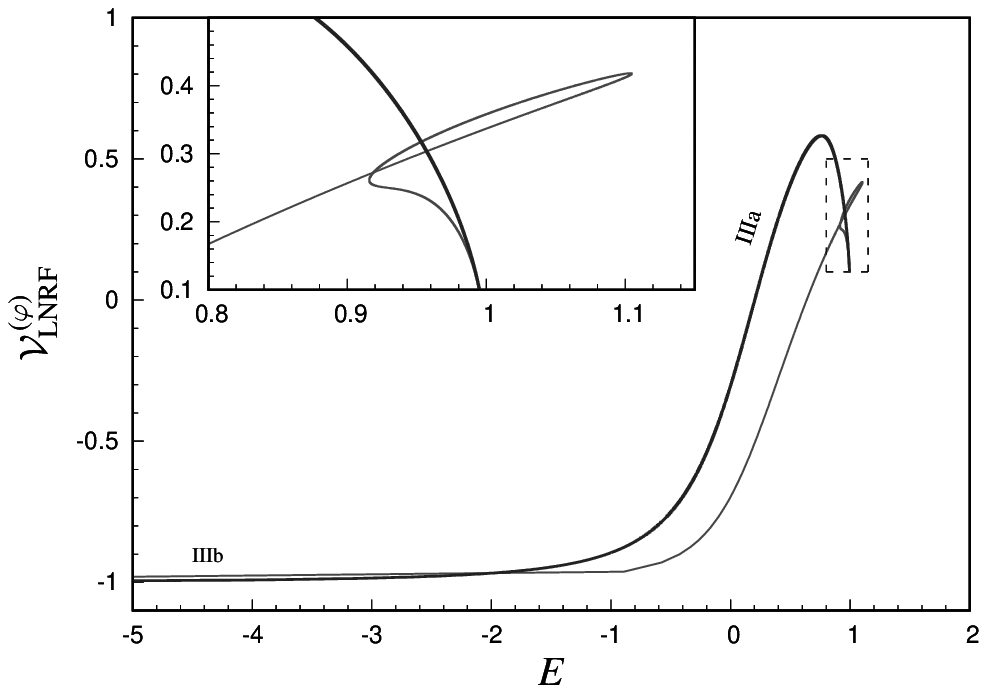}
\caption{\label{VE} The velocity of the observers (frames) orbiting along stable circular geodesics as related to the LNRFs. The CGF velocity is determined for the mining KN naked singularity spacetime of class IIIa ($a=0.7, b=0.6$), and the related spacetime of class IIIb ($a=5, b=0.6$).}
\end{center}
\end{figure}

We give the CGF velocity relative to the LNRFs, exactly calculated according to Eq. (\ref{exact}), in Fig. \ref{VE} for a mining KN naked singularity spacetime of class IIIa and the related KN spacetime of class IIIb, in dependence on the specific energy $E$. 

In order to obtain the first-order correction to the zeroth approximation of the CGF velocity, it is useful to express for the circular geodesics in the mining regime the dependence of the extremely small variation of the circular geodesic radius $\Delta r$ on the specific energy $E$ of the orbit. We thus assume $r=r_{\mathrm{ph}}+\Delta r$, where $\Delta r \ll 1$ satisfies the mining regime conditions. In the linear approximation, we then arrive to the formula relating $\Delta r$ and the specific energy $E$ taking the form  
\begin{equation}
\Delta r = \frac{\left(1-r_{\mathrm{ph}}\right)\left(r_{\mathrm{ph}}-b\right)}{2 E^2 \left(4b-3 r_{\mathrm{ph}}\right)}\, .
\end{equation}
Clearly, the following conditions have to be satisfied for its validity: 
\begin{enumerate}
\item $r_{\mathrm{ph}}<1\ ,$
\item $r_{\mathrm{ph}}>b\, ,$
\item $r_{\mathrm{ph}}<4b/3\, .$
\end{enumerate}
The first condition is well known -- the maximum values of $r_{\mathrm{ph}}$ occur at the mining KN spacetimes with $b=1$ where $r_{\mathrm{phmax}}=1$. The second condition represents the general lower limit on existence of circular geodesics in the KN spacetimes. The third condition is related to the behavior of the radial profile of the angular velocity of circular geodesics. The first-order approximation in terms of $E^2$ as the parameter (or equivalently $\Delta r$ as parameter) of the velocity of the CGF observers relative to the LNRFs can be then given as follows 
\begin{eqnarray}
&&\mathcal{V}^{(\phi)}_{\mathrm{LNRF,m}} \sim \frac{{|r_{\mathrm{ph}}-1|}}{r_{\mathrm{ph}}-1}\\&+&\frac{m^2 \left(2 b^2 - 7 b r_{\mathrm{ph}} + (6 + b) r_{\mathrm{ph}}^2 - 3 r_{\mathrm{ph}}^3 + r_{\mathrm{ph}}^4\right)}{2E^2 r_{\mathrm{ph}}^2(r_{\mathrm{ph}}^2 +3 r_{\mathrm{ph}} -2 b)}\, . \nonumber 
\end{eqnarray}
This formula is correctly behaving as at $r_{\mathrm{ph}}$ there is $E^2 \rightarrow \infty$, the first order term is positive, lowering thus the magnitude of the CGF velocity below the light velocity. 

\subsection{Circular geodesic frames}

The observers orbiting on the circular geodesics of the braneworld KN spacetimes can be equipped with the CGF tetrade that can be created from the LNRF tetrade by the Lorentz boost with the axial CGF velocity calculated above. The CGF tetrade then takes the form (see, e.g., \cite{Stu-Sche:2010:CLAQG:})  
\begin{eqnarray}
	\mathbf{\omega}^{(t)}&=&\frac{r^2-2r+b\pm F}{Z_\pm}\,\mathbf{\omega}^t\mp\frac{(r^2+a^2)F\mp a(2r-b)}{Z_{\pm}}\,\mathbf{\omega}^\phi\, ,\nonumber\\
	\mathbf{\omega}^{(\phi)}&=&\mp\frac{\sqrt{\Delta}F}{Z_{\pm}}\,\mathbf{\omega}+\frac{\sqrt{\Delta}(r^2\pm a F)}{Z_{\pm}}\, ,\nonumber\\
	\mathbf{\omega}^{(r)}&=&\sqrt{\frac{\Sigma}{\Delta}}\,\mathbf{\omega}^r\, ,\nonumber\\
	\mathbf{\omega}^{(\theta)}&=&\sqrt{\Sigma}\,\mathbf{\omega}^\theta\, ,
\end{eqnarray}
where 
\begin{equation}
	F\equiv \sqrt{r-b}
\end{equation}
and 
\begin{equation}	
	Z_{\pm}=r\sqrt{r^2-3r+2b\pm 2a F}.
\end{equation}
The upper (lower) sign refers to upper (lower) family circular geodesic frames; recall that the lower family orbits are purely counter-rotating. We consider in the following only the upper family circular geodesics. 

\subsection{Construction of the local Sky}

We construct the Sky image related to the CMB radiation by using the approach developed in \cite{Sche:Stu:2009b:,Stu-Sche:2010:CLAQG:}, using photons representing all the higher order images. We briefly summarize the main idea of this approach. 

The position of an image on the observer Sky is characterized by the observing angles $\theta_\mathrm{o}, \phi_\mathrm{o}$. The photon 4-momentum relative to the CGF reads 
\begin{equation}
	\mathbf{k}=k_{(t)}\mathbf{\omega}^{(t)} + k_{(r)}\mathbf{\omega}^{(r)}+k_{(\theta)}\mathbf{\omega}^{(\theta)} + k_{(\phi)}\mathbf{\omega}^{(\phi)}.
\end{equation}
The two angles $(\theta_{\mathrm{o}},\phi_{\mathrm{o}})$ that identify an image on the local sky of the observer can be expressed by the formulae
\begin{eqnarray}
	k^{(t)}&=&-k_{(t)}=1,\\
	k^{(r)}&=&k_{(r)}=\sin\theta_\mathrm{o}\cos\phi_\mathrm{o},\\
	k^{(\theta)}&=&k_{(\theta)}=\sin\theta_\mathrm{o}\sin\phi_\mathrm{o},\\
	k^{(\phi)}&=&k_{(\phi)}=\cos\theta_\mathrm{o}.
\end{eqnarray}
For each double $(\theta_\mathrm{o},\phi_\mathrm{o})$ one can determine the corresponding impact parameters of the photon using the formulae 
\begin{equation}
	\lambda \equiv \frac{l}{\gamma},\quad \eta\equiv\frac{k-(l-a \gamma)^2}{\gamma^2}, 
\end{equation}
and we can express the motion constants $l$, $\gamma$ and $k$ by locally measured components of the photon 4-vector $\mathbf{k}$. They read 
\begin{eqnarray}
	l &\equiv &k_\phi=\omega^{(t)}_\phi k^{(t)}+\omega^{(r)}_\phi k^{(r)}+\omega^{(\theta)}_\phi k^{(\theta)}+\omega^{(\phi)}_\phi k^{(\phi)},\nonumber\\
	\gamma&\equiv &-k_t = \omega^{(t)}_t k^{(t)}+\omega^{(r)}_t k^{(r)}+\omega^{(\theta)}_t k^{(\theta)}+\omega^{(\phi)}_t k^{(\phi)},\nonumber\\
	k&= &\frac{1}{\Delta}\left[(\gamma(r^2+a^2)-a l)^2-\Sigma\Delta (k^{(r)})^2\right].
\end{eqnarray}
The next step is to integrate equations of motion to infinity (large distance) to find the intersection of the photon geodesics with the global sky, obtaining corresponding $\theta_\mathrm{s}$ and $\phi_\mathrm{s}$ coordinates corresponding to the source (CMB radiation in the present case), and taking into account the focusing and frequency shift effects. This way we obtain a mapping $(\theta_\mathrm{o},\phi_\mathrm{o})\rightarrow (\theta_\mathrm{s},\phi_\mathrm{s})$ which is used to map a point from global sky to the local one -- for details see \cite{Sche:Stu:2009b:,Stu-Sche:2010:CLAQG:,Stu-Sche:2013:CLAQG:}. 

We have constructed the local sky snapshots for three representative central object characterized by the spin $a$ and the tidal charge $b$. For comparison we first give the standard known images of the CMB created for CGF at close vicinity of the KN black holes in Fig. \ref{skyBH} -- these are of the same character as those obtained for the Kerr black holes (see, e.g., \cite{James-etal:2015:CLAQG:}). We can observe the standard increasing deformation of the flat-space sky as the observer circular geodesic approaches the black hole horizon. The case of so called effectively ordinary KN naked singularities is represented in Fig. \ref{skyNS} demonstrating similarities to the black hole case when the radius of the orbit is small, and the differences demonstrated by existence of some part of the image of the sky in the central dark region, if the orbit radius is large enough. Finally, we represent evolution of the CGF sky in the case of the mining KN naked singularities in Fig. \ref{skyMNS}, demonstrating the crucial differences related to the CGF motion in the mining regime where the local sky shrinks to small circular object with radius decreasing as the circular geodesic approaches the stable photon circular geodesic.

\section{Conclusions}

We considered collisional processes and optical phenomena in the braneworld KN spacetimes, concentrating our attention to the newly discovered \uvozovky{mining} KN naked singularity spacetimes demonstrating possibility of a special, mining regime of accretion processes with very large efficiency and non-standard radial profile of angular velocity having $\mathrm{d}\Omega/\mathrm{d}r > 0$. The vanishing of the angular velocity gradient, $\mathrm{d}\Omega/\mathrm{d}r = 0$, occurs at $r=4b/3$ independently of spin $a$; implications of this effect are discussed in \cite{Stu-Sche:2014:CLAQG:}.  

We have demonstrated explicitly that the ultra-high CM energy collisional processes can be obtained in whole the family of Class IIIa braneworld mining KN naked singularity spacetimes for quite generic incoming particles, if the collisions occur with particles orbiting in the mining regime near the stable photon circular geodesic. This is situation contrasting the collisions in the ordinary KN (Kerr) naked singularity spacetimes where the ultra-high energy collisions are possible at the special radius $r=1$, in the near-extreme naked singularity spacetimes with $a^2+b \sim 1$, but for a relatively wide range of the motion constants of the colliding particles. On the contrary, in the near-extreme KN black hole spacetimes the extremely large collisional energy can be obtained only for finely tuned motion constants of particles approaching from infinity. 

In the braneworld KN naked singularity spacetimes the ultra-high energy collisional processes could be quite frequent as these are not confined to a special choice of spacetime parameters, or the motion constants of colliding particles. 

We have also demonstrated interesting optical phenomena related to the mining KN naked singularity spacetimes, especially in the so called mining regime when an observer is orbiting very close to the limiting stable photon circular geodesic. Then the observer motion is ultra-relativistic that causes an enormous energy supply from the CMB radiation due to enormous blue-shift of the radiation; however, the observer local sky is strongly shrinking to a small circular region in the direction of its motion. Due to this effect, the mining KN naked singularities (superspinars) could be even more efficient sources of energy for orbiting planets in the cold accelerating Universe in comparison with the standard case of planets orbiting black holes that was discussed in \cite{Tho,James-etal:2015:CLAQG:,Tho-etal:2015:,O}. Of course, our results give only indications for such a possibility, requiring more detailed and extended studies. 

\section*{Acknowledgements}

ZS, MB, and JS are supported by the Albert Einstein Centre for Gravitation and Astrophysics financed by the Czech Science Agency Grant No. 14-37086G and by the Silesian University at Opava grant SGS/14/2016. 



\section*{References}

\begin{thebibliography}{10}

\bibitem{Ali-Gal:1981:}
A.~N. {Aliev} and D.~V. {Galtsov}.
\newblock {Radiation from relativistic particles in nongeodesic motion in a
  strong gravitational field}.
\newblock {\em General Relativity and Gravitation}, 13:899--912, October 1981.

\bibitem{Ali-Gum:2005:CLAQG:}
A.~N. {Aliev} and A.~E. {G{\"u}mr{\"u}k{\c c}{\"u}o{\v g}lu}.
\newblock {Charged rotating black holes on a 3-brane}.
\newblock {\em Physical Review D}, 71(10):104027, May 2005.

\bibitem{Ali-Tal:2009:prd:}
A.~N. {Aliev} and P.~{Talazan}.
\newblock {Gravitational effects of rotating braneworld black holes}.
\newblock {\em Physical Review D}, 80(4):044023, August 2009.

\bibitem{Eir-Ama:2012:PHYSR4:}
L.~{Amarilla} and E.~F. {Eiroa}.
\newblock {Shadow of a rotating braneworld black hole}.
\newblock {\em Physical Review D}, 85(6):064019, March 2012.

\bibitem{Ark:Dim:Dva:1998:}
N.~{Arkani-Hamed}, S.~{Dimopoulos}, and G.~{Dvali}.
\newblock {The hierarchy problem and new dimensions at a millimeter}.
\newblock {\em Physics Letters B}, 429:263--272, June 1998.

\bibitem{Bobo:2013:}
F.~{Atamurotov}, B.~{Ahmedov}, and S.~{Shaymatov}.
\newblock {Formation of black holes through BSW effect and black hole-black
  hole collisions}.
\newblock {\em Astrophysics and Space Science}, 347:277--281, October 2013.

\bibitem{Ban-Sil-Wes:2009:}
M.~{Ba{\~n}ados}, J.~{Silk}, and S.~M. {West}.
\newblock {Kerr Black Holes as Particle Accelerators to Arbitrarily High
  Energy}.
\newblock {\em Physical Review Letters}, 103(11):111102, September 2009.

\bibitem{Bal:Haw:1998:}
S.~A. {Balbus} and J.~F. {Hawley}.
\newblock {Instability, turbulence, and enhanced transport in accretion disks}.
\newblock {\em Reviews of Modern Physics}, 70:1--53, January 1998.

\bibitem{Bal-Bic-Stu:1989:BAC:}
V.~{Balek}, J.~{Bi{\v{c}}{\'{a}}k}, and Z.~{Stuchl{\'{\i}}k}.
\newblock {The motion of the charged particles in the field of rotating charged
  black holes and naked singularities. II - The motion in the equatorial
  plane}.
\newblock {\em Bulletin of the Astronomical Institutes of Czechoslovakia},
  40:133--165, June 1989.

\bibitem{Bao-Stu:1992:ApJ:}
G.~{Bao} and Z.~{Stuchl{\'{\i}}k}.
\newblock {Accretion disk self-eclipse - X-ray light curve and emission line}.
\newblock {\em The Astrophysical Journal}, 400:163--169, November 1992.

\bibitem{Bar:1973:BlaHol:}
J.~M. {Bardeen}.
\newblock {Timelike and null geodesics in the Kerr metric.}
\newblock In C.~{Dewitt} and B.~S. {Dewitt}, editors, {\em Black Holes (Les
  Astres Occlus)}, pages 215--239, 1973.

\bibitem{Bar-Pre-Teu:1972:ApJ:}
J.~M. {Bardeen}, W.~H. {Press}, and S.~A. {Teukolsky}.
\newblock {Rotating Black Holes: Locally Nonrotating Frames, Energy Extraction,
  and Scalar Synchrotron Radiation}.
\newblock {\em Astrophysical Journal}, 178:347--370, December 1972.

\bibitem{Nun:2010:PHYSR4:}
A.~Y. {Bin-Nun}.
\newblock {Relativistic images in Randall-Sundrum II braneworld lensing}.
\newblock {\em Physical Review D}, 81(12):123011, June 2010.

\bibitem{Bic-Stu:1976:BAC:}
J.~{Bi{\v{c}}{\'{a}}k} and Z.~{Stuchl{\'{\i}}k}.
\newblock {On the latitudinal and radial motion in the field of a rotating
  black hole}.
\newblock {\em Bulletin of the Astronomical Institutes of Czechoslovakia},
  27:129--133, 1976.

\bibitem{Bic-Stu:1976:MONNRS:}
J.~{Bi{\v{c}}{\'{a}}k} and Z.~{Stuchl{\'{\i}}k}.
\newblock {The fall of the shell of dust on to a rotating black hole}.
\newblock {\em Monthly Notices of the Royal Astronomical Society},
  175:381--393, May 1976.

\bibitem{Bic-Stu-Bal:1989:BAC:}
J.~{Bi{\v{c}}{\'{a}}k}, Z.~{Stuchl{\'{\i}}k}, and V.~{Balek}.
\newblock {The motion of charged particles in the field of rotating charged
  black holes and naked singularities}.
\newblock {\em Bulletin of the Astronomical Institutes of Czechoslovakia},
  40:65--92, March 1989.

\bibitem{Bla-Stu:2016:PHYSR4:}
M.~{Blaschke} and Z.~{Stuchl{\'{\i}}k}.
\newblock {Efficiency of the Keplerian accretion in braneworld Kerr-Newman
  spacetimes and mining instability of some naked singularity spacetimes}.
\newblock {\em Physical Review D}, 94(8):086006, October 2016.

\bibitem{Boh-Har-Lob:2008:CLAQG:}
C.~G. {B{\"o}hmer}, T.~{Harko}, and F.~S.~N. {Lobo}.
\newblock {Solar system tests of brane world models}.
\newblock {\em Classical and Quantum Gravity}, 25(4):045015, February 2008.

\bibitem{Cal-Nob:1979:NUOC2:}
M.~{Calvani} and L.~{Nobili}.
\newblock {Dressing up a Kerr naked singularity}.
\newblock {\em Nuovo Cimento B Serie}, 51:247--261, June 1979.

\bibitem{Car:1968:PRD:}
B.~{Carter}.
\newblock {Global Structure of the Kerr Family of Gravitational Fields}.
\newblock {\em Physical Review}, 174:1559--1571, October 1968.

\bibitem{Car:1973:BlaHol:}
B.~{Carter}.
\newblock {Black hole equilibrium states.}
\newblock In C.~{Dewitt} and B.~S. {Dewitt}, editors, {\em Black Holes (Les
  Astres Occlus)}, pages 57--214, 1973.

\bibitem{Dad-Kal:1977:}
N.~{Dadhich} and P.~P. {Kale}.
\newblock {Equatorial circular geodesics in the Kerr-Newman geometry.}
\newblock {\em Journal of Mathematical Physics}, 18:1727--1728, 1977.

\bibitem{Dad:2000:}
N.~{Dadhich}, R.~{Maartens}, P.~{Papadopoulos}, and V.~{Rezania}.
\newblock {Black holes on the brane}.
\newblock {\em Physics Letters B}, 487:1--6, August 2000.

\bibitem{deF:1974:aap}
F.~{de Felice}.
\newblock {Repulsive Phenomena and Energy Emission in the Field of a Naked
  Singularity}.
\newblock {\em Astronomy and Astrophysics}, 34:15, August 1974.

\bibitem{Ger:Maa:2001:}
C.~{Germani} and R.~{Maartens}.
\newblock {Stars in the braneworld}.
\newblock {\em Physical Review D}, 64(12):124010, December 2001.

\bibitem{Gib-Haw:1977:PRD:}
G.~W. {Gibbons} and S.~W. {Hawking}.
\newblock {Cosmological event horizons, thermodynamics, and particle creation}.
\newblock {\em Physical Review D}, 15:2738--2751, May 1977.

\bibitem{Gim-Hor:2009:PhysLetB:}
E.~G. {Gimon} and P.~{Ho{\v r}ava}.
\newblock {Astrophysical violations of the Kerr bound as a possible signature
  of string theory}.
\newblock {\em Physics Letters B}, 672:299--302, February 2009.

\bibitem{Gri:Pav:2013:}
A.~A. {Grib} and Y.~V. {Pavlov}.
\newblock {On the energy of particle collisions in the ergosphere of the
  rotating black holes}.
\newblock {\em EPL (Europhysics Letters)}, 101:20004, January 2013.

\bibitem{Har-Kim:2011:}
T.~{Harada} and M.~{Kimura}.
\newblock {Collision of an object in the transition from adiabatic inspiral to
  plunge around a Kerr black hole}.
\newblock {\em Physical Review D}, 84(12):124032, December 2011.

\bibitem{Har:Kim:2014}
T.~{Harada} and M.~{Kimura}.
\newblock {Black holes as particle accelerators: a brief review}.
\newblock {\em Classical and Quantum Gravity}, 31(24):243001, December 2014.

\bibitem{Har-etal:2012:}
T.~{Harada}, H.~{Nemoto}, and U.~{Miyamoto}.
\newblock {Upper limits of particle emission from high-energy collision and
  reaction near a maximally rotating Kerr black hole}.
\newblock {\em Physical Review D}, 86(2):024027, July 2012.

\bibitem{Hla-Stu:2011:JCAP:}
J.~{Hlad{\'{\i}}k} and Z.~{Stuchl{\'{\i}}k}.
\newblock {Photon and neutrino redshift in the field of braneworld compact
  stars}.
\newblock {\em Journal of Cosmology and Astroparticle Physics}, 7:012, July
  2011.

\bibitem{Hor:Wit:1996b:}
P.~{Ho{\v r}ava} and E.~{Witten}.
\newblock {Eleven-dimensional supergravity on a manifold with boundary}.
\newblock {\em Nuclear Physics B}, 475:94--114, February 1996.

\bibitem{Hor:Wit:1996:}
P.~{Ho{\v r}ava} and E.~{Witten}.
\newblock {Heterotic and Type I string dynamics from eleven dimensions}.
\newblock {\em Nuclear Physics B}, 460:506--524, February 1996.

\bibitem{James-etal:2015:CLAQG:}
O.~{James}, E.~{von Tunzelmann}, P.~{Franklin}, and K.~S. {Thorne}.
\newblock {Gravitational lensing by spinning black holes in astrophysics, and
  in the movie Interstellar}.
\newblock {\em Classical and Quantum Gravity}, 32(6):065001, March 2015.

\bibitem{Tho-etal:2015:}
O.~{James}, E.~{von Tunzelmann}, P.~{Franklin}, and K.~S. {Thorne}.
\newblock {Visualizing Interstellar's Wormhole}.
\newblock {\em American Journal of Physics}, 83:486--499, June 2015.

\bibitem{Kot-Stu-Tor:2008:CLAQG:}
A.~{Kotrlov{\'a}}, Z.~{Stuchl{\'{\i}}k}, and G.~{T{\"o}r{\"o}k}.
\newblock {Quasiperiodic oscillations in a strong gravitational field around
  neutron stars testing braneworld models}.
\newblock {\em Classical and Quantum Gravity}, 25(22):225016, November 2008.

\bibitem{Kra:2005:CLAQG:}
G.~V. {Kraniotis}.
\newblock {Frame dragging and bending of light in Kerr and Kerr (anti) de
  Sitter spacetimes}.
\newblock {\em Classical and Quantum Gravity}, 22:4391--4424, November 2005.

\bibitem{Kra:2007:CLAQG:}
G.~V. {Kraniotis}.
\newblock {Periapsis and gravitomagnetic precessions of stellar orbits in Kerr
  and Kerr de Sitter black hole spacetimes}.
\newblock {\em Classical and Quantum Gravity}, 24:1775--1808, April 2007.

\bibitem{Kra:2014:GRG:}
G.~V. {Kraniotis}.
\newblock {Gravitational lensing and frame dragging of light in the Kerr-Newman
  and the Kerr-Newman (anti) de Sitter black hole spacetimes}.
\newblock {\em General Relativity and Gravitation}, 46:1818, November 2014.

\bibitem{Gravitation}
C.~W. {Misner}, K.~S. {Thorne}, and J.~A. {Wheeler}.
\newblock {\em {Gravitation}}.
\newblock {San Francisco: W.H.~Freeman and Co., 1973}, 1973.

\bibitem{Nak-etal:2017:}
K.-i. {Nakao}, P.~S. {Joshi}, J.-Q. {Guo}, P.~{Kocherlakota}, H.~{Tagoshi},
  T.~{Harada}, M.~{Patil}, and A.~{Krolak}.
\newblock {On the stability of a superspinar}.
\newblock {\em ArXiv e-prints}, July 2017.

\bibitem{Note1}
Formally the same results are relevant for the KN spacetimes \cite
  {Ali-Gal:1981:}.

\bibitem{Note2}
In the KN black hole spacetimes this condition cannot be satisfied for
  colliding particles incoming in the equatorial plane from infinity, as they
  must be inward moving in close vicinity of the outer black hole horizon;
  turning point of their radial motion has to be located above the photon
  circular geodesic. The outward directed colliding particles can be obtained
  only in the so called cascade collisional processes \cite {Har-Kim:2011:}.

\bibitem{Note3}
Here we correct the misprint presented in \cite {Sche:Stu:2009:}.

\bibitem{O}
T.~{Opatrn{\' y}}, L.~{Richterek}, and P.~{Bakala}.
\newblock {Life under a black sun}.
\newblock {\em {American Journal of Physics}}, 85:14--22, January 2017.

\bibitem{Pat-etal:2016:}
M.~{Patil}, T.~{Harada}, K.-i. {Nakao}, P.~S. {Joshi}, and M.~{Kimura}.
\newblock {Infinite efficiency of the collisional Penrose process: Can a
  overspinning Kerr geometry be the source of ultrahigh-energy cosmic rays and
  neutrinos?}
\newblock {\em Physical Review D}, 93(10):104015, May 2016.

\bibitem{Pat-Jos:2011:}
M.~{Patil} and P.~{Joshi}.
\newblock {Kerr naked singularities as particle accelerators}.
\newblock {\em Classical and Quantum Gravity}, 28(23):235012, December 2011.

\bibitem{Pat-Jos:2010:}
M.~{Patil} and P.~S. {Joshi}.
\newblock {Naked singularities as particle accelerators}.
\newblock {\em Physical Review D}, 82(10):104049, November 2010.

\bibitem{Pug-Que-Ruf:2011:PHYSR4:}
D.~{Pugliese}, H.~{Quevedo}, and R.~{Ruffini}.
\newblock {Equatorial circular motion in Kerr spacetime}.
\newblock {\em Physical Review D}, 84(4):044030, August 2011.

\bibitem{Pug:Que:Ruf:2013:}
D.~{Pugliese}, H.~{Quevedo}, and R.~{Ruffini}.
\newblock {Equatorial circular orbits of neutral test particles in the
  Kerr-Newman spacetime}.
\newblock {\em Physical Review D}, 88(2):024042, July 2013.

\bibitem{Ran:Sun:1999b:}
L.~{Randall} and R.~{Sundrum}.
\newblock {An Alternative to Compactification}.
\newblock {\em Physical Review Letters}, 83:4690--4693, December 1999.

\bibitem{Rem:McC:2006:ARAA:}
R.~A. {Remillard} and J.~E. {McClintock}.
\newblock {X-Ray Properties of Black-Hole Binaries}.
\newblock {\em Annual Review of Astronomy and Astrophysics}, 44:49--92,
  September 2006.

\bibitem{Sche:Stu:2009b:}
J.~{Schee} and Z.~{Stuchl{\'{\i}}k}.
\newblock {Optical Phenomena in the Field of Braneworld Kerr Black Holes}.
\newblock {\em International Journal of Modern Physics D}, 18:983--1024, 2009.

\bibitem{Sche:Stu:2009:}
J.~{Schee} and Z.~{Stuchl{\'{\i}}k}.
\newblock {Profiles of emission lines generated by rings orbiting braneworld
  Kerr black holes}.
\newblock {\em General Relativity and Gravitation}, 41:1795--1818, August 2009.

\bibitem{Schee:Stu:2015:}
J.~{Schee} and Z.~{Stuchl{\'{\i}}k}.
\newblock {Gravitational lensing and ghost images in the regular Bardeen
  no-horizon spacetimes}.
\newblock {\em Journal of Cosmology and Astroparticle Physics}, 6:048, June
  2015.

\bibitem{Sha-etal:2013:}
S.~R. Shaymatov, B.~J. Ahmedov, and A.~A. Abdujabbarov.
\newblock {Particle acceleration near a rotating black hole in a
  Randall-Sundrum brane with a cosmological constant}.
\newblock {\em Phys. Rev. D}, 88:024016, Jul 2013.

\bibitem{Shi:Mae:Sas:2000:}
T.~{Shiromizu}, K.-I. {Maeda}, and M.~{Sasaki}.
\newblock {The Einstein equations on the 3-brane world}.
\newblock {\em Physical Review D}, 62(2):024012, July 2000.

\bibitem{Stu:1980:BAC:}
Z.~{Stuchl{\'{\i}}k}.
\newblock {Equatorial circular orbits and the motion of the shell of dust in
  the field of a rotating naked singularity}.
\newblock {\em Bulletin of the Astronomical Institutes of Czechoslovakia},
  31:129--144, 1980.

\bibitem{Stu:1981b:BAC}
Z.~{Stuchl{\'{\i}}k}.
\newblock {Evolution of Kerr naked singularities}.
\newblock {\em Bulletin of the Astronomical Institutes of Czechoslovakia},
  32:68--72, 1981.

\bibitem{Stu:1981:}
Z.~{Stuchl{\'{\i}}k}.
\newblock {The radial motion of photons in Kerr metric}.
\newblock {\em Bulletin of the Astronomical Institutes of Czechoslovakia},
  32:40--52, 1981.

\bibitem{Stu:1983:BAC:}
Z.~{Stuchl{\'{\i}}k}.
\newblock {The motion of test particles in black-hole backgrounds with non-zero
  cosmological constant}.
\newblock {\em Bulletin of the Astronomical Institutes of Czechoslovakia},
  34:129--149, March 1983.

\bibitem{Stu-etal:1998:PHYSR4:}
Z.~{Stuchl{\'{\i}}k}, G.~{Bao}, E.~{{\O}stgaard}, and S.~{Hled{\'{\i}}k}.
\newblock {Kerr-Newman-de Sitter black holes with a restricted repulsive
  barrier of equatorial photon motion}.
\newblock {\em Physical Review D}, 58(8):084003, October 1998.

\bibitem{Stu-Bic-Bal:1999:GRG:}
Z.~{Stuchl{\'{\i}}k}, J.~{Bi{\v{c}}{\'{a}}k}, and V.~{Balek}.
\newblock {The Shell of Incoherent Charged Matter Falling onto a Charged
  Rotating Black Hole}.
\newblock {\em General Relativity and Gravitation}, 31:53--71, January 1999.

\bibitem{Stu:Cal:1991:}
Z.~{Stuchl{\'{\i}}k} and M.~{Calvani}.
\newblock {Null geodesics in black hole metrics with non-zero cosmological
  constant}.
\newblock {\em General Relativity and Gravitation}, 23:507--519, May 1991.

\bibitem{Stu-Hla-Urb:2012:GRG:}
Z.~{Stuchl{\'{\i}}k}, J.~{Hlad{\'{\i}}k}, and M.~{Urbanec}.
\newblock {Trapping of Neutrinos in Extremely Compact Stars and the Influence
  of Brane Tension on This Process}.
\newblock In {\em Twelfth Marcel Grossmann Meeting on General Relativity}, page
  955, 2012.

\bibitem{Stu-Hle:1999:PHYSR4:}
Z.~{Stuchl{\'{\i}}k} and S.~{Hled{\'{\i}}k}.
\newblock {Some properties of the Schwarzschild-de Sitter and
  Schwarzschild-anti-de Sitter spacetimes}.
\newblock {\em Physical Review D}, 60(4):044006, August 1999.

\bibitem{Stu:Hle:2000:}
Z.~{Stuchl{\'{\i}}k} and S.~{Hled{\'{\i}}k}.
\newblock {Equatorial photon motion in the Kerr-Newman spacetimes with a
  non-zero cosmological constant}.
\newblock {\em Classical and Quantum Gravity}, 17:4541--4576, November 2000.

\bibitem{Stu-Hle:2002:ActaPhysSlov:}
Z.~{Stuchl{\'{\i}}k} and S.~{Hled{\'{\i}}k}.
\newblock {Properties of the Reissner-Nordstr{\" o}m Spacetimes with a Nonzero
  Cosmological Constant}.
\newblock {\em Acta Physica Slovaca}, {52}({5}):{363--407}, {oct} {2002}.

\bibitem{Stu:Hle:Tru:2011:}
Z.~{Stuchl{\'{\i}}k}, S.~{Hled{\'{\i}}k}, and K.~{Truparov{\'a}}.
\newblock {Evolution of Kerr superspinars due to accretion counterrotating thin
  discs}.
\newblock {\em Classical and Quantum Gravity}, 28(15):155017, August 2011.

\bibitem{Stu-Kot:2009:GRG:}
Z.~{Stuchl{\'{\i}}k} and A.~{Kotrlov{\'a}}.
\newblock {Orbital resonances in discs around braneworld Kerr black holes}.
\newblock {\em General Relativity and Gravitation}, 41:1305--1343, June 2009.

\bibitem{Stu-Sche:2010:CLAQG:}
Z.~{Stuchl{\'{\i}}k} and J.~{Schee}.
\newblock {Appearance of Keplerian discs orbiting Kerr superspinars}.
\newblock {\em Classical and Quantum Gravity}, 27(21):215017, November 2010.

\bibitem{Stu-Sche:2012:CLAQG:}
Z.~{Stuchl{\'{\i}}k} and J.~{Schee}.
\newblock {Observational phenomena related to primordial Kerr superspinars}.
\newblock {\em Classical and Quantum Gravity}, 29(6):065002, March 2012.

\bibitem{Stu-Sche:2013:CLAQG:}
Z.~{Stuchl{\'{\i}}k} and J.~{Schee}.
\newblock {Ultra-high-energy collisions in the superspinning Kerr geometry}.
\newblock {\em Classical and Quantum Gravity}, 30(7):075012, April 2013.

\bibitem{Stu-Sche:2014:CLAQG:}
Z.~{Stuchl{\'{\i}}k} and J.~{Schee}.
\newblock {Optical effects related to Keplerian discs orbiting Kehagias-Sfetsos
  naked singularities}.
\newblock {\em Classical and Quantum Gravity}, 31(19):195013, October 2014.

\bibitem{Stu-Sche:2015:IJMPD:}
Z.~{Stuchl{\'{\i}}k} and J.~{Schee}.
\newblock {Circular geodesic of Bardeen and Ayon-Beato-Garcia regular
  black-hole and no-horizon spacetimes}.
\newblock {\em International Journal of Modern Physics D}, 24:1550020--289,
  December 2015.

\bibitem{Tho}
K.~Thorne and C.~Nolan.
\newblock {\em The Science of Interstellar}.
\newblock W. W. Norton, 2014.

\bibitem{Zas:2013:}
O.~B. {Zaslavskii}.
\newblock {High energy collisions of particles inside ergosphere: general
  approach}.
\newblock {\em ArXiv e-prints}, January 2013.

\bibitem{Zas:2014:}
O.~B. {Zaslavskii}.
\newblock {Ultrahigh energy particle collisions near the black hole horizon in
  the strong magnetic field}.
\newblock {\em Modern Physics Letters A}, 29:1450112, June 2014.

\bibitem{Zas:2015:}
O.~B. {Zaslavskii}.
\newblock {Near-horizon circular orbits and extremal limit for dirty rotating
  black holes}.
\newblock {\em Physical Review D}, 92(4):044017, August 2015.

\bibitem{Zio:2005:}
J.~{Zi{\'o}kowski}.
\newblock {Galactic collapsed objects}.
\newblock {\em Nuovo Cimento B Serie}, 120:757, June 2005.

\end{thebibliography}

\end{document}